\documentclass[11pt]{article}
\usepackage[utf8]{inputenc}
\usepackage{tensor}
\usepackage{bbm}
\usepackage{amssymb}
\usepackage{authblk}
\usepackage{bbold}
\usepackage{xcolor}
\usepackage{graphicx}
\usepackage{jheppub}
\usepackage{bm}
\newcommand{\p}{\partial}
\usepackage[multiple]{footmisc}
\usepackage{subcaption}
\usepackage{multicol}
\usepackage{multirow}
\usepackage{rotating}
\usepackage{array}
\usepackage{longtable}
\usepackage{booktabs}

\title{Gepner-like boundary states on $T^4$}
\author[a]{Martin Schnabl,} 
\author[a,b]{Jakub Vo\v{s}mera
} 
\affiliation[a]{CEICO, Institute of Physics of the Czech Academy of Sciences,\\ Na Slovance 2, 182 21 Prague 8, Czech~Republic} \affiliation[b]{Institute of Particle and Nuclear Physics, Charles University,\\ V Hole\v{s}ovi\v{c}k\'{a}ch 2, 180 00 Prague 8, Czech~Republic} 
\emailAdd{schnabl.martin@gmail.com} \emailAdd{vosmera@gmail.com} 
\abstract{We present exact expressions for elementary boundary states which describe \hbox{D-branes} preserving 16 or fewer supercharges in type II superstring compactified on certain self-dual 4-tori. 
While being manifestly superconformal, our boundary states are not a priori required to satisfy the usual free-field gluing conditions along the internal directions of the 4-tori. 
Our calculations 
proceed along the lines of Gepner's construction by recasting the $\mathcal{N}=(2,2)$ worldsheet 
sigma model on the 4-tori in terms of $\mathcal{N}=2$ minimal models. 
Imposing general permutation gluing conditions on the $\mathcal{N}=(2,2)$ generators is shown to yield various stable and unstable D-branes, where the stable ones include the known 1/2-BPS and 1/4-BPS bound states of D$p$-branes, as well as new non-BPS D-branes, which do not carry RR charges.
}
\keywords{boundary states, worldsheet approach to D-branes, bound states of D$p$-branes, stable non-BPS D-branes, permutation D-branes}
\arxivnumber{}

\begin{document}
 \maketitle
\flushbottom

\section{Introduction and summary}

It is well known \cite{Polchinski:1994fq,Polchinski:1995mt} that D-branes in type II superstring are described (at weak string coupling) by superconformal boundary states. These satisfy the gluing conditions 
\begin{subequations}\label{eq:glC}
\begin{align}
(L_n - \overline{L}_{-n})\|b,\eta\rangle\!\rangle 
&=0\,,\label{eq:glL}\\
(G_r + i\eta\,\overline{G}_{-r})\|b,\eta\rangle\!\rangle 
&=0\,,\label{eq:glG}
\end{align}
\end{subequations}
on the
$\mathcal{N}=(1,1)$ superconformal algebra in both matter and ghost sector of the bulk worldsheet theory 
(see \cite{Recknagel:2013uja} for a review of the worldsheet approach to D-branes).
A question, which has been looming in the community for quite some time, is whether there exist stable elementary superconformal boundary states\footnote{Boundary state is elementary if the (GSO-projected) NS vacuum irrep appears as non-degenerate in its open string spectrum. It is stable if its open string spectrum is free of tachyons.} in type~II superstring on flat 
backgrounds (uncompactified or toroidal), which do \emph{not} simultaneously satisfy linear 
gluing conditions on the free fields $\partial X^
\mu$, $\psi^\mu$. While it is true that the gluing conditions \eqref{eq:glC} are implied by 
\begin{subequations}\label{eq:glO}
\begin{align}
(\alpha_n^\mu +\tensor{\Omega}{^\mu_\nu} \overline{\alpha}^\nu_{-n})\|b,\eta\rangle\!\rangle 
&=0\,,\label{eq:glAl}\\
(\psi^\mu_r - i\eta\,\tensor{\Omega}{^\mu_\nu}\overline{\psi}^\nu_{-r})\|b,\eta\rangle\!\rangle 
&=0\,,\label{eq:glPsi}
\end{align}
\end{subequations}
where the gluing automorphism $\Omega$ has to satisfy $\Omega^T g\,\Omega = g$ with respect to the spacetime metric $g$, the converse does not hold.
This means that even in flat backgrounds, one 
should not automatically expect that all elementary boundary states can be written as standard coherent states of the oscillators $\alpha_n^\mu,\psi_r^
\mu$, as it is the case for boundary states describing the conventional \hbox{1/2-BPS}
$\mathrm{D}p$-branes. Superconformal boundary 
states which are elementary, but do not satisfy linear gluing conditions \eqref{eq:glO}, will be henceforth termed \emph{non-conventional}. 

One can show (see e.g.\ \cite{Green:1996um}) that boundary
states for all elementary 1/2-BPS D-branes have to satisfy linear gluing conditions \eqref{eq:glO} with some gluing automorphism $\Omega$. It therefore seems natural that in a search 
for stable non-conventional boundary states, one should 
start by thinking about 1/4-BPS objects, whose tension scales as $g_\mathrm{s}^{-1}$, and, which support open string excitations. Indeed, there 
are known examples of stable bound states of ordinary 
$\mathrm{D}p$-branes
in type II superstring theories on flat 
backgrounds, which saturate the BPS bound while preserving only 8 (or fewer) spacetime supercharges. Description of such objects in terms of boundary states is still missing to date. 
The problem of finding general 
sewing-consistent boundary states which satisfy the gluing conditions~\eqref{eq:glC} in flat backgrounds has been previously addressed by a number of authors~\cite{Gaberdiel:2001zq,Gaberdiel:2001xm,Janik:2001hb,Gaberdiel:2004nv} (see \cite{Gaberdiel:1999ch,Stefanski:2000fp,Quiroz:2001xz,Braun:2002qa} for related work on orbifolds and orientifolds). However, no \emph{stable} 
non-conventional boundary states were reported.

The main subject of this paper is to analyze boundary states which wrap certain self-dual 4-tori and which do not necessarily satisfy the linear gluing conditions~\eqref{eq:glO}.
This will be achieved by employing a 
Gepner-like\footnote{Gepner's original construction cannot be applied literally: instead 
of a toroidal compactification, it would yield a supersymmetric orbifold thereof.} construction~\cite{Gepner:1987qi,Gepner:1989gr}. That is, we 
will recast the SCFT describing the $\mathcal{N}=(2,2)$ 
worldsheet superconfomal sigma model on the 4-tori in terms $\mathcal{N}=2$ superconformal minimal 
models. This will enable us to use rational CFT methods 
(mainly following the ideas of~\cite{Recknagel:1997sb,Recknagel:2002qq}) 
to write down consistent elementary boundary states which respect the minimal model superconformal 
algebras with most general gluing conditions. Note that this procedure does not impose an a priori requirement on the boundary states to satisfy linear gluing conditions \eqref{eq:glO} on the worldsheet bosonic and femionic oscillators $\alpha^m_n,\psi^m_r$ along internal directions of the 4-tori. While it is shown that the simplest gluing conditions yield boundary states for ordinary 1/2-BPS D$p$-branes, more complicated permutation gluing conditions will be 
demonstrated to give stable non-conventional boundary states describing some of the known 1/4-BPS bound states of D$p$-branes, as well as new stable D-branes, which break all spacetime supersymmetries and do not carry any RR charges. A large number of unstable boundary states is also found: these include both the non-BPS D$p$-brane boundary states with $p$ odd/even for type IIA/IIB superstring as well as unstable non-conventional boundary states. None of the unstable boundary states is found to carry RR charges.

The boundary states describing 1/4-BPS D-branes will be shown to reproduce the masses and conserved spacetime supercharges calculated by saturating the BPS bound for the RR charges they carry. Our construction also directly exhibits the exact open string spectra of the boundary states, which turn out to be free of tachyons. We also make it manifest that the boundary states cannot possibly satisfy any linear gluing condition of the type \eqref{eq:glO} on the worldsheet bosonic and fermionic oscillators in the internal directions of the 4-torus. Therefore, being elementary, these boundary states are examples of stable non-conventional boundary states. Motivated by the fact that any configuration of D$p$-brane charges which gives rise to \hbox{1/4-BPS} bound states on a 4-torus can be T-dualized into a D0/D4 system, we show that the number of massless open string excitations of our 1/4-BPS boundary states (minus the center-of-mass moduli) agrees with the dimensionality of the corresponding instanton moduli space. We also extract the leading asymptotic behavior of massless NSNS and RR fields which couple the boundary state describing a D1/D5 bound state in a $B$-field and find precise agreement with the corresponding supergravity solution. 

 
Apart from the boundary states for 1/4-BPS bound sates of D$p$-branes, we also find a number of boundary states which are stable, elementary, do not couple to massless RR states in the bulk and have open string spectra without bose-fermi degeneracy (therefore being necessarily non-supersymmetric). We conclude that these boundary states describe previously unknown D-branes which are stable in the weak coupling limit. We calculate their exact masses and open string spectra. We also compute (at one loop in open string channel) their mutual interaction potential as a function of their distance $r$ in the non-compact external directions. Since these D-branes do not carry any conserved charges, it is not clear if their stability property survives at strong coupling. Also, the fact that they violate the no-force condition means that they are not likely to be associated with a regular supergravity solution.
 
Note that similar ideas were recently employed in \cite{Kudrna:2018} to construct non-conventional boundary states in free boson CFT on a 2-torus, where the internal $c=2$ CFT was resolved in terms 
of Virasoro and $\mathcal{W}_3$ minimal models. 
These boundary states were identified with certain ``intermediate" boundary RG fixed points frequently mentioned in condensed matter literature \cite{PhysRevB.57.R5579,Affleck:2000ws,Oshikawa:2005fh,PhysRevB.79.235321}. They were also successfully matched with exotic solutions of Witten's cubic open string field 
theory obtained numerically in level truncation. 

Our paper is structured as follows. In Section~\ref{sec:BPS} we review the implications of saturating 
the BPS bound for RR charge configurations wrapping a 4-torus. In Section~\ref{sec:ws}, specializing on two 
particular self-dual 4-tori which admit a description in terms of $\mathcal{N}=2$ minimal models, we classify all boundary states which respect the corresponding enhanced chiral algebra with most general gluing conditions. We construct permutation boundary states corresponding to some of the 1/4-BPS bound states 
described in Section~\ref{sec:BPS}. We also find boundary states for previously unknown stable non-supersymmetric D-branes which do not carry any RR charges.
In each case we will expose our calculations in detail for a representative example of (permutation) gluing conditions and only make a brief summary of our results for general gluing conditions. We relegate a complete classification of RR-charged boundary states to an appendix.
Finally, in Section~\ref{sec:disc} we 
discuss a number of directions into which our work can be 
extended. We also provide two additional appendices where we elucidate our conventions.

We will set $\alpha'=1$ in the entire paper.

\section{Bound states of D$p$-branes on $T^4$}
\label{sec:BPS}

In this section we will review relevant topics concerning the bound states of D$p$-branes which wrap general 4-tori (see also~\cite{Polchinski:1996na,Obers:1998fb}). The basic requirement which we will impose 
is that the system settles (flows) down to a state with 
maximum spacetime supersymmetry allowed by the given 
configuration of RR charges (i.e.\ saturating the BPS bound). These are the $1/2^k$-BPS bound
states with $k\geqslant 0$ (where $k=0$ corresponds to the closed string vacuum). Fixing the RR charges, the conditions for the 
state to preserve a given amount of supersymmetry 
together with the corresponding mass-formulae and also the expressions for conserved supercharges for such 
states are derived in Appendix~\ref{app:bps}. 

Let us consider type IIA superstring compactified 
on a ${T}^4$ (analogous results follow for type IIB). The worldsheet-embedding fields $X^\mu,
\psi^\mu$ for $\mu=0,\ldots,9$ can be grouped into five 
complex free bosons and fermions as $X^{0\pm} = (X^1\pm 
X^0)/\sqrt{2}$, $\psi^{0\pm} = (\psi^1\pm \psi^0)/
\sqrt{2}$ and $X^{r\pm} = (X^{2r}\pm i X^{2r+1})/\sqrt{2}
$, \hbox{$\psi^{r\pm} = (\psi^{2r}\pm i\psi^{2r+1})/\sqrt{2}$} 
for $r=1,2,3,4$.
We will assume that the 4-torus wraps the directions 
$\mu=6,7,8,9$. We will denote the internal spacetime 
indices by \hbox{$m,n,\ldots\in\{6,7,8,9\}$.} We will mostly work in the coordinates 
$X^m$ adapted to the cycles of the torus in which the 
closed string metric $g$ has elements $g_{mn} = e_m\cdot e_n $ where $e_{m}$ are the fundamental lattice vectors of the toroidal identification. $R_m=|e_m|$ is the length of cycle $m$ of the 4-torus while, for $m\neq n$, $\theta_{mn}=e_m\cdot e_n/(|e_m||e_n|)$ is the angle between cycles $m$ and $n$. Volume of the face spanned by cycles $e_m$ and $e_n$ will be denoted by $V_{mn}$. The components of the NSNS $2$-form potential in the adapted coordinates will be denoted by~$B_{mn}$. For all D-branes which we consider, we will impose D conditions on all external (non-compact) spacelike directions. This will allow us to encode the corresponding RR charges in terms 
of the \hbox{$D=10$} super-Poincar\'{e} central charges $Z$, $Z^{mn}$, $Z^{mnrs}$ which are integer-valued and give the number of D0-branes, D2-branes wrapping the cycles $m,n$ and D4-branes wrapping the cycles $m,n,r,s$.

\subsection{1/2-BPS D$p$-branes}

Let us first analyze the 1/2-BPS bound states of D$p$-branes on 4-torus. The condition for a D$p$-brane configuration to flow into a $1/2$-BPS bound state reads \hbox{${Z}^{mnrs}{Z}-3{Z}
^{[mn}{Z}^{rs]}=0$} (see Appendix \ref{app:bps}). This clearly holds for ordinary D$p$-branes. More generally, the condition is satisfied if and only if the boundary state describing the D-brane satisfies linear gluing conditions \eqref{eq:glO} on $\p X^m$ and $\psi^m$.\footnote{The equality holds trivially also for the D$p$-branes with $p$ odd/even for type IIA/IIB which are non-BPS.} First, note that any D$p$-brane configuration on $T^4$ can be T-dualized to a different D$p$-brane configuration (on a different $T^4$) for which $Z^{mnrs}\neq 0$. Bound D2-branes can then be realized by switching on a constant distribution of $U(1)$ gauge field-strength on the D4 world-volume. 
The embedding 
of the spacelike component $\Sigma_4$ of the D4-brane 
world-volume in the 4-torus can be expressed as $X^m = 
\tensor{N}{^m_\alpha}\sigma^\alpha$, where $\sigma^\alpha
$ are the worldvolume coordinates on the D4-brane and $
\tensor{N}{^m_\alpha}$ are its wrapping numbers, so that 
\hbox{$Z^{mnrs}=\epsilon^{\alpha\beta\gamma\delta}\tensor{N}
{^m_\alpha}\tensor{N}{^n_\beta}\tensor{N}{^r_\gamma}
\tensor{N}{^s_\delta}$}.
The constant $U(1)$ gauge 
field-strength $F_{\alpha\beta}$ on the world-volume of 
the D4-branes induces D2 charge $Z^{mn}=(1/2)\epsilon^{\alpha\beta\gamma\delta}\tensor{N}{^m_
\alpha}\tensor{N}{^n_\beta}F_{\gamma\delta}$. Setting  \hbox{$F_{mn} = (1/2)\epsilon_{mnrs}Z^{rs}/Z^{6789}$}, we can easily verify that the induced D0 charge $Z=(1/8)\epsilon^{\alpha\beta\gamma\delta}F_{\alpha\beta}
F_{\gamma\delta}$ is indeed equal to $3Z^{[67}Z^{89]}/Z^{6789}$, as required by the 1/2-BPS condition. Since any constant $U(1)$ gauge-field strength configuration can be realized by linear gluing conditions~\eqref{eq:glO} with $\Omega_{mn}=[(g-\mathcal{F})/(g+\mathcal{F})]_{mn}$, where $\mathcal{F}=B+F$, this argument shows that all $1/2$-BPS bound states can be realized by boundary states satisfying linear gluing conditions on $\p X^m, \psi^m$. 
Defining the $B$-deformed central charges $\tilde{Z}=Z + \frac{1}{2}Z^{mn}B_{mn}+\frac{1}{8}Z^{mnrs}B_{mn}B_{rs}$, $\tilde{Z}^{mn}= Z^{mn} + \frac{1}{2}Z^{mnrs}B_{rs}$, $\tilde{Z}^{mnrs}= Z^{mnrs}$, the mass of 1/2-BPS bound states which follows from saturating the BPS bound satisfies (see Appendix \ref{app:bps} for a derivation) 
\begin{equation}
\mathcal{M}_{1/2}^2 = \widetilde{Z}^2  +\frac{1}{2}\widetilde{Z}^{mn}\widetilde{Z}_{mn}+\frac{1}{4!}\widetilde{Z}^{mnrs}\widetilde{Z}_{mnrs}\,,\label{eq:12BPSmassDef}
\end{equation}
where the indices on the central charges are lowered using $g_{mn}$ and we normalize all masses so that $\mathcal{M}_\mathrm{D0}=1$. For configurations with $Z^{mnrs}\neq 0$, this can be shown to agree with the DBI-mass $\mathcal{M}_{\mathrm{DBI}}^2=\mathrm{det}(g+\mathcal{F}
)_{\alpha\beta}$.
Formula \eqref{eq:12BPSmassDef} can be used to show that the mass of a $1/2$-BPS bound state of D$p$-branes is always strictly less than the sum of masses of the constituent D$p$-branes.
This mass defect indicates that a simple 
superposition of such D-branes breaks all supersymmetries and flows into a truly bound state. 

\subsection{1/4-BPS bound states}

When $ZZ^{mnrs}-3Z^{[mn}Z^{rs]}\neq 0$, the bound state is necessarily 1/4-BPS: one would need to allow for the constituent D$p$-branes to wrap more then four directions in order to produce 1/8- or lower BPS bound states (see Appendix \ref{app:bps} and also \cite{Obers:1998fb}). Saturating the BPS condition yields the mass formula
\begin{align}
\mathcal{M}_{1/4}^2&=\mathcal{M}_{1/2}^2+\frac{1}{12}\epsilon_{mnrs}\left|ZZ^{mnrs}-
3Z^{[mn}Z^{rs]}\right|\sqrt{\det g}\,.\label{eq:14BPST22}
\end{align}
The conserved combinations of spacetime supercharges can be determined as zero-eigenvalue eigenvectors of the matrix $\Gamma-\mathcal{M}
_{1/4}$, which is defined by~\eqref{eq:defGamma}. Formula \eqref{eq:14BPST22} always gives mass which is less than or equal to the sum of masses of the constituent D$p$-branes. 1/4-BPS bound states of D$p$-branes therefore fall into two categories: truly bound (those with non-zero mass defect) and marginally bound (those with zero mass defect). 

In the case of the truly bound states, the superposition of the constituent D$p$-branes is generally 
not supersymmetric and contains tachyonic modes in the 
spectrum of stretched open strings. Turning on the corresponding relevant boundary deformation then drives the system into a lower-mass supersymmetric bound state with same RR charges. From the worldsheet 
point of view, this final state should be thought of as 
a new elementary superconformal boundary state. Arguably the simplest 
example is the $\mathrm{D0/D4}$ system in a 
generic constant NSNS background $B$-field extending 
along the D4-brane~\cite{Nekrasov:1998ss,Seiberg:1999vs,David:2000um}. 

Marginally bound states come from those 1/4-BPS 
configurations of RR charges where the strings stretched 
between the constituent D$p$-branes are massless. In such cases, the 
superposition of the constituent D$p$-branes is supersymmetric and marginally stable. 
However, even then one can often form non-trivial bound states, as it may happen that the massless stretched strings give rise to exactly marginal boundary deformations~(\cite{Witten:1995gx,Douglas:1995bn,Billo:2002hm}; see \cite{Mattiello:2018kue,Mattiello:2019gxc,Maccaferri:2018vwo} for some recent developments). These will then turn the initial superposition of D$p$-branes into a new D-brane described by an  
elementary superconformal boundary state (this time with both mass and RR charges being the same as 
those of the original superposition of D$p$-branes). As a concrete example, one can take system containing a number of superposed D0- and D4-branes in vanishing $B$-field.

One can easily think of infinitely many 
configurations of RR charges $Z,Z^{mn},Z^{mnrs}$ on a 4-torus which 
violate 1/2-BPS conditions and therefore yield 1/4-BPS bound states. Note, however, that an $O(4,4;\mathbb{Z})$ duality transformation can always be constructed which takes such a general configuration into a system of $k$ D0-branes and $N$ D4-branes with $k,N\in\mathbb{Z}$ such that
\begin{align}
kN &=\frac{1}{4!}\epsilon_{mnrs}( Z Z^{mnrs} - 3 Z^{[mn}Z^{rs]})\,.
\label{eq:kN}
\end{align}
For a D0/D4 system, the formula \eqref{eq:14BPST22} yields mass
\begin{align}
\mathcal{M}_{\mathrm{D0}/{\mathrm{D4}}}^2 = 
k^2 \mathcal{M}_\mathrm{D0}^2 + N^2\mathcal{M}_\mathrm{D4}^2+2kN \,\mbox{Pf}\,B +2|kN|\sqrt{\det g}\,.
\end{align}
where $\mathcal{M}_\mathrm{D0}=1$ and $\mathcal{M}_\mathrm{D4}^2={\det g +(\mbox{Pf}\, B)^2+(1/8)\epsilon^{mnab}\epsilon^{rscd}g_{mr}g_{ns}B_{ab}B_{cd}}$. Such D0/D4 system is marginally bound whenever $B=0$. More generally, $B$ can be fine-tuned (subject to a constraint) so that one obtains marginally bound states even for $B\neq 0$. For instance, in the case of factorized $T^4=T^2\times T^2$ with factorized $B$-field ($B_{68}=B_{69}=B_{78}=B_{79}=0$), this constraint is that of (anti-)selfduality of the $B$-field, namely
\begin{equation}
\frac{B_{67}}{V_{67}} = (\mbox{sgn}\, kN)\frac{B_{89}}{V_{89}}\,.
\label{eq:margCond}
\end{equation}
The same result was obtained in \cite{Seiberg:1999vs} arguing directly from the stretched string spectrum.

From the low-energy effective point of view, the dynamics of the marginally stable \hbox{1/4-BPS} superpositions of D$p$-branes on 4-tori is described by quiver gauge theories with 8 supercharges~\cite{Douglas:1996sw}. These should be thought of as living on the worldvolume component which is external to the compactification 4-tori. Entering the Higgs branch of these theories then exactly corresponds to the formation of marginally bound states \cite{Witten:1997yu,Seiberg:1999xz}. Couplings to the closed string sector may introduce additional FI parameters into the D-term (e.g.\ due to a non-selfdual NSNS $B$-field \cite{Nekrasov:1998ss,Seiberg:1999vs,David:1999ec}), whose effect is to resolve the singularities inside the Higgs branch which correspond to the emission of one or more of the constituent D$p$-branes from the bound state. 

Also note that the 1/4-BPS bound states admit effective description as gauge instantons on $T^4$.
Moduli spaces of marginal deformations of these bound states can then be related to the moduli spaces of the corresponding instantons. For example, in the case of coincident $k$ D0- and $N$ D4-branes on a $T^4$ with $B=0$, turning on the exactly marginal deformations due to massless strings stretched between the D0- and D4-branes corresponds to passing to a general (finite-size) point in the moduli space of $k$ $U(N)$ instantons~\cite{Witten:1995gx,Douglas:1995bn,Billo:2002hm,Mattiello:2018kue,Mattiello:2019gxc}. Upon turning on a generic $B$-field along the 4-torus, the system becomes truly bound and is described by a finite-size non-commutative instanton~\cite{Nekrasov:1998ss,Seiberg:1999vs}. 
It is a well-known consequence of the ADHM construction that the instanton moduli space $\mathcal{M}_{kN}$ is generally given by a smooth resolution of the symmetric orbifold $(\widetilde{T}^4)^{kN}/S(kN)$, where the $\widetilde{T}^4$ may be different from the compactification torus $T^4$. This gives that $\mathrm{dim}\,\mathcal{M}_{kN}=4kN$. It follows that the (physical) moduli space of a general \hbox{1/4-BPS} bound state of D$p$-branes should be identified with \hbox{$\mathbb{R}^4\times T^4\times\mathcal{M}_{kN}$} where $kN$ is determined by \eqref{eq:kN} and the additional factor of $\mathbb{R}^4\times T^4$ is to account for the centre-of-mass moduli. The physical open string spectra of \hbox{1/4-BPS} bound states must therefore contain 
\hbox{$8+4kN$} exactly marginal modes.\footnote{This reduces to the expected answer for the number of massless physical fields living on 1/2-BPS D-branes because the 1/2-BPS condition together with \eqref{eq:kN} gives \hbox{$kN =0$}.}

No description of 1/4-BPS D-branes in terms of the standard coherent states of the bosonic and fermionic oscillators $\alpha^m_n,\psi^m_r$ is 
generally possible. An exception to this are the 
marginally bound states, for which there exist points in 
their moduli spaces which correspond to stable 
superpositions of the constituent D$p$-branes. Boundary states for the truly 
bound states and marginally bound states at generic 
points in their moduli space, however, seem to admit no 
simple description in terms of worldsheet bosonic and fermionic oscillators $\alpha^m_n,\psi^m_r$ along the 4-torus.
A systematic worldsheet 
description of 1/4-BPS bound states of D$p$-branes lies outside of the scope of this paper. In the following section we will instead construct examples of elementary boundary states for a number of 1/4-BPS bound states in a very specific setting of two particular 4-tori, which admit a resolution in terms of $\mathcal{N}=2$ minimal models. 
Both of them can be factorized as $T^2\times T^2$: first the $SU(3)^2$ torus with 
$R_m=1$, $\theta_{67}=\theta_{89}=\pi/3$, $B_{67}=B_{89}=1/2$ and second the 
$SU(2)^4$ torus with $R_m=1$, $\theta_{67}=\theta_{89}=\pi/2$, 
$B=0$. 
Note that both of these 4-tori are self-dual with respect to $T$-dulities along all of their cycles.

\subsubsection*{$SU(3)^2$ 4-torus}

Substituting the $SU(3)^2$ 4-torus parameters into~\eqref{eq:14BPST22}, we find that 1/4-BPS bound states wrapping the $SU(3)^2$ 4-torus have masses
\begin{align}
\mathcal{M}_{1/4}^2&=Z^2 +(Z^{67})^2+(Z^{89})^2+(Z^{6789})^2 +\frac{1}{2}(Z Z^{6789}+Z^{67}Z^{89})+(Z^{67}+Z^{89})(Z+Z^{6789})
\nonumber\\
&\hspace{-0.5cm}+(Z^{68})^2+(Z^{69})^2+(Z^{78})^2+(Z^{79})^2+\frac{1}{2}(Z^{68}Z^{79}+Z^{69}Z^{78})+(Z^{68}+Z^{79})(Z^{69}+Z^{78})\nonumber\\
&\hspace{9.0cm}+\frac{3}{2}|ZZ^{6789}-3Z^{[67}Z^{89]}|\,.
\label{eq:14BPSSU3}
\end{align}
The lightest 1/4-BPS bound states have mass $
\sqrt{3}$ and are truly bound. These include, for instance, $\mathrm{D0}/
\overline{\mathrm{D4}}$, $\overline{\mathrm{D0}}/
{\mathrm{D2}_{67}}/{\mathrm{D2}}_{89}$, $
{\mathrm{D2}}_{67}/{\mathrm{D2}}_{89}/\overline{\mathrm{D4}}$, $\mathrm{D2}_{67}/\overline{\mathrm{D2}}_{89}$, $
\mathrm{D0}/\overline{\mathrm{D2}}_{89}/\mathrm{D4},
{\mathrm{D0}}/\overline{\mathrm{D2}}_{67}/
{\mathrm{D4}}$, where $\mathrm{D2}_{mn}$ denotes a D2-brane wrapping the cycles $m,n$. 

\subsubsection*{$SU(2)^4$ 4-torus}

Substituting the $SU(2)^4$ 4-torus parameters into~\eqref{eq:14BPST22}, we find that the 1/4-BPS bound 
states wrapping the $SU(2)^4$ 4-torus have masses
\begin{align}
\mathcal{M}_{1/4}^2 &= Z^2 +(Z^{67})^2 +(Z^{89})^2 +(Z^{6789})^2+(Z^{68})^2+(Z^{69})^2+(Z^{78})^2+(Z^{79})^2+\nonumber\\
&\hspace{9cm}+2|ZZ^{6789}-3Z^{[67}Z^{89]}|\,.
\label{eq:14BPSSU22}
\end{align}
As an example, let us consider the 1/4-BPS bound 
states of D$p$-branes with mass $2\sqrt{2}$, such as $\mathrm{D0}/{\mathrm{D2}}_{67}/
{\mathrm{D2}}_{89}/\overline{\mathrm{D4}}$ and $\mathrm{D0}/
\overline{\mathrm{D2}}_{67}/\overline{\mathrm{D2}}_{89}/
\overline{\mathrm{D4}}$. These can be regarded as marginally bound states of the 
1/2-BPS bound states $
\mathrm{D0}/{\mathrm{D2}}_{67}$ with ${\mathrm{D2}}_{89}/
\overline{\mathrm{D4}}$ and $\mathrm{D0}/\overline{\mathrm{D2}}_{67}$ with 
$\overline{\mathrm{D2}}_{89}/\overline{\mathrm{D4}}$, respectively. 

\section{Worldsheet analysis}
\label{sec:ws}

The focus of this section will be on constructing light-cone gauge boundary states (\cite{Green:1996um}; see \cite{Gaberdiel:2000jr,Recknagel:2013uja} for a review) for type IIA superstring\footnote{Analogous results follow for type IIB as well.} compactifications on certain stringy \hbox{4-tori}, namely the $SU(3)^2$ 4-torus and the $SU(2)^4$ 4-torus. We will first describe the way the bulk spectrum of the two respective $\mathcal{N}=(2,2)$ worldsheet sigma models can be organized into irreducible representations of a number of copies of certain $\mathcal{N}=2$ minimal models. For some basic 
information about the representations of the $\mathcal{N}=2$ superconformal algebra in two dimensions and the conventions we follow, see Appendix~\ref{app:scft}. 
Unless we specify otherwise, we will assume the D-branes to satisfy Dirichlet conditions in all spacelike directions external to the compactification 4-tori. For the light-cone gauge boundary states, this corresponds to taking A-type gluing conditions in the $X^{2}\pm i X^{3}$ free field theory and B-type conditions in the $X^{4}\pm i X^{5}$ free field theory. Also note that we will generally drop the omnipresent factors coming from external non-compact 
worldsheet bosons as they will only play spectator role 
and can be easily reattached when needed.
For a fixed gluing condition on the rational chiral currents, the boundary states will then be computed by either directly applying the results of \cite{Recknagel:1997sb} and \cite{Recknagel:2002qq} (for simpler gluing conditions) or by requiring the open string spectra to contain only integer multiplicities (for more complicated cases). We expose explicit calculations for a number of representative cases of gluing conditions and summarize our results for all admissible gluing conditions at the end. In a number of cases, we also check that the spectra of strings stretched between boundary states satisfying different gluing conditions automatically contain only integer multiplicities. In light of the papers \cite{Behrend:1999bn,Fuchs:1999zi,Fuchs:1999xn,Fuchs:2000vg}, this is not an unexpected result. 

\subsection{$SU(3)^2$ 4-torus}

Let us deal with the $SU(3)^2$ case first. Before analyzing the boundary states, we will establish description of the bulk spectrum in terms of irreducible representations of six copies of the $k=1$ minimal model of the $\mathcal{N}=2$ superconformal algebra.   

\subsubsection*{Bulk theory}

We can write the $SU(3)^2$ 4-torus as a product of two 
$SU(3)$ 2-tori $T^4=T^2\times {T^2}$ which extend in the 
67 and 89 planes, respectively.
It has been known for some time \cite{Chun:1991js,Recknagel:1997sb,Gutperle:1998hb} that the bulk spectrum of 
the $\mathcal{N}=(2,2)$ worldsheet sigma model on the $SU(3)$ 2-torus can be given in terms of the irreps of three $
\mathcal{N}=2$ minimal models with $k_a=1$ for $a=1,2,3$. 
The fusion algebra of these minimal models has $
\mathbb{Z}_3\times\mathbb{Z}_2$ symmetry generated by 
$g_a\Phi^{l_a}_{m_a,s_a}= e^{\frac{2\pi i }{3}m_a}
\Phi^{l_a}_{m_a,s_a}$ and $h_a\Phi^{l_a}_{m_a,s_a}= e^{-i
\pi s_a}\Phi^{l_a}_{m_a,s_a}$. This, in particular, 
extends to a diagonal $\mathbb{Z}_3$ symmetry $G$ of the 
$(k=1)^3$ tensor product fusion algebra, where $G$ is 
generated by $g_1g_2g_3$. The $\mathcal{N}=(2,2)$ worldsheet sigma model 
on the $SU(3)$ 2-torus can then be obtained as the $G$-orbifold of the direct product of three copies of $k=1$ 
minimal models with $U(1)$ charge-conjugate 
modular invariant.\footnote{The choice of the charge-conjugate modular invariant (rather than the diagonal modular invariant) will turn out to be more convenient for our purposes.} Also note that projecting onto 
$G$-invariant states $\sum_a m_a \in 3\mathbb{Z}$ can be equally well characterized as 
projecting onto states having integer $U(1)$ charge in 
the NS sector and half-integer $U(1)$ charge in the R 
sector. Given this 
information, it is now straightforward to write down the 
GSO-unprojected spectrum of the $\mathcal{N}=(2,2)$ worldsheet sigma model on the $SU(3)$ 2-torus in the NSNS and RR sector
\begin{align}
Z_{SU(3)}(q,\overline{q})&=\sum_{l_a,m_a,s_a}\,\sum_{\substack{t\in\mathbb{Z}_3}}\sum_{\overline{s}_a}
\prod_{a=1}^{3}\chi^{l_a}_{m_a,s_a}(q)\,\overline{\chi}
^{l_a}_{-m_a+2t,\overline{s}_a}(\overline{q})\,, 
\label{eq:ZSU3}
\end{align}
where the $l_a,m_a,s_a$ sum on the RHS again runs over distinct 
$G$-invariant states in the NS and R sector, respectively, such that $s_1-s_a\in 2\mathbb{Z}$ 
and $l_a+m_a+s_a\in 2\mathbb{Z}$ together with $\overline{s}_a\in\mathbb{Z}_4$ such that $s_a-\overline{s}_a\in 2\mathbb{Z}$ for all $a=1,2,3$. Here $\chi^{l_a}_{m_a,s_a}$ are the $\mathcal{N}=2$ maximal bosonic subalgebra characters (see Appendix \ref{app:scft} for their $q$-expansions).
The sum over $t$ gives twisted sectors. 
It can be indeed shown (e.g.\ order by order in the ($q,\overline{q}$)-expansion) that
\begin{align}
Z_{SU(3)}(q,\overline{q}) = \left|\frac{\theta_{i}(q)}{\eta(q)^3}\right|^2\sum_{M,N,R,S\in\mathbb{Z}}q^{h^\mathrm{L}_{M,N,R,S}} \overline{q}^{{h}^\mathrm{R}_{M,N,R,S}}\,,
\end{align}
where we take $i=3$ and $i=2$ in the NSNS and RR sector, respectively ($\theta_i$ being the standard Jacobi theta functions), and
\begin{subequations}
\begin{align}
h^{\mathrm{L}}_{M,N,R,S} &= \frac{1}{4}k^\mathrm{L}_{m}g^{mn}k^\mathrm{L}_{n} = \frac{1}{4}[(M+R)^2+\frac{1}{3}(M+2N-R+2S)^2]\,,\\
h^{\mathrm{R}}_{M,N,R,S} &= \frac{1}{4}k^\mathrm{R}_{m}g^{mn}k^\mathrm{R}_{n} = \frac{1}{4}[(M-R+S)^2+\frac{1}{3}(M+2N-R-S)^2]\,.
\end{align}
\end{subequations}
Here we have introduced $k_m^\mathrm{L} = p_m+E_{mn}w^n$, $k_m^\mathrm{R} = p_m-E^T_{mn}w^n$ and $E_{mn} = g_{mn}+B_{mn}$. In the adapted coordinates $X^{m}$, $m=6,7$, these have components $p_m=(M,N)$, $w^m=(R,S)$, $M,N,R,S\in\mathbb{Z}$ and $g_{66}=g_{77}=1$, $g_{67}=-B_{67}=-1/2$. 

We will now use the results for the $SU(3)$ 2-torus to 
describe the light-cone gauge spectrum of type II superstring compactified on the $SU(3)^2$ 4-torus. Let us introduce the 6-component vectors $\bm{l}=(l_1,\ldots,l_6)$, $\bm{m}=(m_1,\ldots,m_6)$, $\bm{s}=(s_1,\ldots,s_6)$, which encode the internal $\mathcal{N}=2$ minimal model data. For later convenience, we also define the (2+6)-component vectors $\bm{\lambda} = (0,0;\bm{l})$, $\bm{\mu}=(0,0;\bm{m})$ together with $
{\bm{\sigma}}=(\sigma_1,\sigma_2;\bm{s})$ which also include information about the external transverse fermionic representations. We will refer to their components by 
$l_a,m_a,s_a$ for $a=1,\ldots,6$ and $\lambda_p,\mu_p,\sigma_p$ for $p=1,\ldots,8$. Note that $a=1,2,3$ belong to one of the two constituent $SU(3)$ 2-tori while $a=4,5,6$ pertain to the other. We then define
\begin{equation}
    \chi^{\bm{\lambda}}_{\bm{\mu},\bm{\sigma}} (q)= \prod_{r=1}
    ^{2}\chi_{\sigma_r}(q)\prod_{a=1}^6 \chi^{l_a}
    _{m_a,s_a}(q)\,.
\end{equation}
Here $\chi_{\sigma_r}$, $r=1,2$ are the $\widehat{\mathfrak{so}}(2)_1$ 
characters for the two transverse external complex 
fermions, where $\sigma_r=0,2$ gives the $o$ and $v$ 
irreps and $\sigma_r=\pm 1$ gives the $s$ and $c$ irreps. Denoting $\bm{\beta}_1 = (1,0;\bm{0})$ and $\bm{\beta}_2=(0,1;\bm{0})$ we therefore have $\Phi^{\bm{0}}_{\bm{0},2\bm{\beta}_r} = \psi^{r\pm} = (\psi^{2r}\pm i \psi^{2r+1})/
\sqrt{2}$ for $r=1,2$. Introducing also the vectors $\bm{\beta}_{3}=(0,0;1,1,1,0,0,0)$ and $\bm{\beta}_{4}=(0,0;0,
0,0,1,1,1)$, the fields $\bm{\lambda}=\pm\bm{\mu}=\bm{\beta}_r$, $\bm{\sigma}=\bm{0}$ then correspond to the internal complex fermions $\psi^{r\pm}$ for $r=3,4$. In the R sector, the fields with $\bm{\lambda}=\bm{0}$, $\bm{\mu}=2\sum_{r=3}^4\tau_r \bm{\beta}_r$ and $\bm{\sigma}=2\sum_{r=1}^4\tau_r \bm{\beta}_r$ give the spin fields with $SO(8)$ spins $\tau_r=\pm 1/2$ for $r=1,2,3,4$.
The GSO-unprojected spectrum of the $
\mathcal{N}=(2,2)$ worldsheet sigma model involving the $SU(3)^2$ 4-torus and four non-compact external directions then reads (NSNS and RR parts)
\begin{align}
    Z(q,\overline{q})&=\sum_{\bm{\lambda},\bm{\mu},\bm{\sigma}}
    \sum_{\substack{t_r\in\mathbb{Z}_3}}\sum_{\overline{\bm{\sigma}}}\chi^{\bm{\lambda}}
    _{\bm{\mu},\bm{\sigma}}(q)\overline{\chi}^{\bm{\lambda}}_{-
    \bm{\mu}+ 2t_3\bm{\beta}_3+ 2t_4\bm{\beta}_4,
    \overline{\bm{\sigma}}}(\overline{q})\,,
    \label{eq:partSU3typeII}
\end{align}
where the $\bm{\lambda},\bm{\mu},\bm{\sigma}$ sum on the RHS runs over 
distinct $G\times G$-invariant states in the NS and R sector such that $\sigma_1-\sigma_p\in 2\mathbb{Z}$ and $l_a
+m_a+s_a\in 2\mathbb{Z}$ together with $\overline{\sigma}_p\in
\mathbb{Z}_4$ such that $\sigma_p-\overline{\sigma}_p\in 
2\mathbb{Z}$ for all $a=1,\ldots,6$ and 
$p=1,\ldots,8$. Note that $G\!\times\! G$ invariant states are precisely those with $\bm{\beta}_r\cdot \bm{\mu}\in 3\mathbb{Z}$ for $r=3,4$. We also recall that the $U(1)$ charge of a primary field with 
labels $\bm{\lambda},\bm{\mu},\bm{\sigma}$ can be expressed as $q(\bm{\mu},\bm{\sigma}) = (\frac{\bm{\mu}}{3}-\frac{\bm{\sigma}}{2})\cdot \sum_{r=1}^4\bm{\beta}_r$.
It follows that all states 
appearing in~\eqref{eq:partSU3typeII} have integer left-and right-moving total $U(1)$ charges $q,\overline{q}$. Further projecting onto states with $q,\overline{q}\in 
2\mathbb{Z}+1$ in both NSNS and RR sectors gives us the 
spacetime-bosonic fields of type IIB superstring while taking instead $\overline{q}\in 2\mathbb{Z}$ in the RR sector gives us the spacetime-bosonic fields of type IIA superstring.

\subsubsection*{$\mathrm{D}p$-brane boundary 
states}


Let us start by considering ordinary B-type gluing conditions 
for all internal SCFTs. We will denote\footnote{See Appendix \ref{app:scft} for an explanation of our notation.} these by $\omega_0 \equiv (1_\mathrm{B})(2_\mathrm{B})(3_\mathrm{B})(4_\mathrm{B})(5_\mathrm{B})(6_\mathrm{B})$. Such boundary states are guaranteed to describe 1/2-BPS D-branes: $\omega_0$ gluing conditions imply conservation of the free field \hbox{$\mathcal{N}=(2,2)$} worldsheet SCFT currents for the two constituent 2-tori which in turn was shown \cite{Gaberdiel:2004nv} to imply linear gluing conditions \eqref{eq:glO} on the oscillators $\alpha^m_n,\psi^m_r$. The charge conjugation property of 
the partition function~\eqref{eq:partSU3typeII} gives 
that the allowed Ishibashi states are labelled as $\left|
\bm{\lambda},\bm{\mu},\bm{\sigma}\rangle\!\rangle_{\omega_0}\right.
$, where $\bm{\lambda},\bm{\mu},\bm{\sigma}$ run over the $G\!\times\! G$-invariant states in the NS and R sector such that $\sigma_1-\sigma_p\in 
2\mathbb{Z}$, $l_a+m_a+s_a\in 2\mathbb{Z}$ for all $a=1,\ldots,6$, $p=1,\ldots,8$, and we only allow 
states with $q(\bm{\mu},\bm{\sigma})\in 2\mathbb{Z}+1$.
In order to simplify algebraic 
manipulations involving the boundary states, we introduce the following projector onto allowed Ishibashi states
\begin{align}
    \delta^{\omega_0}_{\bm{\lambda},\bm{\mu},\bm{\sigma}}& = 2^{-6} 
    \frac{1}{12}\sum_{\zeta\in\mathbb{Z}_{12}}(-1)^\zeta e^{i\pi 
    q(\bm{\mu},\bm{\sigma})\zeta }\prod_{r=3}^4\frac{1}{3}\sum_{t_r\in
    \mathbb{Z}_{3}}e^{\frac{2i\pi}{3} t_r\bm{\beta}_r\cdot\bm{\mu}}
     \prod_{p=1}
    ^8\frac{1}{2}\sum_{\nu_p\in\mathbb{Z}_2}e^{i\pi\nu_p 
    (\sigma_1 - \sigma_p)}\nonumber\\
   &\hspace{8.4cm} \times \prod_{a=1}^{6}\frac{1}{2}
   \sum_{\xi_a\in\mathbb{Z}_2}e^{i\pi \xi_a(l_a
   +m_a+s_a)}\,,\label{eq:deltaSU3}
\end{align}
Following~\cite{Recknagel:1997sb}, we then write
\begin{equation}
    \|\alpha\rangle\!\rangle_{\omega_0} \equiv \|\bm{\Lambda},
    \bm{M},\bm{\Sigma}\rangle\!\rangle_{\omega_0} = \frac{1}
    {\kappa_{\omega_0}^\alpha} \sum_{\bm{\lambda},\bm{\mu},
    \bm{\sigma}}\delta^{\omega_0}_{\bm{\lambda},\bm{\mu},\bm{\sigma}}
    B^{\alpha,{\omega_0}}_{\bm{\lambda},\bm{\mu},\bm{\sigma}}|\bm{\lambda},
    \bm{\mu},\bm{\sigma}\rangle\!\rangle_{\omega_0}\,,
\end{equation}
where we define
\begin{align}
    B^{\alpha,\omega_0}_{\bm{\lambda},\bm{\mu},\bm{\sigma}} = 
  (-1)^{\frac{\sigma_1^2}{2}} e^{-\frac{i\pi}{2}\bm{\sigma}\cdot\bm{\Sigma}}e^{\frac{i\pi}{3}\bm{\mu}\cdot \bm{M}}
    \prod_{a=1}^6\frac{\sin [ \frac{\pi}{3}(l_a+1)(L_a
    +1)]}{\sin^\frac{1}{2} [ \frac{\pi}{3}(l_a+1)]}\,.
    \label{eq:BSU3coef}
\end{align}
Here $\kappa_{\omega_0}^\alpha$ is a normalisation to be fixed below. 
The boundary state 
labels $\bm{\Lambda},\bm{M},\bm{\Sigma}$ must satisfy the minimal model 
constraints $L_a=0,1$, $L_a+M_a+S_a\in 2\mathbb{Z}$, $M_a
\in\mathbb{Z}_6$, $\Sigma_p\in\mathbb{Z}_4$ modulo the 
field identification $(L_a,M_a,S_a)\sim(1-L_a,M_a+3,S_a
+2)$. Alignment of spin structures further requires $S_a \in 2\mathbb{Z}$ for all $a$ (see \cite{Fuchs:2000gv}	for details). To ensure that the labeling by $
\bm{\Lambda},\bm{M},\bm{\Sigma}$ gives distinct boundary 
states, we should take the labels modulo the action of $G
\times G$. 
In order to fix the normalization $\kappa_{\omega_0}^\alpha$ 
let us calculate the overlap 
\begin{align}
    \widetilde{Z}^{\omega_0}_{\alpha\widetilde{\alpha}}(\tilde{q}) &=\, _{\omega_0}\! \langle\!\langle \Theta\widetilde{\alpha}\| \tilde{q}^{\frac{1}
    {2}(L_0+\overline{L}_0 - \frac{c}{12})}\| 
    {\alpha}\rangle\!\rangle_{\omega_0} =\frac{1}
    {\kappa_{\omega_0}^\alpha\kappa_{\omega_0}^{\tilde{\alpha}}} \sum_{\bm{\lambda},\bm{\mu},
    \bm{\sigma}}\delta^{\omega_0}_{\bm{\lambda},\bm{\mu},\bm{\sigma}} 
    B^{\widetilde{\alpha},\omega_0}_{\bm{\lambda},-\bm{\mu},-\bm{\sigma}}
    B^{{\alpha},\omega_0}_{\bm{\lambda},\bm{\mu},\bm{\sigma}} 
    \chi^{\bm{\lambda}}_{\bm{\mu},\bm{\sigma}}(\tilde{q})\,.
    \label{eq:ovCS}
\end{align}
Let us denote by $\delta^{(p)}$ the Dirac delta function on $
\mathbb{Z}_p$. By applying the modular $S$-transformation, we can express the open string partition function as
\begin{align}
  Z^{\omega_0}_{\alpha\widetilde{\alpha}}(q)&=
  \frac{3^{3}2^{-8}}{\kappa_{\omega_0}^\alpha\kappa_
  {\omega_0}^{\widetilde{\alpha}}}\sum_{\bm{\lambda}',\bm{\mu}',
  \bm{\sigma}'}\null\!\!\!\!\!^\mathrm{ev} \sum_{\zeta\in
  \mathbb{Z}_{12}}\sum_{\nu_p\in\mathbb{Z}_2} \sum_{t_r
  \in\mathbb{Z}_{3}}(-1)^{\sigma_1'+\Sigma_1-
  \widetilde{\Sigma}_1}
 \delta^{(4)}_{\sigma_1'+\Sigma_1-\widetilde{\Sigma}_1+
 \zeta-2\sum_{p=2}^8\nu_p+2}\prod_{a=1}^6\delta^{(2)}
 _{l_a'+L_a-\widetilde{L}_a}
   \nonumber\\
   &\hspace{2mm}\times\prod_{p=2}^{8}\delta^{(4)}_{{\sigma_p'+
   \Sigma_p-\widetilde{\Sigma}_p}+{2\nu_{p}+\zeta}}
   \prod_{a=1}^{3}\delta^{(6)}_{m_a'+M_a-\widetilde{M}_a+
   \zeta+2t_{3}}\prod_{a=4}^{6}\delta^{(6)}_{m_a'+M_a-
   \widetilde{M}_a+\zeta+2t_{4}}\chi^{\bm{\lambda}'}_{\bm{\mu}',
   \bm{\sigma}'}(q)\,,\label{eq:nopexplSU3}
\end{align}
where the sum $\sum_{\bm{\lambda}',\bm{\mu}',\bm{\sigma}'}^
\mathrm{ev}$ on the RHS runs over $l_a'=0,1$, 
$m_a'\in\mathbb{Z}_6$, $\sigma_p'\in\mathbb{Z}_4$ with 
$l_a'+m_a'+s_a'\in 2\mathbb{Z}$. For instance, in the case $\alpha=\tilde{\alpha}$, one can expand the summations in \eqref{eq:nopexplSU3} to obtain
\begin{align}
Z_{\alpha\alpha}^{\omega_0}(q) &= \frac{3^{4} 2^{-6}}{(\kappa_{\omega_0}^{\alpha})^2}\bigg(\chi^{\bm{0}}_{\bm{0},2\bm{\beta}_1}+\chi^{\bm{0}}_{\bm{0},2\bm{\beta}_2}+\chi^{\bm{0}}_{\bm{0},(0,0;2,0,0,0,0,0)}+\ldots\nonumber\\
&\hspace{5cm} -\sum_{\tau_1\tau_2\tau_3\tau_4<0} \chi^{\bm{0}}_{2\sum_{r=3}^4\tau_r\bm{\beta}_r,2\sum_{r=1}^4\tau_r\bm{\beta}_r}-\ldots\bigg)\,,\label{eq:nopexplSU3char}
\end{align}
where all characters in the sum have the same coefficient (up to a minus sign). Indeed, the minimal normalization which yields integer multiplicities of states of a string stretched between general $\omega_0$ D-branes $\alpha,\tilde{\alpha}$ is $\kappa_{\omega_0}^
\alpha=3^2 2^{-3}$ for all $\alpha$. As it is apparent from \eqref{eq:nopexplSU3char},
this normalization also makes all $\omega_0$ boundary states elementary.\footnote{Boundary state is elementary if and only if the (GSO-projected) open-string NS vacuum irrep is non-degenerate.} \eqref{eq:nopexplSU3} implies the selection rule $q'+Q-\widetilde{Q}\in 2\mathbb{Z}+1$, where we define $q'=q(\bm{\mu}',\bm{\sigma}')$, $Q=q(\bm{M},\bm{\Sigma})$ and $\widetilde{Q}=q(\widetilde{\bm{M}},\widetilde{\bm{\Sigma}})$. 
Since $h\geqslant |q|/2$ for all unitary 
representations of $\mathcal{N}=2$ SCAs, it is clear that the lowest states with $q'\in 2\mathbb{Z}+1$ have mass squared greater than or equal to zero, meaning that all spectra with $Q-
\widetilde{Q}\in 2\mathbb{Z}$ are tachyon-free. In particular, this holds for $\alpha=\widetilde{\alpha}$, so that our 
boundary states describe stable D-branes. 

Let us write $|\bm{\lambda},\bm{\mu},
\bm{\sigma}\rangle_{\omega_0}$ for the leading term 
of the Ishibashi state $|\bm{\lambda},\bm{\mu},\bm{\sigma}\rangle
\!\rangle_{\omega_0}$, so that \hbox{$_{\omega_0} \langle\bm{\lambda},\bm{\mu},
\bm{\sigma}|\bm{\lambda}',\bm{\mu}',\bm{\sigma}'\rangle\!\rangle_
{\omega_0} = \delta_{\bm{\lambda}\bm{\lambda}'}\delta_{\bm{\mu}\bm{\mu}'}
\delta_{\bm{\sigma}\bm{\sigma}'}$}.
Probing the boundary states with the closed string state $|\bm{\lambda},\bm{\mu},
\bm{\sigma}\rangle_{\omega_0}$ will therefore yields its NSNS 
and RR couplings. Setting $\Sigma_1=0$, $\Sigma_2=0$ corresponds to taking N conditions for $\psi^3$ and D conditions for 
$\psi^2,\psi^4,\psi^5$ (a valid choice for light-cone gauge boundary states describing D-branes with Dirichlet boundary conditions on all external spacelike directions). 
We also set $S_a=0$ for all $a=1,\ldots,6$, thus fixing a particular parity for the D-branes. The massless NSNS Ishibashi states which provide couplings to the closed string states $\psi^{\pm r}_{-1/2}\bar{\psi}^{\mp r}_{-1/2}$ along the internal 4-torus are precisely $|\bm{\beta}_r,\pm\bm{\beta}_r,\bm{0}\rangle\!\rangle_{\omega_0}$ for $r=3,4$. In particular, if a boundary state $\|\alpha\rangle\!\rangle$ is to satisfy linear gluing conditions on the worldsheet oscillators $\alpha^m_n,\psi^m_r$ with gluing automorphism ${\Omega}$, we need to have
\begin{align}
\|\alpha\rangle\!\rangle\supset i g_\alpha\, \tensor{\Omega}{_{mn}}\psi^m_{-\frac{1}{2}}{\bar{\psi}}{^n_{-\frac{1}{2}}}|0\rangle_\mathrm{NSNS}\,,
\label{eq:linBS}
\end{align}
where $g_\alpha$ is an overall normalization proportional to the mass of the corresponding D-brane. Couplings to the massless RR sector give the RR charges carried by the boundary state. The massless RR Ishibashi states can be parametrized by the corresponding $SO(8)$ spins $
\tau_r=\pm 1/2$ for $r=1,2,3,4$ as $\bm{\lambda}=\bm{0}$, $\bm{\mu}=2\sum_{r=3}^4\tau_r\bm{\beta}_r$ and $\bm{\sigma}=2\sum_{r=1}^4\tau_r\bm{\beta}_r$, subject to $\tau_1\tau_2\tau_3\tau_4<0$. To save some space, we will denote them by $|\tau_1\tau_2\tau_3\tau_4\rangle\!\rangle_{\omega_0}$.
The formula~\eqref{eq:BSU3coef} yields 9 distinct forms for massless RR and NSNS 
boundary state coefficients meaning that $\omega_0$ boundary states describe 9 distinct types of D$p$-branes. 
Note that $g_{\alpha}^{\omega_0}=1/\sqrt{3}$ for all $\alpha$, so that all $\omega_0$ boundary states describe D-branes with identical masses. 
Boundary state $\|\bm{0},\bm{0},
\bm{0}\rangle\!\rangle_{\omega_0}$ clearly 
describes a $\mathrm{D0}$-brane. We will further adopt a convention that $\|\bm{0},\bm{0},
\bm{0}\rangle\!\rangle_{\omega_0}$ represents the positive 
parity D0-brane and call the boundary state $
\|\mathrm{D0}\rangle\!\rangle$.\footnote{Here and in the following, the notation $\|\mathrm{D}p_1/\mathrm{D}p_2/\ldots\rangle\!\rangle$ refers to boundary states describing BPS D-branes with the respective RR charges at a concrete, but unspecified, point in their moduli space.} The D$p$-brane 
interpretation and parity of the rest of the $\omega_0$ 
boundary states follows from \eqref{eq:linBS} and also by considering various relations between massless RR boundary state coefficients. We obtain the following 9 representatives:
\begingroup
\allowdisplaybreaks
\begin{subequations}\label{eq:SU3Dpbr}
\begin{align}
 \|\bm{0},\bm{0},\bm{0}\rangle\!\rangle_{\omega_0} &=\|
 \mathrm{D0}\rangle\!\rangle \\
 \|L_1\!=\!+M_1\!=\!1\rangle\!\rangle_{\omega_0} &=\|
\mathrm{D0}/\overline{\mathrm{D2}}_{67}\rangle\!\rangle   \\
\|L_1\!=\!-M_1\!=\!1\rangle\!\rangle_{\omega_0} &= \|\mathrm{D2}_{67}\rangle\!\rangle
\\
\|L_4\!=\! +M_4\!=\!1\rangle\!\rangle_{\omega_0} &=\|
\mathrm{D0}/\overline{\mathrm{D2}}_{89}\rangle\!\rangle\\
\|L_4\!=\! -M_4\!=\!1\rangle\!\rangle_{\omega_0} &=\|
\mathrm{D2}_{89}\rangle\!\rangle\\
 \|L_{1,4}\!=\! +M_{1,4}\!=\!1\rangle\!\rangle_{\omega_0} &=\|
\mathrm{D0}/\overline{\mathrm{D2}}_{67}/
\overline{\mathrm{D2}}_{89}/\mathrm{D4}\rangle\!\rangle   \\
\|L_{1,4}\!=\! -M_{1,4}\!=\!1\rangle\!\rangle_{\omega_0} &=\|
 {\mathrm{D4}}\rangle\!\rangle\\
 \|L_{1,4}\!=\! +M_{1}\!=\!-M_4\!=\!1\rangle\!\rangle_{\omega_0} 
 &=\|\mathrm{D2}_{89}/\overline{\mathrm{D4}}\rangle\!\rangle\\
\|L_{1,4}\!=\!-M_{1}\!=\!+M_4\!=\!1\rangle\!\rangle_{\omega_0} &=\|\mathrm{D2}_{67}/\overline{\mathrm{D4}}\rangle\!\rangle
\,.
\end{align}
\end{subequations}
\endgroup
The rest of $\omega_0$ boundary states do not contribute new RR charges and describe the above 9 D-branes at different points in their moduli space.
Note that the masses of 1/2-BPS D-branes with these RR charges (computed from \eqref{eq:12BPSmassDef}) are indeed all identical and equal to the mass of the D0-brane. Furthermore, using the explicit $q$-series \eqref{eq:qseries}
for the $\mathcal{N}=2$ characters, it can be shown 
that\footnote{Here we explicitely include the thus far suppressed contribution $\eta(q)^{-4}$ from the external bosonic oscillators.} 
\begin{subequations}
\label{eq:omega0bos}
\begin{align}
\eta(q)^{-4}Z^{\omega_0}_{\alpha\alpha}(q)_\mathrm{NS} &= 8 + 224 q + 2976 q^2+\mathcal{O}(q^3)=Z_{\mathrm{f,NS}}(q) Z_{\mathrm{b}}(q)\,,\label{eq:omega0bosNS}\\[+4pt]
\eta(q)^{-4}\, Z^{\omega_0}_{\alpha\alpha}(q)_\mathrm{R} &= 8 + 224 q + 2976 q^2+\mathcal{O}(q^3)=Z_{\mathrm{f,R}}(q)\,Z_{\mathrm{b}}(q)\,,
\end{align}
\end{subequations}
where we have denoted $Z_{\mathrm{f,NS}}(q)=(1/2)\eta(q)^{-4}[\theta_3(q)^4-\theta_4(q)^4]$, $Z_{\mathrm{f,R}}(q)=(1/2)\eta(q)^{-4}\theta_2(q)^4$ and
$Z_{\mathrm{b}}(q)=\eta(q)^{-8}\big(\sum_{m,n\in\mathbb{Z}} q^{m^2+n^2+mn}\big)^2$.
This is consistent with the statement that $\omega_0$ boundary states describe D0-, D2- and D4-branes wrapping the $SU(3)^2$ 4-torus. We also recover the expected 8 massless physical modes in both the NS and R sector of the open string spectrum. It is clear from \eqref{eq:omega0bos} that the spectra are exactly bose-fermi degenerate: $Z^{\omega_0}_{\alpha{\alpha}}(q)=Z^{\omega_0}_{\alpha{\alpha}}(q)_\mathrm{NS}-Z^{\omega_0}_{\alpha{\alpha}}(q)_\mathrm{R}=0$, as required by spacetime supersymmetry. 
Indeed, it can be explicitely shown that the $\omega_0$ boundary states preserve the correct combinations of spacetime supercharges, which are specified by their RR charges and the requirement that they saturate the BPS bound. Starting from the matrix $\Gamma -\mathcal{M}$ defined in~\eqref{eq:defGamma}, it is straightforward to show that if the $\omega_0$ boundary states are to be (1/2-)BPS, they should satisfy
\begin{align}
(Q^\mathrm{L}_{\tau_1,\tau_2,\tau_3,\tau_4}
+e^{-\frac{2i\pi}{3}\bm{M}\cdot\sum_{r=1}^4\tau_r\bm{\beta}_r} Q^\mathrm{R}_{-
\tau_1,\tau_2,\tau_3,\tau_4})\| \bm{\Lambda},
\bm{M},\bm{0}\rangle\!\rangle_{\omega_0} =0\,,
\label{eq:consSCSU3}
\end{align}
where
$Q^\mathrm{L,R}
_{\tau_1,\tau_2,\tau_3,\tau_4}$ are the spacetime supercharges, $
\tau_r=\pm 1/2$, $r=1,2$ being the $SO(8)$ spins and $
\mathrm{L,R}$ being the $SO(1,9)$ chirality. 
To show that \eqref{eq:consSCSU3} indeed holds, one uses the fact \cite{Gutperle:1998hb} that the supercharges act on the Ishibashi states by spectral flow with parameters $\eta_r=-\tau_r$ for $r=1,2,3,4$.

To summarize, we observe that $\omega_0$ boundary states describe D0, $\mathrm{D2}_{67}$, $\mathrm{D2}_{89}$ and D4-branes together with some of their 1/2-BPS bound states. In order to obtain the rest of type IIA fundamental stable D$p$-brane boundary states (namely $\mathrm{D2}_{68}$, $\mathrm{D2}_{69}$, $\mathrm{D2}_{78}$, $\mathrm{D2}_{79}$), one should start, for instance, with gluing conditions $(1_\mathrm{A}2_\mathrm{A})(3_\mathrm{A})(4_\mathrm{A}5_\mathrm{A})(6_\mathrm{A})$.
Some of the higher-mass 1/2-BPS bound states of D$p$-branes can be obtained by considering more complicated gluing conditions: see Appendix \ref{app:class}, Table \ref{tab:classSU32} for a complete classification. Considering gluing conditions with
A-type for one of the constituent 2-tori and B-type for the other (such as $(1_\mathrm{A}2_\mathrm{A})(3_\mathrm{A})(4_\mathrm{B})(5_\mathrm{B})(6_\mathrm{B})$) yields the unstable non-BPS D$p$-branes of type IIA (with $p$ odd).

\subsubsection*{1/4-BPS D-brane boundary states}

Let us now consider the gluing automorphism \hbox{$\omega_1\equiv (1_\mathrm{B})(2_\mathrm{B})(3_\mathrm{B}4_\mathrm{B})(5_\mathrm{B})(6_\mathrm{B})$}. The resulting boundary states clearly cannot be factorized into boundary states separately wrapping the two $SU(3)$ 2-tori. The allowed Ishibashi states carry the same labels as in the $\omega_0$ case where, in addition, we also require that $l_3=l_4$ and $m_3=m_4$. This already fixes half of the overcounting for the $a=3,4$ minimal models so the projector now reads
\begin{align}
     \delta^{\omega_1}_{\bm{\lambda},\bm{\mu},\bm{\sigma}}& = 
     2^{-5} \frac{1}{12}\sum_{\zeta\in\mathbb{Z}_{12}}(-1)^{\zeta}e^{i
     \pi q(\bm{\mu},\bm{\sigma})\zeta}\prod_{r=3}^4\frac{1}{3}
     \sum_{t_r\in\mathbb{Z}_{3}}e^{\frac{2i\pi}{3} 
     t_r\bm{\beta}_r\cdot\bm{\mu}}\prod_{p=1}^8 \frac{1}{2}\sum_{\nu_p\in
     \mathbb{Z}_2}e^{i\pi\nu_p (\sigma_1 - \sigma_p)} 
     \nonumber\\
   &\hspace{1.7cm}\times
  \frac{1}{2}
   \sum_{\rho\in\mathbb{Z}_2} e^{i\pi\rho (l_3-l_4)}
   \frac{1}{6}\sum_{\rho'\in\mathbb{Z}_6} e^{\frac{i\pi}
   {3}\rho' (m_3-m_4)} \prod_{a=1}^{6}\frac{1}{2}\sum_{\xi_a\in\mathbb{Z}
   _2}e^{i\pi \xi_a(l_a+m_a+s_a)}\,.\label{eq:deltapSU3}
\end{align}
The construction of boundary states for the $\omega_1$ gluing conditions follows by the prescription given in~\cite{Recknagel:2002qq}. We write
\begin{equation}
    \|\alpha\rangle\!\rangle_{\omega_1} \equiv \|
    \bm{\Lambda},\bm{M},\bm{\Sigma}\rangle\!\rangle_{\omega_1} = 
   \frac{1}{\kappa_{\omega_1}^\alpha} \sum_{\bm{\lambda},\bm{\mu},\bm{\sigma}}\delta^{\omega_1}
    _{\bm{\lambda},\bm{\mu},\bm{\sigma}}B^{\alpha,\omega_1}
    _{\bm{\lambda},\bm{\mu},\bm{\sigma}}|\bm{\lambda},\bm{\mu},
    \bm{\sigma}\rangle\!\rangle_{\omega_1}\,,
\end{equation}
where
\begin{align}
    B^{\alpha,\omega_1}_{\bm{\lambda},\bm{\mu},\bm{\sigma}} &= 
    (-1)^\frac{\sigma_1^2}{2}e^{-\frac{i\pi}{2}\bm{\sigma}\cdot\bm{\Sigma}}
    \prod_{a=1,2,5,6}\frac{\sin [ \frac{\pi}{3}(l_a+1)
    (L_a+1)]}{\sin^\frac{1}{2} [ \frac{\pi}{3}(l_a+1)]}
    e^{\frac{i\pi}{3}m_a M_a}\times\nonumber\\
    &\hspace{6.5cm}
    \times \frac{\sin [ \frac{\pi}{3}(l_{3}+1)(L_{3}+1)]}
    {\sin [ \frac{\pi}{3}(l_{3}+1)]}e^{\frac{i\pi}{3}
    m_{3} M_{3}}\,.
    \label{eq:BSU3pcoef}
\end{align}
The range of the boundary state labels $\bm{\Lambda},\bm{M},
\bm{\Sigma}$ is the same as for the $\omega_0$ boundary 
states except that now there is only 
one $L$ and $M$ label associated with the cycle $(34)$.
The open string partition function can be evaluated as
\begin{align}
   Z_{\alpha\widetilde{\alpha}}^{\omega_1}(q)&=
  \frac{3^1 2^{-7}}{\kappa_{\omega_1}^\alpha
  \kappa_{\omega_1}^{\widetilde{\alpha}}}\!
  \sum_{\bm{\lambda}',\bm{\mu}',\bm{\sigma}'}\null\!\!\!\!\!^
  \mathrm{ev} \! \sum_{\zeta\in\mathbb{Z}_{12}}\sum_{\nu_p
  \in\mathbb{Z}_2}\sum_{t_r\in\mathbb{Z}_{3}}\! 
  (-1)^{\sigma_1'+\Sigma_1-\widetilde{\Sigma}_1}
 \delta^{(4)}_{\sigma'_1+ \Sigma_1-\widetilde{\Sigma}_1+
 \zeta-2\sum_{p=2}^8\nu_p+2} \!\!\prod_{\substack{{a=1}\\ a
 \neq 3,4}}^{6}\!\!\delta^{(2)}_{{l_a' +L_a-\widetilde{L}
 _a}}\nonumber\\[-3mm]
   &\hspace{1.8cm}\times\prod_{p=2}^8\delta^{(4)}
   _{{\sigma_p'+\Sigma_p-\widetilde{\Sigma}_p}+\zeta
   +2\nu_{p}}
 \prod_{a=1}^{2}\delta^{(6)}_{m_a'+M_a-\widetilde{M}_a+
 \zeta+2t_3} \prod_{a=5}^{6}\delta^{(6)}_{m_a'+M_a-
 \widetilde{M}_a+\zeta+2t_4}\nonumber\\[+2.0mm]
    &\hspace{4.5cm}\times
    \delta^{(2)}_{{l_3'+l_4'+{L}_3-\widetilde{L}_3}}
    \delta^{(6)}_{m_3'+m_4'+M_3-\widetilde{M}
    _3+2t_3+2t_4+2\zeta}\chi^{\bm{\lambda}'}_{\bm{\mu}',\bm{\sigma}'}
    (q) \,.\label{eq:nopexplSU3p}
\end{align}
Expanding the sums in \eqref{eq:nopexplSU3p} using computer algebra software, it is easy to establish that the minimal normalization which yields integer multiplicities is $\kappa_{\omega_1}^\alpha=3^{1}2^{-2}$. It also makes all $\omega_1$ D-branes elementary. As 
we will see below, this normalization also ensures that the 
RR charges of $\omega_1$ D-branes belong to the lattice 
generated by the RR charges of $\omega_0$ D-branes. 
\eqref{eq:nopexplSU3p} also gives the selection rule 
\hbox{$q'+Q-\widetilde{Q}-(M_3-\widetilde{M}_3)/3\in 
2\mathbb{Z}+1$}.  In particular, for $\alpha=
\tilde{\alpha}$ we have $q'\in 2\mathbb{Z}+1$ so that all $
\omega_1$ D-branes are stable. 
When calculating spectra of strings stretched between $\omega_0$ and $\omega_1$ branes, the corresponding closed string channel overlap contains characters of $\mathcal{N}=2$ tensor product representations twisted by the symmetric group element $(34)$ (as detailed in  \cite{Recknagel:2002qq} for general permutations). Considering also the additional phase appearing on the RHS of eq.~(5.7) in ref.~\cite{Brunner:2005fv}, we obtain
\begin{align}
\widetilde{Z}^{\omega_1\omega_0}_{\alpha\tilde{\alpha}}(\tilde{q})&=\null_{\omega_0}\!\langle\!\langle \Theta\widetilde{\alpha}\| \tilde{q}^{\frac{1}{2}(L_0+\overline{L}_{0}-\frac{c}{12})}\|{\alpha}\rangle\!\rangle_{\omega_1}\nonumber\\[+2mm]
&=\frac{1}{\kappa_{\omega_0}^{\widetilde{\alpha}}\kappa_{\omega_1}^{{\alpha}}}\sum_{\bm{\lambda},\bm{\mu},\bm{\sigma}}\delta^{\omega_1}_{\bm{\lambda},\bm{\mu},\bm{\sigma}}\delta_{s_3, s_4} e^{i\pi\left(\frac{m_3}{3}-\frac{s_3}{2}\right)}B^{\widetilde{\alpha},\omega_0}_{\bm{\lambda},\bm{-\mu},\bm{-\sigma}} B^{{\alpha},\omega_1}_{\bm{\lambda},\bm{\mu},\bm{\sigma}}\times\nonumber\\[-2mm]
&\hspace{2.2cm}\times
\chi^{l_1}_{m_1,s_1}(\tilde{q})\chi^{l_2}_{m_2,s_2}(\tilde{q})\chi^{l_3}_{m_3,s_3}(\tilde{q}^2)\chi^{l_5}_{m_5,s_5}(\tilde{q})\chi^{l_6}_{m_6,s_6}(\tilde{q})\prod_{r=1}
    ^{2}\chi_{\sigma_r}(\tilde{q})\,.
\end{align}
S-transforming into the open string channel and substituting for $\kappa_{\omega_0}^{\widetilde{\alpha}},\kappa_{\omega_1}^{{\alpha}}$, we have
\begin{align}
Z_{\alpha\widetilde{\alpha}}^{\omega_1,\omega_0}(q) &=\frac{1}{48}\!\sum_{\bm{\lambda}',\bm{\mu}',\bm{\sigma}'}\null\!\!\!\!\!^
  \mathrm{ev,\bar{4}} \! \sum_{\zeta\in\mathbb{Z}_{12}}\sum_{t_r\in\mathbb{Z}_{3}}\sum_{\nu_p\in
     \mathbb{Z}_2}\sum_{\xi_4\in\mathbb{Z}_2}(-1)^{\sigma_1'+\Sigma_1-\widetilde{\Sigma}_1}
 \delta^{(4)}_{\sigma'_1+ \Sigma_1-\widetilde{\Sigma}_1+
 \zeta-2\sum_{p=2}^8\nu_p+2}\nonumber\\
 &\hspace{-1.1cm}\times\delta^{(6)}_{m_3'+M_3-\widetilde{M}_3-\widetilde{M}_4+1+2\zeta+2t_3+2t_4+3\xi_4} \prod_{a=1}^{2}\delta^{(6)}_{m_a'+M_a-\widetilde{M}_a+
 \zeta+2t_3} \prod_{a=5}^{6}\delta^{(6)}_{m_a'+M_a-
 \widetilde{M}_a+\zeta+2t_4}\nonumber\\ 
   &\hspace{2.3cm}\times\delta^{(4)}
   _{{s_3'+S_3+S_4-\widetilde{S}_3-\widetilde{S}_4}+2\zeta
   +2\nu_{5}+2\nu_{6}+1+2\xi_4}\prod_{\substack{p=2\\ p\neq 5,6}}^8\delta^{(4)}
   _{{\sigma_p'+\Sigma_p-\widetilde{\Sigma}_p}+\zeta
   +2\nu_{p}}\nonumber\\[-3.5mm]
   &\hspace{6.5cm}\times \delta^{(2)}_{l_3'+L_3-\widetilde{L}_3-\widetilde{L}_4+\xi_4}\prod_{\substack{{a=1}\\ a
 \neq 3,4}}^{6}\!\!\delta^{(2)}_{{l_a' +L_a-\widetilde{L}
 _a}}\nonumber\\[-3mm]
  &\hspace{2cm}\times \chi^{l'_1}_{m'_1,s'_1}({q})\chi^{l'_2}_{m'_2,s'_2}({q})\chi^{l'_3}_{m'_3,s'_3}({q}^\frac{1}{2})\chi^{l'_5}_{m'_5,s'_5}({q})\chi^{l'_6}_{m'_6,s'_6}({q})\prod_{r=1}
    ^{2}\chi_{\sigma_r}({q})\,,\label{eq:om0om1}
\end{align}
where the $\bm{\lambda}',\bm{\mu}',\bm{\sigma}'$ sum does not run over $l_4',m_4',s_4'$. Expanding the sums in \eqref{eq:om0om1} using computer algebra software, it is easy to verify that the coefficients in front of the characters are integers. We have therefore automatically obtained consistent spectra for open strings stretched between $\omega_0$ and $\omega_1$ boundary states.

Let us now compute the couplings of the $\omega_1$ 
boundary states to massless closed string states. In 
order to be able to compare these with the couplings 
computed for the $\omega_0$ boundary states, we will probe 
the $\omega_1$ boundary states with the same massless closed 
string states as we probed the $\omega_0$ boundary states in 
the previous subsection, namely with $|\bm{\lambda},\bm{\mu},
\bm{\sigma}\rangle_{\omega_0}$. Keeping the relative phase of~\cite{Brunner:2005fv} in mind, we obtain
\begin{align}
    \null_{\omega_0}\!\langle\bm{\lambda},\bm{\mu},\bm{\sigma}\|\alpha
    \rangle\!\rangle_{\omega_1} = \frac{1}{\kappa_{\omega_1}^\alpha} \delta_{\bm{\lambda},\bm{\mu},\bm{\sigma}}^{\omega_1}\,\delta_{s_3, s_4}\, e^{i\pi
    \left(\frac{m_3}{3}-\frac{s_3}{2}\right)} B^{\alpha,
    \omega_1}_{\bm{\lambda},\bm{\mu},\bm{\sigma}}\,.
    \label{eq:permPhase}
\end{align}
Noting that the Ishibashi states which 
provide couplings to the massless NSNS closed string states 
along the internal 4-torus are not allowed by the $
\omega_1$ gluing conditions, we conclude that \eqref{eq:linBS} must violated. Since the $\omega_1$ boundary states are elementary, it follows that they are examples of stable non-conventional boundary 
states. Also note that $g_\alpha^{\omega_1}=1=\sqrt{3}\,g_{\alpha}^{\omega_0}$ meaning that
the mass of all $\omega_1$ D-branes is $\sqrt{3}$ 
times the mass of $\omega_0$ D-branes. The allowed massless RR Ishibashi states can be parametrized as $\bm{\lambda}=\bm{0}$, $\bm{\mu}=2\sum_{r=3}^4 \tau_r\bm{\beta}_r$ and \hbox{$\bm{\sigma}=2\sum_{r=1}^4 \tau_r\bm{\beta}_r$} with $\tau_1 
\tau_2 \tau_3 \tau_4<0$ and $\tau_3=
\tau_4$. 
Starting from~\eqref{eq:BSU3pcoef}, we 
therefore obtain
\begin{align}
     _{\omega_0}\!\langle \tau_1\tau_2\tau_3
     \tau_4\|\bm{\Lambda},\bm{M},\bm{0}\rangle\!
     \rangle_{\omega_1}  &= 
          i\,\delta_{\tau_3,\tau_4}\, e^{\frac{2i\pi}{3}\tau_3(-\frac{1}{2}+
          \sum_{a\neq 4}M_a)}  \,.
\end{align}
We find that $\omega_1$ boundary states can carry 3 distinct sets of RR charges. The corresponding representatives can be 
chosen as $\|\bm{0},\bm{0},\bm{0}\rangle\!
\rangle_{\omega_1}$, $\|L_1\!=\! M_1\!=\!1\rangle\!
\rangle_{\omega_1}$, $\|L_1\!=\! -M_1\!=\!1\rangle\!
\rangle_{\omega_1}$. Their RR 
couplings satisfy the relations
\begin{subequations}
\label{eq:RRcouplings}
\begin{align}
     _{\omega_0}\!\langle \tau_1\tau_2\tau_3
     \tau_4\|\bm{0},\bm{0},\bm{0}\rangle\!
     \rangle_{\omega_1} 
    \! &=  \null_{\omega_0}\!\langle 
    \tau_1\tau_2\tau_3
     \tau_4\|\mathrm{D2_{67}}
     \rangle\!\rangle+\null_{\omega_0}\!\langle\tau_1\tau_2\tau_3
     \tau_4\|\mathrm{D2}_{89}\rangle\!
     \rangle+ \nonumber\\
     &\hspace{5.8cm}+\null_{\omega_0}\!\langle 
    \tau_1\tau_2\tau_3
     \tau_4\|
     \overline{\mathrm{D4}}\rangle\!\rangle\\[4pt]
     _{\omega_0}\!\langle \tau_1\tau_2\tau_3
     \tau_4\|L_1\!=\!+M_1\!=\!1\rangle\!
     \rangle_{\omega_1}\! &= \null_{\omega_0}\!\langle\tau_1\tau_2\tau_3
     \tau_4\|\mathrm{D0}\rangle\!\rangle+ 
     \null_{\omega_0}\!\langle \tau_1\tau_2\tau_3
     \tau_4\|\overline{\mathrm{D4}}\rangle\!\rangle\\[4pt]
       _{\omega_0}\!\langle  \tau_1\tau_2\tau_3
     \tau_4\|L_1\!=\!-M_1\!=\!1\rangle\!
       \rangle_{\omega_1} 
     \!&=  
     \null_{\omega_0}\!\langle  \tau_1\tau_2\tau_3
     \tau_4\|{\mathrm{D2}_{67}}\rangle\!\rangle
    +\null_{\omega_0}\!\langle  
    \tau_1\tau_2\tau_3
     \tau_4\|
    {\mathrm{D2}}_{89}\rangle\!\rangle +\nonumber\\
    &\hspace{5.8cm}+\null_{\omega_0}\!\langle  \tau_1\tau_2\tau_3
     \tau_4\|\overline{\mathrm{D0}}\rangle\!\rangle
\end{align}
\end{subequations}
This means that \hbox{$\|\bm{0},\bm{0},\bm{0}\rangle\!\rangle_{\omega_1} $}, \hbox{$\|L_1\!=\!+M_1\!=\!1\rangle\!\rangle_{\omega_1}$} and \hbox{$\|L_1\!=\!-M_1\!=\!1\rangle\!\rangle_{\omega_1}$} carry the same RR charges as do the 1/4-BPS bound states $\mathrm{D2}_{67}/\mathrm{D2}_{89}/\overline{\mathrm{D4}}$, $\mathrm{D0}/\overline{\mathrm{D4}}$ and $\overline{\mathrm{D0}}/\mathrm{D2}_{67}/\mathrm{D2}_{89}$, respectively. Also note that the mass $\sqrt{3}$ obtained from the boundary states agrees with the mass obtained from saturating the BPS bound for the corresponding RR charges (formula~\eqref{eq:14BPSSU3}). It also follows from \eqref{eq:nopexplSU3p} (upon substituting the $q$-expansions \eqref{eq:qseries}) that
\begin{subequations}
\label{eq:omega1bos}
\begin{align}
\eta(q)^{-4} Z^{\omega_1}_{\alpha\alpha}(q)_\mathrm{NS} &=12 + 16 q^{\frac{1}{3}} + 48 q^{\frac{2}{3}} + 368 q + 368 q^{\frac{4}{3}} + 864 q^{\frac{5}{3}}+\mathcal{O}(q^{2})\,,\label{eq:omega1bosNS}\\
\eta(q)^{-4}Z^{\omega_1}_{\alpha\alpha}(q)_\mathrm{R} &=12 + 16 q^{\frac{1}{3}} + 48 q^{\frac{2}{3}} + 368 q + 368 q^{\frac{4}{3}} + 864 q^{\frac{5}{3}} +\mathcal{O}(q^{2})\,,
\end{align}
\end{subequations}
for all $\alpha$. 
Also, we have shown to $\mathcal{O}(q^{1000})$ that {$Z^{\omega_1}_{\alpha\alpha}(q)=Z^{\omega_1}_{\alpha\alpha}(q)_\mathrm{NS}-Z^{\omega_1}_{\alpha\alpha}(q)_\mathrm{R}=0$}, i.e.\ that the open string spectrum is bose-fermi degenerate, which suggests that the $\omega_1$ boundary states preserve some amount of spacetime supersymmetry. Saturating the BPS bound for the above-computed RR charges, we find that if the $\omega_1$ boundary states are to be (1/4-)BPS, we need
\begin{align}
(Q^\mathrm{L}_{\tau_1,\tau_2,\tau_3,
     \tau_4}+ 
e^{-\frac{2i\pi}{3}\tau_3(-\frac{1}{2}+\sum_{a\neq 4}
M_a)} Q^\mathrm{R}_{-\tau_1,\tau_2,\tau_3,
     \tau_4})\| \bm{\Lambda},\bm{M},\bm{0}\rangle\!
\rangle_{\omega_1} =0\,,\quad \tau_3=\tau_4\,.
\label{eq:consSCSU3p}
\end{align}
The action of $Q^\mathrm{L,R}_{\tau_1,\tau_2,\tau_3,\tau_4}$ on the $\omega_1$ boundary states adheres to similar rules as for the $\omega_0
$ boundary states, where, in addition, the phase appearing in~\eqref{eq:permPhase} produces an extra factor
of $ e^{-\frac{i\pi}{3}\tau_3}$ upon acting with right-moving supercharges. It follows that~\eqref{eq:consSCSU3p} holds true.
Note that the $q$-expansion \eqref{eq:omega1bos} shows that there are 12 massless modes in both NS and R sector. 
This seems to be in agreement with 
the instanton description of the bound states, as the identification \eqref{eq:kN} for the respective RR charges together with the ADHM formula $8+4kN$ indeed requires 12 exactly marginal operators. We leave the proof of exact marginality of the 12 massless NS boundary operators living on the $\omega_1$ boundary states for future work. Note that the spectra of all $\omega_1$ boundary states are identical as the corresponding bound states are T-dual to one another.

Finally, let us exhibit the correspondence between the $\|L_1\!=\!M_1\!=\! 1\rangle\!\rangle_{\omega_1}$ boundary state and the D1/D5 supergravity solution obtained by \cite{Dhar:1999ax}. Since the boundary state satisfies the no-force condition\footnote{At one loop in the open string channel, this is expressed as $Z_{\alpha\alpha}^{\omega_1}(q)=0$. Spacetime supersymmetry ensures that this 1-loop result can be extended to arbitrary loop level.}, it should yield the same asymptotic behavior of massless NSNS and RR fields as a regular solution of supergravity. Here we will T-dualize the boundary state in the $X^5$ direction, so that it looks like a type IIB D-string from the viewpoint of the six-dimensional non-compact space. In the NSNS sector, the boundary state only couples to massless closed string states in the six non-compact dimensions. Denoting $\Xi_{\mu\nu}=\mbox{diag}\,[1,1,1,1,1,-1,0,0,0,0]_{\mu\nu}$ (in the cartesian coordinates $x^\mu$, $\mu=0,\ldots,10$) and using the results of \cite{DiVecchia:1997vef}, one obtains\footnote{Now we also double-wick rotate back to the usual picture so that we impose N conditions on $X^0,X^5$ and D conditions on $X^{1},X^{2},X^{3},X^{4}$}
\begin{align}
\delta g_{\mu\nu}(r) =   Kr^{-2}\Xi_{\mu\nu}\,,\qquad
\delta B_{\mu\nu}(r) =  0\,,\qquad
\delta \phi(r) = 0\,,
\label{eq:metD1D5}
\end{align}
for the asymptotic deviations of NSNS fields from their background values. We have also denoted\footnote{Here $\kappa_{10}^2=8\pi G_{10}$ with $G_{10}=8\pi^6 g_\mathrm{s}^2$ being the ten-dimensional Newton's constant.}
\begin{align}
K =  \frac{\kappa_{10}\sqrt{3}N T_1 }{ 4\pi^2 } = {8\sqrt{3}\pi^4 g_\mathrm{s} N}\,.
\label{eq:KSU3}
\end{align}
where $T_1=\sqrt{\pi}(2\pi)^2$ is the D-string tension and $N$ is the number of superposed \hbox{$\|L_1\!=\!M_1\!=\!1\rangle\!\rangle_{\omega_1}$} boundary states. 
Let us now describe the supergravity solution found in \cite{Dhar:1999ax}. Let us assume that only $B_{67}$ and $B_{89}$ are non-zero and that $T^4 = T^2\times T^2$. We will parametrize the solution by $\mu_1$, $\mu_5$, $\varphi$, $\psi$, so that the asymptotic values $B_{\mu\nu}^\infty$ of the $B$-field components are given by
\begin{subequations}
\begin{align}
B^{\infty}_{67}&=\frac{\mu_5 \sin\varphi \cos\psi -\mu_1\cos\varphi\sin\psi}{\mu_5 \cos\varphi\cos\psi+\mu_1\sin\varphi\sin\psi}\\
B^{\infty}_{89}&=\frac{\mu_5\cos\varphi\sin\psi -\mu_1 \sin\varphi\cos\psi}{\mu_5\cos\varphi\cos\psi+\mu_1\sin\varphi\sin\psi}
\end{align}
\end{subequations}
while the charges of the source D1- and D5-branes are given by
\begin{subequations}
\begin{align}
Q_5 &= \beta(\mu_5 \cos\varphi\cos\psi + \mu_1 \sin\varphi\sin\psi)\,,\\
Q_1 &=  V_{67}V_{89} [\beta(\mu_1 \cos \varphi\cos\psi +\mu_5 \sin\varphi\sin\psi)-{B_{67}^\infty} {B_{89}^\infty} Q_5]\,,
\end{align}
\end{subequations}
where $\beta = \pi /(8 G_{10})$. Let us denote
\begin{subequations}
\begin{align}
\mu_\varphi &= |\mu_1| \sin^2\varphi + |\mu_5|\cos^2\varphi\,,\\
\mu_\psi &=|\mu_1| \sin^2\psi + |\mu_5|\cos^2\psi\,,
\end{align}
\end{subequations}
together with
\begin{align}
f_{1,5}(r) &= 1+ \frac{|\mu_{1,5}|}{2r^2}\,,\qquad
Z_{\varphi,\psi}(r) = 1+\frac{|\mu_{\varphi,\psi}|}{2r^2}\,,
\end{align}
and also
\begin{subequations}
\begin{align}
K^{(3)}&= -\mu_5 f_5^{-2}r^{-3}dr\wedge dt\wedge dx^5 +\mu_1\epsilon_3\\
\widetilde{K}^{(3)}&= -\mu_1 f_1^{-2}r^{-3}dr\wedge dt\wedge dx^5 +\mu_5\epsilon_3\,,
\end{align}
\end{subequations}
where $\epsilon_3$ is the volume-form on the 3-sphere surrounding the D-brane.
The solution then reads
\begin{subequations}
\label{eq:sugra}
\begin{align}
ds^2 &= (f_1 f_5)^{-1/2}[-(dx^0)^2+(dx^5)^2]+(f_1 f_5)^{1/2}(dr^2 + r^2d\Omega_3^2)+\nonumber\\
&\hspace{1.5cm}+(f_1 f_5)^{1/2}\{Z_\varphi^{-1}[(dx^6)^2 +(dx^7)^2]+Z_\psi^{-1}[(dx_8)^2 +(dx_9)^2]\}\,,\\
e^{2\phi} &=f_1 f_5 / (Z_\varphi Z_\psi)\,,\\[+2pt]
B&=[B_{67}^\infty+Z_{\varphi}^{-1}(f_1-f_5)\sin\varphi\cos\varphi]dx^6\wedge dx^7+\nonumber\\
&\hspace{1.5cm}+[B_{89}^\infty+Z_{\psi}^{-1}(f_1-f_5)\sin\psi\cos\psi]dx^8\wedge dx^9\,,\\
F^{(3)} &=\widetilde{K}^{(3)}\cos\varphi\cos\psi +K^{(3)}\sin\varphi\sin\psi\,,\\
F^{(5)} &= Z_\varphi^{-1}(-f_5K^{(3)}\cos\varphi \sin\psi +f_1\widetilde{K}^{(3)}\cos\psi\sin\varphi) \wedge dx^6 \wedge dx^7\nonumber\\[-2pt]
&\hspace{1.5cm}+Z_{\psi}^{-1}(-f_5 K^{(3)}\cos\psi\sin\varphi+f_1 \widetilde{K}^{(3)}\cos\varphi\sin\psi)\wedge dx^{8}\wedge dx^{9}\,.
\end{align}
\end{subequations}
Setting $\mu_1=-\mu_5\equiv\mu$, $V_{67}=V_{89}=\sqrt{3}/2$, $\tan(\varphi+\psi)=1/\sqrt{3}$, we get \hbox{$f_1=f_5=Z_\varphi=Z_\psi\equiv f$}, $K^{(3)}=-\widetilde{K}^{(3)}$ and  $B^{\infty}_{67}=\tan(\varphi+\psi)=B^{\infty}_{89}={1}/{\sqrt{3}}$ (which yields fluxes \hbox{$b_{67}=b_{89}=+1/2$}, as appropriate for the $SU(3)^2$ 4-torus). We also have $Q_5 = -(\sqrt{3}/2)\beta\mu=-Q_1$, which is always satisfied by superpositions of the boundary state $\|L_1\!=\! M_1=1\rangle\!\rangle_{\omega_1}$. Using the ADM mass formula to identify $N/(2\pi g_\mathrm{s})=|Q_1|/(V_{67}V_{89})$, the solution \eqref{eq:sugra} then turns into
\begin{subequations}
\begin{align}
ds^2 &= f(r)^{-1}[-(dx^0)^2+(dx^5)^2]+f(r)(dr^2 + r^2d\Omega_3^2)+\nonumber\\[+2pt]
&\hspace{5cm}+(dx^6)^2 +(dx^7)^2+(dx^8)^2 +(dx^9)^2\,,\label{eq:metSU3}\\[-4pt]
e^{2\phi} &=1\,,\label{eq:dilSU3}\\[2.7mm]
B&=+\frac{1}{\sqrt{3}}(dx^6\wedge dx^7+dx^8\wedge dx^9)\,,\\
F^{(3)} &=-\frac{\sqrt{3}}{2}\mu[f(r)^{-2} r^{-3} dr\wedge dt\wedge dx^5 +\epsilon_3]\,,\label{eq:F3SU3}\\[+1.9mm]
F^{(5)} &=-\frac{1}{2}\mu [f(r)^{-2} r^{-3} dr\wedge dt\wedge dx^5 +\epsilon_3] \wedge( dx^6 \wedge dx^7+dx^8\wedge dx^9)\,.\label{eq:F5SU3}
\end{align}
\end{subequations}
with
\begin{align}
f(r) = 1+\frac{|Q_{1}|/(\sqrt{3}\beta)}{ r^2} =1+ \frac{8\sqrt{3}\pi^4 g_\mathrm{s} N}{ r^2}\,.
\label{eq:fSU3}
\end{align}
It is then easy to see that eqs. \eqref{eq:metSU3}, \eqref{eq:dilSU3} together with \eqref{eq:fSU3} are in a precise agreement with \eqref{eq:metD1D5} and \eqref{eq:KSU3} as $r\to \infty $. In addition, given that the RR couplings \eqref{eq:RRcouplings} are expressed as sums of the RR couplings for the constituent D$p$-branes, the asymptotic supergravity profiles of the massless RR fields computed from the boundary state are necessarily given by superpositions of the corresponding profiles for elementary D$p$-brane solutions. This is indeed the property of \eqref{eq:F3SU3} and \eqref{eq:F5SU3}.

The evidence presented in this subsection leads us to conclude that the $\omega_1$ 
boundary states describe stable elementary D-branes with masses, RR charges and other properties matching those of the truly bound 1/4-BPS states of D$p$-branes. We therefore identify
\begin{subequations}
\begin{align}
\|\bm{0},\bm{0},\bm{0}\rangle\!\rangle_{\omega_1} &= \|
{\mathrm{D2}_{67}}/{\mathrm{D2}}_{89}/\overline{\mathrm{D4}}
\rangle\!\rangle \\
\|L_1\!=\! +M_1\!=\!1\rangle\!\rangle_{\omega_1}  &=\|
\mathrm{D0}/\overline{\mathrm{D4}}\rangle\!\rangle\\
\|L_1\!=\! -M_1\!=\!1\rangle\!\rangle_{\omega_1}  &= \|
\overline{\mathrm{D0}}/{\mathrm{D2}}_{67}/
{\mathrm{D2}}_{89}\rangle\!\rangle 
\end{align}
\end{subequations}
It can be shown (by repeating the above steps with some 
minor modifications) that rational boundary states can be used to describe more 1/4-BPS bound states with mass $\sqrt{3}$. For instance, setting the gluing automorphism to $(1_\mathrm{B})(2_\mathrm{B})(3_\mathrm{A}4_\mathrm{A})(5_\mathrm{B})(6_\mathrm{B})$, one obtains the \hbox{1/4-BPS} bound states $\mathrm{D0}/
\overline{\mathrm{D2}}_{89}/\mathrm{D4}$, $\mathrm{D2}_{67}/
\overline{\mathrm{D2}}_{89}$, ${\mathrm{D0}}/\overline{\mathrm{D2}}_{67}/
{\mathrm{D4}}$. Boundary states for some of the higher-mass 1/4-BPS bound states can be constructed by considering more complicated gluing automorphisms. See Appendix \ref{app:class}, Table \ref{tab:classSU32} for a complete list of results.

\subsubsection*{Stable non-BPS boundary states}

Here we consider the gluing automorphism \hbox{$\omega_2\equiv (1_\mathrm{B}4_\mathrm{B})(2_\mathrm{A}5_\mathrm{A})(3_\mathrm{B})(6_\mathrm{B})$}. Again, the resulting boundary states clearly cannot be factorized into boundary states wrapping the two $SU(3)$ 2-tori.
The structure of the closed string spectrum dictates that the allowed Ishibashi state labels must obey $l_1=l_4$, $m_1=m_4$ and $l_2=l_5$, $m_2=-m_5$. The corresponding projector reads
\begin{align}
     \delta^{\omega_2}_{\bm{\lambda},\bm{\mu},\bm{\sigma}}& = 
     2^{-4} \frac{1}{12}\sum_{\zeta\in\mathbb{Z}_{12}}(-1)^{\zeta}e^{i
     \pi q(\bm{\mu},\bm{\sigma})\zeta}\prod_{r=3}^4\frac{1}{3}
     \sum_{t_r\in\mathbb{Z}_{3}}e^{\frac{2i\pi}{3} 
     t_r\bm{\beta}_r\cdot\bm{\mu}}\prod_{p=1}^8 \frac{1}{2}\sum_{\nu_p\in
     \mathbb{Z}_2}e^{i\pi\nu_p (\sigma_1 - \sigma_p)} 
     \nonumber\\
   &\hspace{4mm}\times
  \frac{1}{2}
   \sum_{\rho_1\in\mathbb{Z}_2} e^{i\pi\rho (l_1-l_4)}
   \frac{1}{6}\sum_{\rho_1'\in\mathbb{Z}_6} e^{\frac{i\pi}
   {3}\rho' (m_1-m_4)}  \frac{1}{2}
   \sum_{\rho_2\in\mathbb{Z}_2} e^{i\pi\rho (l_2-l_5)}
   \frac{1}{6}\sum_{\rho_2'\in\mathbb{Z}_6} e^{\frac{i\pi}
   {3}\rho' (m_2+m_5)}\nonumber\\[-3mm]
   &\hspace{8.2cm}\times \prod_{a=1}^{6}\frac{1}{2}\sum_{\xi_a\in\mathbb{Z}
   _2}e^{i\pi \xi_a(l_a+m_a+s_a)}\,.\label{eq:deltapSU3E}
\end{align}
The $\omega_2$ boundary states then satisfy
\begin{equation}
    \|\alpha\rangle\!\rangle_{\omega_2} \equiv \|
    \bm{\Lambda},\bm{M},\bm{\Sigma}\rangle\!\rangle_{\omega_2} = 
   \frac{1}{\kappa_{\omega_2}^\alpha} \sum_{\bm{\lambda},\bm{\mu},\bm{\sigma}}\delta^{\omega_2}
    _{\bm{\lambda},\bm{\mu},\bm{\sigma}}B^{\alpha,\omega_2}
    _{\bm{\lambda},\bm{\mu},\bm{\sigma}}|\bm{\lambda},\bm{\mu},
    \bm{\sigma}\rangle\!\rangle_{\omega_2}\,,
\end{equation}
where
\begin{align}
    B^{\alpha,\omega_2}_{\bm{\lambda},\bm{\mu},\bm{\sigma}} &= 
    (-1)^\frac{\sigma_1^2}{2}e^{-\frac{i\pi}{2}\bm{\sigma}\cdot\bm{\Sigma}}
    \prod_{a=3,6}\frac{\sin [ \frac{\pi}{3}(l_a+1)
    (L_a+1)]}{\sin^\frac{1}{2} [ \frac{\pi}{3}(l_a+1)]}
    e^{\frac{i\pi}{3}m_a M_a}\times\nonumber\\
    &\hspace{6cm}
    \times \prod_{a=1,2}\frac{\sin [ \frac{\pi}{3}(l_{a}+1)(L_{a}+1)]}
    {\sin [ \frac{\pi}{3}(l_{a}+1)]}e^{\frac{i\pi}{3}
    m_{a} M_{a}}\,.
    \label{eq:BSU3pcoefE}
\end{align}
with a single $L$ and $M$ label for the cycles $(14)$ and $(25)$.
The open string partition function can be evaluated as
\begin{align}
   Z_{\alpha\widetilde{\alpha}}^{\omega_2}(q)&=
  \frac{2^{-6}3^{-1}}{\kappa_{\omega_2}^\alpha
  \kappa_{\omega_2}^{\widetilde{\alpha}}}\!
  \sum_{\bm{\lambda}',\bm{\mu}',\bm{\sigma}'}\null\!\!\!\!\!^
  \mathrm{ev} \! \sum_{\zeta\in\mathbb{Z}_{12}}\sum_{\nu_p
  \in\mathbb{Z}_2}\sum_{t_r\in\mathbb{Z}_{3}}\! \delta^{(2)}_{{l_1'+l_4'+{L}_1-\widetilde{L}_1}}\delta^{(2)}_{{l_2'+l_5'+{L}_2-\widetilde{L}_2}} \delta^{(2)}_{{l_3' +L_3-\widetilde{L}
 _3}}\delta^{(2)}_{{l_6' +L_6-\widetilde{L}
 _6}}
  \!\!\nonumber\\[-3mm]
   & \hspace{2.5cm}\times(-1)^{\sigma_1'+\Sigma_1-\widetilde{\Sigma}_1}
 \delta^{(4)}_{\sigma'_1+ \Sigma_1-\widetilde{\Sigma}_1+
 \zeta-2\sum_{p=2}^8\nu_p+2}\prod_{p=2}^8\delta^{(4)}
   _{{\sigma_p'+\Sigma_p-\widetilde{\Sigma}_p}+\zeta
   +2\nu_{p}}\nonumber\\
     &\hspace{4cm}\times    
    \delta^{(6)}_{m_1'+m_4'+M_1-\widetilde{M}
    _1+2t_3+2t_4+2\zeta}
    \delta^{(6)}_{m_2'-m_5'+M_2-\widetilde{M}
    _2+2t_3-2t_4}\nonumber\\[+3mm]
    &\hspace{4.8cm}\times\delta^{(6)}_{m_3'+M_3-\widetilde{M}_3+
 \zeta+2t_3}\delta^{(6)}_{m_6'+M_6-
 \widetilde{M}_6+\zeta+2t_4}\chi^{\bm{\lambda}'}_{\bm{\mu}',\bm{\sigma}'}
    (q)\,.\label{eq:nopexplSU3pE}
\end{align}
Expanding the sums in \eqref{eq:nopexplSU3pE} using Mathematica, we find that the minimal normalization which yields consistent open 
string spectra (integer multiplicities) is $\kappa_{\omega_2}^\alpha=2^{-1}$. It also makes all $\omega_2$ boundary states elementary. Let us also check the consistency of stretched string spectra between $\omega_0$ D-branes (which include some of the fundamental $1/2$-BPS D$p$-branes) and the $\omega_2$ D-branes. Here we encounter characters of $\mathcal{N}=2$ tensor product representations twisted by the two transpositions $(14)$, $(25)$.\footnote{Note that in spite of imposing A-type gluing conditions on both minimal models in the $(25)$ cycle, the mirror-twisted ($\sigma: J\to -J$) representations do not appear as only the states containing an even number of $J^{(2)}$ and $J^{(5)}$ oscillators in both left- and right-moving sector survive twisting by the transposition $(25)$. However, if we were to calculate overlaps of $(2_\mathrm{B})(5_\mathrm{B})$ boundary states with $(2_\mathrm{A}5_\mathrm{B})$ or $(2_\mathrm{B}5_\mathrm{A})$ boundary states, mirror-twisted characters would show up.}
Including the relative phases of \cite{Brunner:2005fv}, we obtain
\begin{align}
Z_{\alpha\widetilde{\alpha}}^{\omega_2,\omega_0}(q) &=\frac{1}{192}\!\sum_{\bm{\lambda}',\bm{\mu}',\bm{\sigma}'}\null\!\!\!\!\!^
  \mathrm{ev,\bar{4}\bar{5}} \! \sum_{\zeta\in\mathbb{Z}_{12}}\sum_{t_r\in\mathbb{Z}_{3}}\sum_{\nu_p\in
     \mathbb{Z}_2}\sum_{\xi_4\in\mathbb{Z}_2}\sum_{\xi_5\in\mathbb{Z}_2}(-1)^{\sigma_1'+\Sigma_1-\widetilde{\Sigma}_1}
 \delta^{(4)}_{\sigma'_1+ \Sigma_1-\widetilde{\Sigma}_1+
 \zeta-2\sum_{p=2}^8\nu_p+2}\nonumber\\[-3mm]
 & \hspace{-12mm}\times \delta^{(2)}_{{l_3' +L_3-\widetilde{L}_3}}\delta^{(2)}_{{l_6' +L_6-\widetilde{L}_6}}\delta^{(6)}_{m_3'+M_3-\widetilde{M}_3+
 \zeta+2t_3} \delta^{(6)}_{m_6'+M_6-
 \widetilde{M}_6+\zeta+2t_4}\prod_{\substack{p=2\\ p\neq 3,4,6,7}}^8\delta^{(4)}
   _{{\sigma_p'+\Sigma_p-\widetilde{\Sigma}_p}+\zeta
   +2\nu_{p}}\nonumber\\[-2.5mm]
 &\hspace{6.6cm}\times \delta^{(2)}_{l_1'+L_1-\widetilde{L}_1-\widetilde{L}_4+\xi_4}\delta^{(2)}_{l_2'+L_2-\widetilde{L}_2-\widetilde{L}_5+\xi_5}\nonumber\\[2.5mm]
 &\hspace{1.6cm}\times\delta^{(6)}_{m_1'+M_1-\widetilde{M}_1-\widetilde{M}_4+1+2\zeta+2t_3+2t_4+3\xi_4} \delta^{(6)}_{m_2'+M_2-\widetilde{M}_2+\widetilde{M}_5+1+2t_3-2t_4-3\xi_5} \nonumber\\[2.5mm] 
   &\hspace{1.2cm}\times\delta^{(4)}
   _{{s_1'+S_1+S_4-\widetilde{S}_1-\widetilde{S}_4}+2\zeta
   +2\nu_{3}+2\nu_{6}+1+2\xi_4}\delta^{(4)}
   _{{s_2'+S_2-S_5-\widetilde{S}_2+\widetilde{S}_5}
   +2\nu_{4}-2\nu_{7}+1-2\xi_5}\nonumber\\[0.5mm]
  &\hspace{3.6cm}\times \chi^{l'_1}_{m'_1,s'_1}({q}^{\frac{1}{2}})\chi^{l'_2}_{m'_2,s'_2}({q}^{\frac{1}{2}})\chi^{l'_3}_{m'_3,s'_3}({q})\chi^{l'_6}_{m'_6,s'_6}({q})\prod_{r=1}
    ^{2}\chi_{\sigma_r}({q})\,,\label{eq:om0om2}
\end{align}
where the $\bm{\lambda}',\bm{\mu}',\bm{\sigma}'$ sum does not run over $l_4',m_4',s_4'$ and $l_5',m_5',s_5'$. Expanding the summations in \eqref{eq:om0om1} using Mathematica, it is easy to verify that the coefficients in front of the characters are indeed integers. 

\begin{figure}[htpb!]
\centering
\begingroup
  \inputencoding{cp1250}%
  \makeatletter
  \providecommand\color[2][]{%
    \GenericError{(gnuplot) \space\space\space\@spaces}{%
      Package color not loaded in conjunction with
      terminal option `colourtext'%
    }{See the gnuplot documentation for explanation.%
    }{Either use 'blacktext' in gnuplot or load the package
      color.sty in LaTeX.}%
    \renewcommand\color[2][]{}%
  }%
  \providecommand\includegraphics[2][]{%
    \GenericError{(gnuplot) \space\space\space\@spaces}{%
      Package graphicx or graphics not loaded%
    }{See the gnuplot documentation for explanation.%
    }{The gnuplot epslatex terminal needs graphicx.sty or graphics.sty.}%
    \renewcommand\includegraphics[2][]{}%
  }%
  \providecommand\rotatebox[2]{#2}%
  \@ifundefined{ifGPcolor}{%
    \newif\ifGPcolor
    \GPcolorfalse
  }{}%
  \@ifundefined{ifGPblacktext}{%
    \newif\ifGPblacktext
    \GPblacktexttrue
  }{}%
  \let\gplgaddtomacro\g@addto@macro
  \gdef\gplbacktext{}%
  \gdef\gplfronttext{}%
  \makeatother
  \ifGPblacktext
    \def\colorrgb#1{}%
    \def\colorgray#1{}%
  \else
    \ifGPcolor
      \def\colorrgb#1{\color[rgb]{#1}}%
      \def\colorgray#1{\color[gray]{#1}}%
      \expandafter\def\csname LTw\endcsname{\color{white}}%
      \expandafter\def\csname LTb\endcsname{\color{black}}%
      \expandafter\def\csname LTa\endcsname{\color{black}}%
      \expandafter\def\csname LT0\endcsname{\color[rgb]{1,0,0}}%
      \expandafter\def\csname LT1\endcsname{\color[rgb]{0,1,0}}%
      \expandafter\def\csname LT2\endcsname{\color[rgb]{0,0,1}}%
      \expandafter\def\csname LT3\endcsname{\color[rgb]{1,0,1}}%
      \expandafter\def\csname LT4\endcsname{\color[rgb]{0,1,1}}%
      \expandafter\def\csname LT5\endcsname{\color[rgb]{1,1,0}}%
      \expandafter\def\csname LT6\endcsname{\color[rgb]{0,0,0}}%
      \expandafter\def\csname LT7\endcsname{\color[rgb]{1,0.3,0}}%
      \expandafter\def\csname LT8\endcsname{\color[rgb]{0.5,0.5,0.5}}%
    \else
      \def\colorrgb#1{\color{black}}%
      \def\colorgray#1{\color[gray]{#1}}%
      \expandafter\def\csname LTw\endcsname{\color{white}}%
      \expandafter\def\csname LTb\endcsname{\color{black}}%
      \expandafter\def\csname LTa\endcsname{\color{black}}%
      \expandafter\def\csname LT0\endcsname{\color{black}}%
      \expandafter\def\csname LT1\endcsname{\color{black}}%
      \expandafter\def\csname LT2\endcsname{\color{black}}%
      \expandafter\def\csname LT3\endcsname{\color{black}}%
      \expandafter\def\csname LT4\endcsname{\color{black}}%
      \expandafter\def\csname LT5\endcsname{\color{black}}%
      \expandafter\def\csname LT6\endcsname{\color{black}}%
      \expandafter\def\csname LT7\endcsname{\color{black}}%
      \expandafter\def\csname LT8\endcsname{\color{black}}%
    \fi
  \fi
    \setlength{\unitlength}{0.0500bp}%
    \ifx\gptboxheight\undefined%
      \newlength{\gptboxheight}%
      \newlength{\gptboxwidth}%
      \newsavebox{\gptboxtext}%
    \fi%
    \setlength{\fboxrule}{0.5pt}%
    \setlength{\fboxsep}{1pt}%
\begin{picture}(8502.00,5668.00)%
    \gplgaddtomacro\gplbacktext{%
      \csname LTb\endcsname
      \put(946,704){\makebox(0,0)[r]{\strut{}$-1.0$}}%
      \put(946,1178){\makebox(0,0)[r]{\strut{}$-0.9$}}%
      \put(946,1653){\makebox(0,0)[r]{\strut{}$-0.8$}}%
      \put(946,2127){\makebox(0,0)[r]{\strut{}$-0.7$}}%
      \put(946,2601){\makebox(0,0)[r]{\strut{}$-0.6$}}%
      \put(946,3075){\makebox(0,0)[r]{\strut{}$-0.5$}}%
      \put(946,3550){\makebox(0,0)[r]{\strut{}$-0.4$}}%
      \put(946,4024){\makebox(0,0)[r]{\strut{}$-0.3$}}%
      \put(946,4498){\makebox(0,0)[r]{\strut{}$-0.2$}}%
      \put(946,4973){\makebox(0,0)[r]{\strut{}$-0.1$}}%
      \put(946,5447){\makebox(0,0)[r]{\strut{}$0.0$}}%
      \put(1078,484){\makebox(0,0){\strut{}$0$}}%
      \put(1850,484){\makebox(0,0){\strut{}$1$}}%
      \put(2622,484){\makebox(0,0){\strut{}$2$}}%
      \put(3395,484){\makebox(0,0){\strut{}$3$}}%
      \put(4167,484){\makebox(0,0){\strut{}$4$}}%
      \put(4939,484){\makebox(0,0){\strut{}$5$}}%
      \put(5711,484){\makebox(0,0){\strut{}$6$}}%
      \put(6483,484){\makebox(0,0){\strut{}$7$}}%
      \put(7256,484){\makebox(0,0){\strut{}$8$}}%
      \put(8028,484){\makebox(0,0){\strut{}$9$}}%
    }%
    \gplgaddtomacro\gplfronttext{%
      \csname LTb\endcsname
      \put(198,3075){\rotatebox{-270}{\makebox(0,0){\strut{}$V_p(r)$}}}%
      \put(4591,154){\makebox(0,0){\strut{}$r$}}%
      \csname LTb\endcsname
      \put(7118,1317){\makebox(0,0)[r]{\strut{}$p = 2$}}%
      \csname LTb\endcsname
      \put(7118,1097){\makebox(0,0)[r]{\strut{}$p = 1$}}%
      \csname LTb\endcsname
      \put(7118,877){\makebox(0,0)[r]{\strut{}$p = 0$}}%
    }%
    \gplbacktext
    \put(0,0){\includegraphics{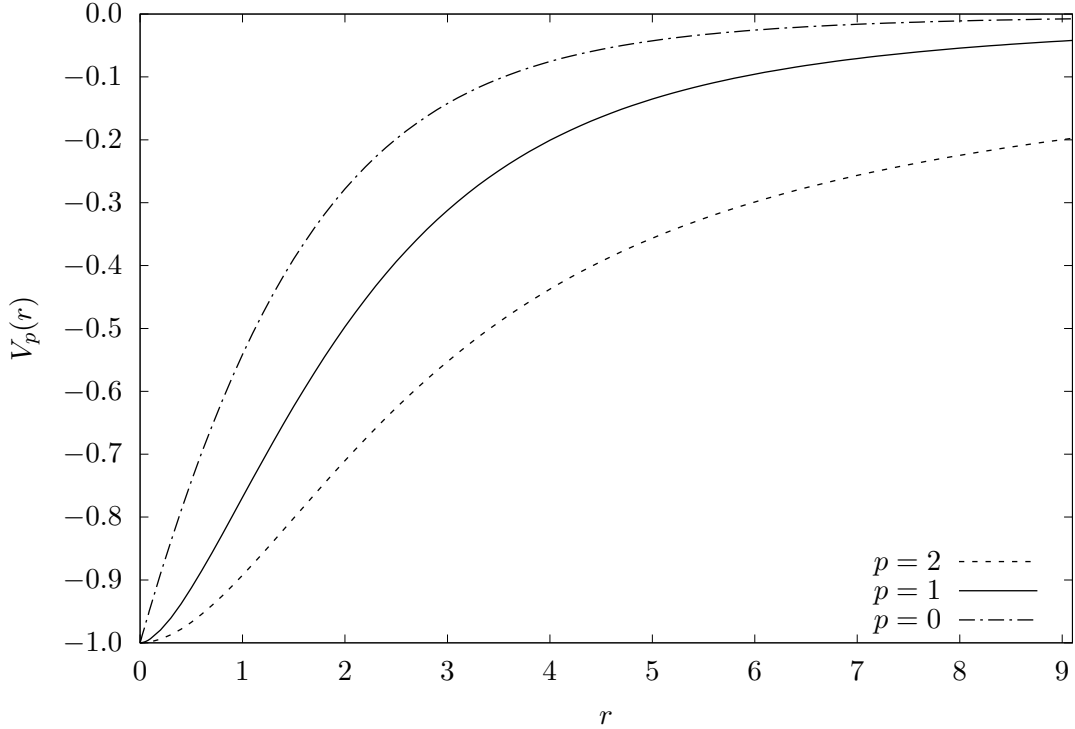}}%
    \gplfronttext
  \end{picture}%
\endgroup
\caption{Interbrane potential $V_p(r)$ for two identical $\omega_2$ D-branes (normalized so that $V_p(0)=-1$).}
\label{fig:fig1}
\end{figure}

It follows from \eqref{eq:BSU3pcoefE} that the mass of each $\omega_2$ D-brane is 3 times the mass of the D0-brane. Using the $q$-series \eqref{eq:qseries} for the $k=1$ characters, we have computed that\footnote{It is a curious fact (which we have checked up to $\mathcal{O}(q^{1000})$), that ${Z}_{\alpha\alpha}^{\omega_2}(q)\equiv Z^{\omega_2}_{\alpha\alpha}(q)_\mathrm{NS}-Z^{\omega_2}_{\alpha\alpha}(q)_\mathrm{R}={Z}_{\alpha\alpha}^{\omega_2}(\widetilde{q})$, namely that the integer coefficients in the $q$-expansion of the open string partition function are the same as the integer coefficients in the $\tilde{q}$-expansion of the closed string channel overlap.}
\begin{subequations}
\label{eq:omega2bos}
\begin{align}
\eta(q)^{-4}Z^{\omega_2}_{\alpha\alpha}(q)_\mathrm{NS} &=28 + 44 q^{\frac{1}{3}} + 192 q^{\frac{2}{3}} + 884 q + 1288 q^{\frac{4}{3}} + 3456 q^{\frac{5}{3}} +\mathcal{O}(q^{2})\,,\label{eq:omega2bosNS}\\[+2pt]
\eta(q)^{-4}Z^{\omega_2}_{\alpha\alpha}(q)_\mathrm{R} &=16 + 80 q^{\frac{1}{3}} + 192 q^{\frac{2}{3}} + 704 q + 1648 q^{\frac{4}{3}} + 3456 q^{\frac{5}{3}} +\mathcal{O}(q^{2})\,,
\end{align}
\end{subequations}
for all $\alpha$. This shows that in the case of $\omega_2$ boundary states, the bose-fermi degeneracy is absent and the corresponding D-branes are non-supersymmetric. At the same time, there are no NS tachyons in \eqref{eq:omega2bosNS}, so the $\omega_2$ boundary states are stable. The gluing automorphism $\omega_2$ does not permit couplings to any massless RR fields and massless NSNS fields along the 4-torus, so the $\omega_2$ boundary states are non-conventional and do not carry any RR charges. We therefore conclude that the $\omega_2$ boundary states describe new non-BPS and uncharged D-branes in type IIA superstring which are stable at weak coupling. They should not be confused with the standard stable non-BPS D$p$-branes of \cite{Sen:1998rg,Sen:1998ii,Bergman:1998xv,Sen:1998tt,Sen:1999mg} which are constructed on type II orbifolds and orientifolds and carry conserved (twisted) RR charges. As it is apparent from \eqref{eq:omega2bos}, the low-energy effective theory on $\omega_2$ \hbox{D-branes} contains 28 spacetime bosons and 16 spacetime fermions. In order to conclusively determine the dimension of the \hbox{D-brane} moduli space, as well as the vertices of the low-energy effective action on the D-brane, boundary and bulk-boundary structure constants would have to be found first. We hope to report on this in the future. Since we have $ {Z}_{\alpha\alpha}^{\omega_2}(q)\equiv Z^{\omega_2}_{\alpha\alpha}(q)_\mathrm{NS}-Z^{\omega_2}_{\alpha\alpha}(q)_\mathrm{R}\neq 0$,
the no-force is in general violated. Indeed, we were unable to find a regular solution of supergravity involving only the massless fields which couple to the boundary state and whose asymptotic behavior would match the prediction from the boundary state (see \cite{Bertolini:2000jy} for a similar analysis). With the help of the results of \cite{DiVecchia:1997vef} and \cite{Zhou:1999nm}, we obtain
\begin{subequations}
\begin{align}
	ds^2 &=B(r)^2 \eta_{\alpha\beta}dx^\alpha dx^\beta+F(r)^2\delta_{ij}dx^{i}dx^{j}+\nonumber\\
	&\hspace{5cm}+(dx^6)^2+(dx^7)^2+(dx^8)^2+(dx^9)^2\,,\\
	e^{\phi} &= B(r)^{2\sqrt{2}\frac{p-1}{3-p}}\,,\label{eq:nonBPSdil}
\end{align}
\end{subequations}
where $\alpha,\beta = 0,\ldots,p$, $i,j=p+1,\ldots,5$ with $0\leqslant p <3$ the number of Neumann conditions imposed on the external non-compact coordinates (T-dualizing the $\omega_2$ D-branes if necessary). We have also denoted
\begin{subequations}
\begin{align}
B(r)^2 &=\left[\frac{f_-(r)}{f_+(r)}\right]^{\frac{1}{2}\sqrt{(3-p)(4-p)}}\,, \\
F(r)^2 &= f_+(r)^{\frac{2}{3-p}+\frac{p+1}{2}\sqrt{\frac{4-p}{3-p}}}f_-(r)^{\frac{2}{3-p}-\frac{p+1}{2}\sqrt{\frac{4-p}{3-p}}}\,,
\end{align}
\end{subequations}
where $f_{\pm}(r)=1\pm K/r^{3-p}$ with $K = 3\times 2^{4-p} \pi^{(9-p)/2} g_\mathrm{s}/\sqrt{(3-p)(4-p)}$.
Computing the Kretschmann scalar, it can be verified that the solution suffers a curvature singularity at $r_p=K^{1/(3-p)}$. Violation of the no-force condition also means that even at rest, two identical $\omega_2$ D-branes separated by a distance $r$ in the non-compact coordinates exert a non-zero force on each other due to closed string exchange. The corresponding static interaction potential can be computed as 
\begin{align}
V_p(r) &=- \int_0^\infty\frac{dt}{t}e^{-\frac{r^2t}{2\pi}}{(8\pi^2 t)^{-\frac{p+1}{2}}}\eta(q)^{-4}Z_{\alpha\alpha}^{\omega_2}(q)\,,
\label{eq:om2potOp}
\end{align}
where $q = e^{-2\pi t}$. While the potential \eqref{eq:om2potOp} is exact in $\alpha'$, it is merely leading order in $g_\mathrm{s}$ (namely $\mathcal{O}(g_\mathrm{s}^0)=\mathcal{O}(1)$).
It can be computed numerically for each $r$ by truncating the expansion \eqref{eq:omega2bos} up to a finite order, replacing it with the corresponding Padé approximant $[M,N]$ and integrating 
\eqref{eq:om2potOp} 
numerically. 
See Figure~\ref{fig:fig1} for our results, where we have expanded up to $\mathcal{O}(q^{1000})$ and set $M=N=500$. Identical results were obtained by instead performing the whole computation in the closed string channel.
%
Note that we recover the expected behavior \hbox{$V_p(r)\propto r^{p-3}$} as $r\to\infty$. Fitting the potential around $r=0$, we also find that $V_2(r)\propto r^{2}$, $V_1(r)\propto r^{2}$ and $V_0(r)\propto r$ as $r\to 0$.
However, only in the case $p=1$, we can be sure that the computed potential is trustworthy for all $r$: the boundary state does not (classically) source the dilaton for $p=1$, so that one can consistently keep $g_\mathrm{s}$ to be small for all $r$. For $p=2$, \eqref{eq:nonBPSdil} gives decreasing $e^{\phi}$ as $r$ becomes small, so it is likely that the $p=2$ potential can also be trusted for all $r$. For $p=0$, on the other hand, $e^{\phi}$ blows up at small $r$, so the linear dependence of $V_0(r)$ as $r\to 0$ is likely to receive substantial corrections.

\subsubsection*{General permutation gluing conditions}

Since the rational bulk partition function \eqref{eq:partSU3typeII} is not diagonal, we were unable to write down closed formulae for boundary states satisfying general permutation gluing conditions. Instead we have developed a computer algorithm which, for each gluing condition, automatically selects Ishibashi states allowed by the bulk partition function and the gluing automorphism, and assigns boundary state coefficients so that after S-transforming, one obtains consistent (integer) open string spectra. It was also checked (for a large number of cases) that integer multiplicities are automatically obtained for stretched string spectra between boundary states with different gluing conditions.\footnote{For a general pair of gluing conditions, the mirror-twisted ($\sigma: J\to -J$) characters enter the calculation of relative overlaps. See \cite{Eguchi:2001ip} for their modular properties and $q$-expansions.}
It is a special feature of the rational boundary states on the $SU(3)^2$ 4-torus that for a fixed gluing condition, all boundary states turn out to have identical masses and spectra of open string excitations. It is easy to see that in total there are \hbox{$N_\text{g}=6!\times 2^6 = 46\,080$} ways how to choose gluing conditions on the chiral generators of the six $\mathcal{N}=2$ superconformal algebras with $c=1$. Out of these, 3384 gluing conditions give stable D-branes, while the rest of the gluing conditions give unstable D-branes. It is also true that all of the unstable D-branes do not carry any RR charges. Out of the 3384 gluing conditions which give stable D-branes, 2736 give boundary states which carry non-zero RR charges. The rest yields stable, non-conventional, uncharged and non-supersymmetric boundary states, all of which have the same mass and open string spectrum as the $\omega_2$ \hbox{D-branes}. Also, none of the permutation gluing conditions which give the stable non-BPS boundary states is ``purely A-type" or ``purely B-type". All of the 2736 RR-charged gluing conditions give elementary supersymmetric D-branes which are either 1/2-BPS or 1/4-BPS, where the 1/4-BPS ones are described by stable non-conventional boundary states. We also found that none of the stable non-conventional boundary states (RR-charged or not) can be factorized into boundary states, which would separately wrap the two $SU(3)$ 2-tori. Existence of such boundary states therefore seems to be a distinctive feature of 4-tori. In the case of 1/4-BPS bound states, this is a simple consequence of saturating the BPS bound for given RR charges, while for the stable non-BPS D-branes the reason for this is not clear.
The gluing conditions which produce RR-charged boundary states are summarized in Appendix \ref{app:class}, Table \ref{tab:classSU32}.
Note that all masses agree with the corresponding 1/2-BPS or 1/4-BPS masses for given RR charges. Also note that for all RR-charged D-branes, the number of massless open string modes in either the NS or the R sector computed from the boundary state  agrees with the ADHM formula $4kN+8$ where $kN$ is given in terms of the RR charges by \eqref{eq:kN}.
    
\subsection{$SU(2)^4$ 4-torus}

We will now give a summary of our results for the $SU(2)^4$ \hbox{4-torus}. Here the bulk spectrum of the corresponding $\mathcal{N}=(2,2)$ 
worldsheet sigma model can be rewritten in terms of four 
copies of the $k=2$ minimal model. 
Unlike in the $SU(3)^2$ case, non-trivial multiplicities appear in the bulk partition 
function, so that additional 
care is needed to resolve the associated fixed 
points.

\subsubsection*{Bulk theory}

We can write the $SU(2)^4$ 4-torus as a product of two 
$SU(2)^2$ 2-tori $T^4=T^2\times {T^2}$ which 
extend in the 67 and 89 planes.
The bulk spectrum of the $\mathcal{N}=(2,2)$ worldsheet sigma 
model on the $SU(2)^2$ 2-torus can be given \cite{Chun:1991js,Gutperle:1998hb} in terms of 
the irreps of two copies of $\mathcal{N}=2$ minimal 
models with $k_a=2$ for $a=1,2$. 
The fusion algebra of these minimal models has $
\mathbb{Z}_4\times\mathbb{Z}_2$ symmetry generated by 
$g_a, h_a$ where $g_a\Phi^{l_a}_{m_a,s_a}= e^{\frac{\pi i 
}{2}m_a}\Phi^{l_a}_{m_a,s_a}$ and $h_a\Phi^{l_a}
_{m_a,s_a}= e^{-i\pi s_a}\Phi^{l_a}_{m_a,s_a}$. This extends to a diagonal $\mathbb{Z}_4$ symmetry 
$G$ of the $(k=2)^2$ tensor product fusion algebra, where 
$G$ is generated by $\pm g_1g_2$ where we take the plus 
sign in the NS sector and minus sign in the R sector. It 
is easy to see that the $G$ invariant states in the NS 
sector ($m_1+m_2\in 4\mathbb{Z}$) are precisely those which have integer $U(1)$ 
charge, while the $G$-invariant states in the R sector ($m_1+m_2 \in 4\mathbb{Z}+2$) 
are precisely those which have half-integer $U(1)$ 
charge. The simple currents associated to $G$ are $J_\pm=
\Phi^{2}_{\pm 2,0}\Phi^{2}_{\pm 2,0}$ (order 4) and $J_0=
\Phi^2_{0,0}\Phi^2_{0,0}$ (order 2). Note that the fixed points of $G$ are precisely the fields with $l_1=l_2=1$ with stabilizer $\{\mathbb{1},J_0\}\cong 
\mathbb{Z}_2$. $J_\pm$ have no fixed 
points. The $\mathcal{N}=(2,2)$ worldsheet sigma model on the $SU(2)^2$ 2-torus can then be obtained as the $G$-orbifold of the direct product of two copies of $k=2$ minimal models with diagonal modular invariant. The unprojected spectrum reads
\begin{align}
Z_{SU(2)^2}(q,\overline{q})&=\sum_{l_a,m_a,s_a}\,\sum_{\substack{t\in\mathbb{Z}_4}}\sum_{\overline{s}_a}
\prod_{a=1}^{2}\chi^{l_a}_{m_a,s_a}(q)\overline{\chi}
^{l_a}_{-m_a+2t,\overline{s}_a}(\overline{q})\,, 
\label{eq:SU22}
\end{align}
where the $l_a,m_a,s_a$ sum on the RHS runs over distinct 
$G$-invariant states in the NS and R sectors, respectively, such that $s_1-s_a\in 2\mathbb{Z}$, $l_a+m_a+s_a\in 2\mathbb{Z}$ and $\overline{s}_a\in\mathbb{Z}_4$ are such that $s_a-
\overline{s}_a\in 2\mathbb{Z}$ for all $a=1,2$. 
We note that since the $G$ action has non-trivial fixed 
points, non-trivial multiplicities will 
appear in~\eqref{eq:SU22} as a consequence of summing over 
$t\in\mathbb{Z}_4$. We also have
\begin{align}
Z_{SU(2)^2}(q,\overline{q}) = \left|\frac{\theta_{i}(q)}{\eta(q)^3}\right|^2\sum_{M,N,R,S\in\mathbb{Z}}q^{\frac{1}{4}[(M+R)^2+(N+S)^2]} \overline{q}^{\frac{1}{4}[(M-R)^2+(N-S)^2]}\,,
\end{align}
where we take $i=3$ and $i=2$ for the NSNS and RR sector, respectively. Following the logic of the notation introduced in the case of the $SU(3)^2$ 4-torus, 
the GSO-unprojected spectrum of the $\mathcal{N}=(2,2)$ worldsheet sigma model involving the $SU(2)^4$ 4-torus and 4 non-compact directions can be written as
\begin{align}
    Z(q,\overline{q})&=\sum_{\bm{\lambda},\bm{\mu},\bm{\sigma}}
    \sum_{\substack{t_r\in\mathbb{Z}_4}}    \sum_{\overline{\bm{\sigma}}}\chi^{\bm{\lambda}}
    _{\bm{\mu},\bm{\sigma}}(q)\overline{\chi}^{\bm{\lambda}}_{-
    \bm{\mu}+ 2 t_3\bm{\beta}_3+ 2 t_4\bm{\beta}_4,
    \overline{\bm{\sigma}}}(\overline{q})\,,
    \label{eq:partSU2typeII}
\end{align}
where the $\bm{\lambda},\bm{\mu}$ sum on the RHS runs over distinct $G\!\times\!G$-invariant states (i.e.\ those with $\bm{\beta}_r\cdot\bm{\mu}\in4\mathbb{Z}$ in NS sector and $\bm{\beta}_r\cdot\bm{\mu}\in 4\mathbb{Z}+2$ in R sector for $r=3,4$) such that $\sigma_1-\sigma_p\in 2\mathbb{Z}$ and $l_a+m_a+s_a\in 2\mathbb{Z}$ and $\overline{\sigma}_p\in\mathbb{Z}_4$ such that $\sigma_p-\overline{\sigma}_p\in 
2\mathbb{Z}$, for all $a=1,\ldots,4$ and $p=1,\ldots,6$. It follows that only states with integer total left- and right-moving $U(1)$ charge $q(\bm{\mu},\bm{\sigma})=(\frac{\bm{\mu}}{4}-\frac{\bm{\sigma}}{2})\cdot \sum_{r=1}^4\bm{\beta}_r$ appear in~\eqref{eq:partSU2typeII}. The 
fields with $\bm{\lambda}=\pm\bm{\mu}=2\bm{\beta}_r$, $
\bm{\sigma}=\bm{0}$, $r=3,4$ are identified with the internal complex 
fermions $\psi^{r\pm}=(\psi^{2r}\pm i\psi^{2r+1})/
\sqrt{2}$, while the spin fields are again represented by 
$\bm{\lambda}=\bm{0}$, $\bm{\mu} = 2\sum_{r=3}^4\tau_r
\bm{\beta}_r$, $\bm{\sigma} = 2\sum_{r=1}^4\tau_r
\bm{\beta}_r$.

\subsubsection*{$\mathrm{D}p$-brane boundary states}

Let us first consider the gluing conditions $\omega_0 = (1_\mathrm{B})(2_\mathrm{B})(3_\mathrm{B})(4_\mathrm{B})$.
Let us denote by $\mathcal{S}^{\bm{\lambda}}_{\bm{\mu},\bm{\sigma}}$ the $G\!\times\! G$ stabilizer of the state with labels $\bm{\lambda},\bm{\mu},\bm{\sigma}$. Clearly, $
\mathcal{S}^{\bm{\lambda}}_{\bm{\mu},\bm{\sigma}}$ can be either trivial, or $\mathbb{Z}_2$ (generated by $(J_0,
\mathbb{1})$ \emph{or} $(\mathbb{1},J_0)$), or $\mathbb{Z}
_2\times\mathbb{Z}_2$ (generated by $(J_0,\mathbb{1})$ 
\emph{and} $(\mathbb{1},J_0)$). The Ishibashi 
states are now labelled as $\left|\bm{\lambda},\bm{\mu},\bm{\sigma},J\rangle\!\rangle_{\omega_0}\right.$, where the allowed labels can again be encoded by means of a projector. The additional label $J
\in\mathcal{S}^{\bm{\lambda}}_{\bm{\mu},\bm{\sigma}}$ is to distinguish 
Ishibashi states which correspond to bulk fields with 
non-trivial multiplicities. The boundary states are labeled by  $\bm{\Lambda},\bm{M},\bm{\Sigma}$ and the characters $\psi
$ of the stabilizer $\mathcal{S}^{\bm{\Lambda}}_{\bm{M},
\bm{\Sigma}}$. It turns out that we can restrict ourselves on boundary states with trivial $\mathcal{S}^{\bm{\Lambda}}_{\bm{M},
\bm{\Sigma}}$ without missing any new types of massless NSNS and RR couplings. Such boundary states satisfy\footnote{$\mathrm{id}$ denotes the identity character.}
\begin{equation}
   \|\bm{\Lambda},
    \bm{M},\bm{\Sigma},\mathrm{id}\rangle\!\rangle_{\omega_0} 
    = \frac{1}{\kappa_{\omega_0}^{\alpha}} \sum_{\bm{\lambda},
    \bm{\mu},\bm{\sigma}}\delta^{\omega_0}_{\bm{\lambda},\bm{\mu},
    \bm{\sigma}}B^{\alpha,\omega_0}_{\bm{\lambda},\bm{\mu},
    \bm{\sigma},\mathbb{1}}|\bm{\lambda},\bm{\mu},\bm{\sigma},
    \mathbb{1}\rangle\!\rangle_{\omega_0}\,,
\end{equation}
where
\begin{equation}
     B^{\alpha,\omega_0}_{\bm{\lambda},\bm{\mu},\bm{\sigma},
     \mathbb{1}} =(-1)^\frac{\sigma_1^2}{2}e^{-\frac{i\pi}{2}\bm{\sigma}\cdot\bm{\Sigma}}e^{\frac{i\pi}{4}\bm{\mu}\cdot\bm{M}}\prod_{a=1}^4 \frac{\sin [ 
     \frac{\pi}{4}(l_a+1)(L_a+1)]}{\sin^{\frac{1}{2}} [ 
     \frac{\pi}{4}(l_a+1)]}\,.
     \label{eq:BSU22coefTriv}
\end{equation}
All calculations then proceed in the same manner as in the $SU(3)^2$ case. We find that the the $\omega_0$ boundary states describe stable supersymmetric 1/2-BPS D-branes with the following 16 representatives
%
\begingroup
\allowdisplaybreaks
\begin{subequations}
\begin{align}
\|\bm{0},\bm{0},\bm{0}\rangle\!\rangle_{\omega_0}
&=\|\mathrm{D0}\rangle\!\rangle   \\
 \|L_1\!=\! M_1\!=\!2\rangle\!\rangle_{\omega_0}
 &=\|\overline{\mathrm{D2}}_{67}\rangle\!\rangle   \\
\|L_1\!=\! +M_1\!=\!1\rangle\!\rangle_{\omega_0} 
&=\|\mathrm{D0}/\overline{\mathrm{D2}}_{67}\rangle\!\rangle\\
\|L_1\!=\! -M_1\!=\!1\rangle\!\rangle_{\omega_0}
&=\|\mathrm{D0}/{\mathrm{D2}}_{67}\rangle\!\rangle
\\
\|L_3\!=\! M_3\!=\!2\rangle\!\rangle_{\omega_0} &=
\|\overline{\mathrm{D2}}_{89}\rangle\!\rangle\\
\|L_3\!=\! +M_3\!=\!1\rangle\!\rangle_{\omega_0} 
&=\|\mathrm{D0}/\overline{\mathrm{D2}}_{89}\rangle\!\rangle\\
\|L_3\!=\! -M_3\!=\!1\rangle\!\rangle_{\omega_0} 
&=\|\mathrm{D0}/{\mathrm{D2}}_{89}\rangle\!\rangle\\
 \|L_{1,3}\!=\! M_{1,3}\!=\!2\rangle\!\rangle_{\omega_0} &=\|{\mathrm{D4}}\rangle\!\rangle \\
\|L_{1,3}\!=\! +M_{1,3}\!=\!1\rangle\!\rangle_{\omega_0} &=\|\mathrm{D0}/\overline{\mathrm{D2}}_{67}/\overline{\mathrm{D2}}_{89}/
\mathrm{D4}\rangle\!\rangle\\
\|L_{1,3}\!=\! -M_{1,3}\!=\!1\rangle\!\rangle_{\omega_0} &=\|\mathrm{D0}/{\mathrm{D2}}_{67}/
{\mathrm{D2}}_{89}/\mathrm{D4}\rangle\!\rangle\\
\|L_{1,3}\!=\! +M_{1}\!=\!-M_3 \!=\!1\rangle\!
\rangle_{\omega_0} &=\|\mathrm{D0}/\overline{\mathrm{D2}}_{67}/
{\mathrm{D2}}_{89}/\overline{\mathrm{D4}}\rangle\!
\rangle\\
\|L_{1,3}\!=\! -M_{1}\!=\!+M_3 \!=\!1\rangle\!
\rangle_{\omega_0} &=\|\mathrm{D0}/{\mathrm{D2}}_{67}/
\overline{\mathrm{D2}}_{89}/\overline{\mathrm{D4}}\rangle\!\rangle\\
\|L_{1}\!=\! M_{1}\!=\!2,L_3\!=\!+M_3\!=\!1\rangle\!
\rangle_{\omega_0} &=\|\overline{\mathrm{D2}}_{67}/{\mathrm{D4}}\rangle\!
\rangle  \\
  \|L_{1}\!=\! +M_{1}\!=\!1,L_3\!=\!M_3\!=\!2\rangle\!
  \rangle_{\omega_0} &=\|\overline{\mathrm{D2}}_{89}/{\mathrm{D4}}\rangle\!
  \rangle\\
 \|L_{1}\!=\! M_{1}\!=\!2,L_3\!=\!-M_3\!=\!1\rangle\!
 \rangle_{\omega_0} &=\|\overline{\mathrm{D2}}_{67}/\overline{\mathrm{D4}}
 \rangle\!\rangle   \\
   \|L_{1}\!=\! -M_{1}\!=\!1,L_3\!=\!M_3\!=\!2\rangle
   \!\rangle_{\omega_0} &=\|\overline{\mathrm{D2}}_{89}/
   \overline{\mathrm{D4}}\rangle\!\rangle  
\end{align}
\end{subequations}
\endgroup
Note that the 
boundary coefficient formula~\eqref{eq:BSU22coefTriv} 
also yields the correct masses the boundary states: we obtain $\mathcal{M}=1$ for the D0, $\mathrm{D2}_{67}$, $\mathrm{D2}_{89}$ and $\mathrm{D4}$-branes, $\mathcal{M}=\sqrt{2}$ for the 1/2-BPS bound states $\mathrm{D0}/\overline{\mathrm{D2}}_{67}$, $\mathrm{D0}/{\mathrm{D2}}_{67}$, $\mathrm{D0}/\overline{\mathrm{D2}}_{89}$, $\mathrm{D0}/{\mathrm{D2}}_{89}$, $\overline{\mathrm{D2}}_{67}/{\mathrm{D4}}$, $\overline{\mathrm{D2}}_{89}/{\mathrm{D4}}$, $\overline{\mathrm{D2}}_{67}/\overline{\mathrm{D4}}$, $\overline{\mathrm{D2}}_{89}/
   \overline{\mathrm{D4}}$ and $\mathcal{M}=2$ for the 1/2-BPS bound states $\mathrm{D0}/\overline{\mathrm{D2}}_{67}/\overline{\mathrm{D2}}_{89}/
\mathrm{D4}$, $\mathrm{D0}/{\mathrm{D2}}_{67}/
{\mathrm{D2}}_{89}/\mathrm{D4}$, $\mathrm{D0}/\overline{\mathrm{D2}}_{67}/
{\mathrm{D2}}_{89}/\overline{\mathrm{D4}}$, $\mathrm{D0}/{\mathrm{D2}}_{67}/
\overline{\mathrm{D2}}_{89}/\overline{\mathrm{D4}}$. 
The rest of the fundamental D$p$-branes (i.e.\ the $\mathrm{D2}_{68}$, $\mathrm{D2}_{69}$, $\mathrm{D2}_{78}$, $\mathrm{D2}_{79}$ branes) together with some of the higher-mass 1/2-BPS bound states can be obtained by considering more complicated gluing conditions. See Appendix \ref{app:class}, Table \ref{tab:classSU24} for a summary.

\subsubsection*{1/4-BPS boundary states}

We will now consider the permutation gluing conditions $
\omega_1=(1_\mathrm{B})(2_\mathrm{B}3_\mathrm{B})(4_\mathrm{B})$. Restricting again on the boundary states with trivial $\mathcal{S}^{\bm{\Lambda}}_{\bm{M},\bm{\Sigma}}$, we obtain stable elementary supersymmetric boundary states describing   1/4-BPS bound states with the following four representatives
\begin{subequations}
\label{eq:BSSU24}
\begin{align}
\|\bm{0},\bm{0},\bm{0}\rangle\!\rangle_{\omega_1} &= \|{\mathrm{D0}}/{\mathrm{D2}}_{67}/{\mathrm{D2}}_{89}/\overline{\mathrm{D4}}\rangle\!\rangle\\
\|L_1\!=\! M_1\!=\!2\rangle\!\rangle_{\omega_1}  &= \|{\mathrm{D0}}/\overline{\mathrm{D2}}_{67}/\overline{\mathrm{D2}}_{89}/\overline{\mathrm{D4}}\rangle\!\rangle \\
\|L_1\!=\! +M_1\!=\!1\rangle\!\rangle_{\omega_1}  &= \|2\mathrm{D0}/2\overline{\mathrm{D4}}\rangle\!\rangle \\
\|L_1\!=\! -M_1\!=\!1\rangle\!\rangle_{\omega_1}  &= \|2{\mathrm{D2}}_{67}/2{\mathrm{D2}}_{89}\rangle\!\rangle\,. 
\end{align}
\end{subequations}
Couplings to the 
massless NSNS Ishibashi states along internal directions of the 4-torus are not allowed by the $
\omega_1$ gluing conditions, so the $\omega_1$ 
boundary states are non-conventional.
It is also easy to check that the boundary state coefficients give mass $\mathcal{M}=2\sqrt{2}$ for the boundary states $\|\bm{0},\bm{0},\bm{0}\rangle\!\rangle_{\omega_1}$ and $\|L_1\!=\! M_1\!=\!2\rangle\!\rangle_{\omega_1}$ and $\mathcal{M}=4$ for the boundary states $\|L_1\!=\! +M_1\!=\!1\rangle\!\rangle_{\omega_1}$ and $\|L_1\!=\! -M_1\!=\!1\rangle\!\rangle_{\omega_1}$, which is in agreement with~\eqref{eq:14BPSSU22}. 
Since the boundary states \eqref{eq:BSSU24} are elementary, the D-branes described by the $\omega_1$ boundary states can be thought of as marginal 1/4-BPS bound states of 1/2-BPS D$p$-branes at some non-trivial points in their moduli space (where the massless strings stretched between the constituent 1/2-BPS D$p$-branes acquire non-zero vevs).
Furthermore, we computed that (denoting by $\alpha_{1}$ and $\alpha_{2}$ the boundary states with mass $2\sqrt{2}$ and 4, respectively)
\begin{subequations}
\label{eq:omega1bosSU2}
\begin{align}
\eta(q)^{-4} Z^{\omega_1}_{\alpha_{1}\alpha_{1}}(q)_\mathrm{NS} &=16 + 16 q^{\frac{1}{4}} + 64 q^{\frac{1}{2}} + 64 q^{\frac{3}{4}} + 640 q + 352 q^{\frac{5}{4}}+\mathcal{O}(q^{\frac{3}{2}})\,,\label{eq:omega1bosNSSU21}\\[+2pt]
\eta(q)^{-4} Z^{\omega_1}_{\alpha_2\alpha_2}(q)_\mathrm{NS} &=24 + 32 q^{\frac{1}{4}} + 192 q^{\frac{1}{2}} + 128 q^{\frac{3}{4}} + 960 q + 704 q^{\frac{5}{4}}+\mathcal{O}(q^{\frac{3}{2}})\,,\label{eq:omega1bosNSSU22}
\end{align}
\end{subequations}
The same expansions were found in the R sector: indeed, using Mathematica, we have shown to $\mathcal{O}(q^{1000})$ that the open string spectrum is bose-fermi degenerate for both $\alpha_1$ and $\alpha_2$ boundary states (as it should be, given the fact that $\omega_1$ boundary states preserve 8 spacetime supercharges). 
For the boundary states $\|2\mathrm{D0}/2\overline{\mathrm{D4}}\rangle\!\rangle$ and $\|2{\mathrm{D2}}_{67}/2{\mathrm{D2}}_{89}\rangle\!\rangle$, we obtain 24 massless modes in both the NS and R sector. For the boundary states $\|{\mathrm{D0}}/{\mathrm{D2}}_{67}/{\mathrm{D2}}_{89}/\overline{\mathrm{D4}}\rangle\!\rangle$ and $\|{\mathrm{D0}}/\overline{\mathrm{D2}}_{67}/\overline{\mathrm{D2}}_{89}/\overline{\mathrm{D4}}\rangle\!\rangle$ we have 16 massless modes in both the NS and R sector. 
It is straightforward to check that these results agree with the ADHM formula $8+4kN$, where $kN$ can be calculated (for the given RR charges) using \eqref{eq:kN}. Again, this observation is only meaningful subject to the assumption that all massless NS boundary operators are exactly marginal.
We also obtain a complete agreement between the asymptotic profiles of the massless NSNS and RR fields extracted from the superposition of $N$ boundary states $\|L_1\!=\! +M_1\!=\!1\rangle\!\rangle_{\omega_1}$ (T-dualized along $X^5$) and the supergravity solution \eqref{eq:sugra} with $\mu_1=-\mu_5$, $\varphi=\psi=0$ and $V_{67}=V_{89}=1$. More 1/4-BPS bound states of D$p$-branes wrapping the $SU(2)^4$ 4-torus can be constructed by considering other permutation gluing conditions. See Appendix \ref{app:class}, Table \ref{tab:classSU24} for the complete list of results.

\subsubsection*{Stable non-BPS boundary states}

Now we consider the gluing automorphism \hbox{$\omega_2=(1_\mathrm{B}3_\mathrm{B})(2_\mathrm{A}4_\mathrm{A})$}. Here we were unable to write down a projector on the allowed Ishibashi states in a simple closed form. Instead, a computer algorithm was developed to seek out Ishibashi states compatible with $\omega_2$ among the states permitted by the partition function \eqref{eq:partSU2typeII} and the type IIA GSO projection.
We found that $\omega_2$ gluing conditions generate two types of boundary states: unstable ones with mass $4\sqrt{2}$ and stable ones with mass 4 and open string spectrum
\begin{align}
  \eta(q)^{-4} Z_{\alpha{\alpha}}^{\omega_2}(q)&=24 - 64q^{\frac{1}{4}}+64q^{\frac{1}{2}}-256q^{\frac{3}{4}}+960q -1408q^{\frac{5}{4}}+\mathcal{O}(q^{\frac{3}{2}})\,.
\end{align}
Note that the bose-fermi degeneracy is lost so that these boundary states necessarily break all spacetime supersymmetries. The gluing automorphism $\omega_2$ does not permit Ishibashi states in neither the massless NSNS sector internal to the $SU(2)^4$ 4-torus nor the massless RR sector. We therefore conclude that the $\omega_2$ gluing conditions yield stable non-conventional boundary states which are associated with new non-BPS D-branes. These carry no RR charges and yet seem to be stable at weak coupling. The low-energy effective theory on these D-branes contains 40 spacetime bosons and 16 spacetime fermions. Our results regarding the corresponding supergravity solution and interbrane potential are completely analogous to those of the $\omega_2$ boundary states on the $SU(3)^2$ 4-torus.

\subsubsection*{General permutation gluing conditions}

In total there are $N_\mathrm{g} = 4!\times 2^{4}=384$ ways of choosing gluing conditions on the four copies of the $\mathcal{N}=2$ superconformal chiral generators with $c=3/2$. Out of these, 96 were found to yield at least one stable boundary state. All unstable boundary states were found to carry no RR charge. On the other hand, not all stable boundary states were found to couple to massless RR sector: 16 out of the 96 gluing conditions which yield stable boundary states do not permit massless RR Ishibashi states. The spectrum of boundary states for each of these 16 gluing conditions is structurally identical to the $\omega_2$ boundary states. Moreover, all of these stable RR-neutral boundary states are non-conventional, non-supersymmetric, their mass is equal to 4 and their open string spectrum is identical to that of the stable $\omega_2$ boundary states. The remaining 80 gluing conditions, which allow for massless RR Ishibashi states, all yield stable RR-charged boundary states, which are exclusively either 1/2-BPS or 1/4-BPS. See Appendix~\ref{app:class}, Table~\ref{tab:classSU24} for a classification of gluing conditions which yield RR-charged boundary states. All such boundary states are found to produce masses which agree with the BPS formula \eqref{eq:14BPST22} for the corresponding RR charges. Also, their open string spectra always contain $4kN+8$ massless modes, where $kN$ is calculated using \eqref{eq:kN}.

\section{Discussion and outlook}
\label{sec:disc}

In this paper we have performed a rational brane-scan of certain 4-tori which admit Gepner-like description. Having successfully recovered boundary states for 1/2-BPS D$p$-branes, we have exemplified stable non-conventional boundary states describing certain 1/4-BPS bound states of D$p$-branes, as well as new  non-BPS D-branes which are neutral under the massless RR forms. While a more systematic approach for calculating non-conventional boundary states would be highly desirable (enabling worldsheet description of the corresponding \hbox{D-branes} at general points in both bulk and boundary moduli space, as well as a description of arbitrarily charged 1/4-BPS bound states of D$p$-branes), our findings represent concrete evidence that there are stable consistent superconformal boundary states which do not satisfy gluing conditions \eqref{eq:glO} and that the family of type II superstring D-branes which are stable at weak string coupling is richer than previously thought.

It would be of great interest to completely characterize the boundary conformal field theory of the D-branes described in this paper. This would require solving the Cardy-Lewellen sewing relations (along the lines of \cite{Brunner:1999jq,Brunner:2000wx}) for boundary and bulk-boundary structure constants in the rational superconformal theories which we used to describe the two toroidal compactifications. Armed with these results, one should be able to compute correlators involving insertions of vertex operators corresponding to the massless open string modes on the D-branes as well as the massless closed string modes which couple to the D-branes. This would facilitate direct investigation of the low-energy effective physics of the \hbox{D-branes}, their moduli space geometry and also their behavior under bulk deformations. In the case of the 1/4-BPS bound states of D$p$-branes, one should recover the results already established by other means over the years. On the other hand, in the case of the new stable non-BPS \hbox{D-branes}, one could use this framework to find the non-supersymmetric action which lives on their worldvolumes, to study their moduli spaces and to learn whether they remain stable as we change the parameters of the compactification 4-torus.

In the case of the 1/4-BPS bound states of D$p$-branes, which are often associated with either ``purely A-type" or ``purely B-type" permutation gluing conditions, it could be beneficial to first address some of the above questions in a simplified setting by studying their analogues in topologically twisted theories. In particular, for the more complicated cases of Calabi-Yau manifolds, it has proven very fruitful to study matrix factorizations of the superpotential in suitable B-twisted Landau-Ginzburg orbifolds 
\cite{Kapustin:2002bi,Brunner:2003dc,Kapustin:2003ga}.
For those LG superpotentials which induce RG flows to the $\mathcal{N}=(2,2)$ worldsheet sigma models on the $SU(3)^2$ and $SU(2)^4$ 4-tori, one could try and identify matrix factorizations which correspond to the topological analogues of the 1/4-BPS bound states of D$p$-branes. 
A similar analysis comparing matrix factorizations with tensor product and permutation boundary states was performed in~\cite{SchmidtColinet:2007vi} for D-branes on the $SU(2)$ 2-torus and in~\cite{Brunner:2006tc} for D-branes on a continuous family of $T^4/\mathbb{Z}^4$ orbifolds. It might also be instructive to write the theory on each of the two $SU(3)$ 2-tori in terms of one $k=1$ and one $k=4$ minimal model and check if the generalized permutation boundary states of \cite{Fredenhagen:2005an,Fredenhagen:2006qw} yield new branes for the $SU(3)^2$ 4-torus.

The property that a boundary state does not satisfy linear gluing conditions \eqref{eq:glO} implies that on the doubled 4-torus\footnote{I.e.\ including the dual coordinates: translations along these correspond to turning on Wilson lines.}, it is not possible 
to choose a hyperplane of co-dimension four along which the worldvolume of the D-brane would be translationally invariant (in contrast to the usual D$p$-branes). The energy density and RR charge profiles induced by the non-conventional boundary states constructed in this paper could be straightforwardly (but tediously) computed by first rewriting the rational $\mathcal{N}=(2,2)$ chiral algebra generators and corresponding superprimaries in terms of the free boson and free fermion fields and then calculating the couplings of the boundary state to the closed string vertex operators level by level up to some (finite) momentum fourier mode. Similar calculation was performed in \cite{Kudrna:2018} on a 2-torus for the bosonic non-conventional boundary states, whose energy density profiles were found to have finite width and height.

\acknowledgments 

We would like to thank Matthias Gaberdiel, Ondra Hul\'{i}k, Mat\v{e}j Kudrna, Rennan Lipinski-Jusinskas, Ashoke Sen and Bogdan Stefa\'{n}ski for useful discussions and correspondence. This research has been supported by the Czech Science Foundation (GA\v{C}R) grant 17-22899S.

\appendix

\section{Derivation of BPS formulae}
\label{app:bps}

Here we analyze the BPS bound for D-branes wrapping a 4-torus. For more details see e.g.\ \cite{Obers:1998fb} whose conventions we largely follow.
Let us start with the Grassman-odd sector of the $D=10$ $\mathcal{N}=(1,1)$ (i.e.\ type IIA) super-Poincaré algebra
\begin{align}
\{Q_{\alpha},Q_{\beta}\} &= (C\Gamma^\mu)_{\alpha\beta}P_\mu+ (C\Gamma^{11})_{\alpha\beta}Z+\frac{1}{2}(C\Gamma_{\mu\nu})_{\alpha\beta}Z^{\mu\nu}+(C\Gamma_\mu\Gamma^{11})_{\alpha\beta}Z^\mu+\nonumber\\
&\hspace{4cm}+\frac{1}{4!}(C\Gamma_{\mu\nu\rho\sigma}\Gamma^{11})_{\alpha\beta}Z^{\mu\nu\rho\sigma}+\frac{1}{5!}(C\Gamma_{\mu\nu\rho\sigma\lambda})_{\alpha\beta}Z^{\mu\nu\rho\sigma\lambda}\,.
\label{eq:IIAsugra}
\end{align}
$Z,Z^{\mu},Z^{\mu\nu},Z^{\mu\nu\rho\sigma},Z^{\mu\nu\rho\sigma\lambda}$ denote totally antisymmetric central charges while $P_i$ denotes momentum. We will only consider D0-, D2- and D4-branes at rest wrapping a 4-torus which extends along the axes $m,n,\ldots\in \{6,7,8,9\}$ with closed string metric $g_{mn}$ and $B$-field $B_{mn}$ which we, for now, will assume to vanish. The only non-vanishing central charges will therefore be $Z,Z^{mn},Z^{mnrs}$. Note that normalizations are chosen such that in the coordinates adapted to the cycles of the 4-torus, the central charges are integer valued and express the wrapping numbers of given D-brane. Fixing a set of such charges and working in chiral basis, we can derive that for any multiplet $|\lambda\rangle$ we must have\footnote{In going from~\eqref{eq:IIAsugra} to~\eqref{eq:defGamma} we have to use the Majorana condition $Q^\dagger = B Q$ and also the fact that $BC=\Gamma^0$ in the chiral basis.}
\begin{equation}
\langle\lambda|\{Q^\dagger,Q\}|\lambda\rangle = \mathcal{M}-Z\Gamma_0\Gamma^{11}  -\frac{1}{2}Z^{mn}\Gamma_0\Gamma_{mn} -\frac{1}{4!}Z^{mnrs}\Gamma_0\Gamma_{mnrs}\Gamma^{11}\equiv \mathcal{M}-\Gamma\geqslant 0\,,
\label{eq:defGamma}
\end{equation}
which gives a bound on $\mathcal{M}$. Here $\mathcal{M}\equiv -P_0$ is the mass (normalised so that the mass of D0-brane is equal to 1). When this bound is saturated, we have 
$\det (\Gamma-\mathcal{M})=0$ and $(\Gamma-\mathcal{M})^{\alpha\beta}Q_{\beta}|\lambda\rangle=0$. The multiplet therefore preserves a fraction of vacuum supersymetries, which is given by the number of zero eigenvalues of $\Gamma-\mathcal{M}$. The conserved combinations of supercharges are then given by zero-eigenvalue eigenvectors of $\Gamma-\mathcal{M}$. Squaring the eigenvalue equation $\Gamma\epsilon=\mathcal{M}\epsilon$, we obtain (remembering that the indices run over a 4-torus)
\begin{equation}
\Gamma^2\epsilon=\left(Z^2  +\frac{1}{2}Z^{mn}Z_{mn}+\frac{1}{4!}Z^{mnrs}Z_{mnrs}+\frac{1}{12}(ZZ^{mnrs}-3Z^{[mn}Z^{rs]})\Gamma_{mnrs}\right)\epsilon\,.
\end{equation}
Hence, if 
\begin{equation}
k^{mnrs}\equiv ZZ^{mnrs} - 3 Z^{[mn}Z^{rs]}=0\,,\label{eq:12BPS}
\end{equation}
we find that $\Gamma$ squares to a multiple of identity, so it has eigenvalues $\pm\mathcal{M}_{1/2}$, where we denoted
\begin{equation}
\mathcal{M}_{1/2}^2 = Z^2  +\frac{1}{2}Z^{mn}Z_{mn}+\frac{1}{4!}Z^{mnrs}Z_{mnrs}\,.\label{eq:12BPSmass}
\end{equation}
Moreover, since $\Gamma$ is traceless, we see that half of the eigenvalues is positive and half negative. This means that for $\mathcal{M}=\mathcal{M}_{1/2}$, the matrix $\Gamma-\mathcal{M}$ has 16 zero eigenvalues, so exactly half of the vacuum supersymmetries are preserved. We call such multiplets 1/2-BPS. On the other hand, if $k^{mnrs}\neq 0$, less supersymmetry is preserved. If we denote $\Gamma' \equiv \frac{1}{12}k^{mnrs}\Gamma_{mnrs}$, we have
\begin{equation}
(\mathcal{M}^2-\mathcal{M}^2_{1/2})^2\epsilon = \Gamma'^2\epsilon = \frac{1}{6}k^{mnrs}k_{mnrs}\epsilon\,.
\end{equation}
That is, on a 4-torus, the matrix $\Gamma'$ always squares to a multiple of identity and, being traceless, its eigenvalues are $\pm\Delta\mathcal{M}^2\equiv\pm 2\sqrt{(1/4!)k^{mnrs}k_{mnrs}}$ with half of them positive and half of them negative. This in turn gives that $\Gamma^2 = \mathcal{M}_{1/2}^2 +\Gamma'$ has eigenvalues equal to $\mathcal{M}_{1/2}^2\pm\Delta\mathcal{M}^2$, each of them with 16-fold degeneracy. Since $\Gamma$ must be traceless, its\\[-2pt]
eigenvalues must be precisely $\pm(\mathcal{M}_{1/2}^2 \pm\Delta\mathcal{M}^2)^{1/2}$, each with eight-fold degeneracy. Hence, setting $\mathcal{M}=\mathcal{M}_{1/4}$, where we define
\begin{equation}
\mathcal{M}_{1/4}^2 \equiv\mathcal{M}_{1/2}^2+\Delta\mathcal{M}^2 =  Z^2  +\frac{1}{2}Z^{mn}Z_{mn}+\frac{1}{4!}Z^{mnrs}Z_{mnrs} +2\sqrt{\frac{1}{4!}k^{mnrs}k_{mnrs}}\,,
\label{eq:14BPSmass}
\end{equation}
we obtain that the matrix $\Gamma-\mathcal{M}$ has 8 zero eigenvalues, so a quarter of spacetime supersymmetries are preserved. We call such multiplets 1/4-BPS. We also observe that all supersymmetric multiplets on a 4-torus are at least 1/4-BPS, since $\Gamma'$ always squares to a multiple of identity on a 4-torus: we would have to consider non-trivial charges wrapping along more than four spacetime directions to obtain states preserving a smaller fraction of spacetime supersymmetry. Finally, note that when $B_{mn}\neq 0$, the above equations must be modified by substituting the $B$-deformed central charges $\tilde{Z}=Z + \frac{1}{2}Z^{mn}B_{mn}+\frac{1}{8}Z^{mnrs}B_{mn}B_{rs}$, $\tilde{Z}^{mn}= Z^{mn} + \frac{1}{2}Z^{mnrs}B_{rs}$, $\tilde{Z}^{mnrs}= Z^{mnrs}$. 

\section{Representations of $\mathcal{N}=2$ superconformal algebras in two dimensions}
\label{app:scft}

Here we collect some basic information on the $\mathcal{N}=2$ superconformal theories in two dimensions. More details can be found e.g.\ in~\cite{Gepner:1989gr}. 

\subsection*{Introduction}
The $\mathcal{N}=2$ super-Virasoro algebra reads
\begingroup
\allowdisplaybreaks
\begin{subequations}
\begin{align}
    [L_n,L_m] &= (n-m)L_{n+m}+\frac{c}{12}(n^3-n)\delta_{n+m,0}\,,\\[1pt]
    [L_n,J_m] &= -m J_{n+m}\,,\\[3pt]
    [L_n,G_{r\pm a}^\pm] &= (\frac{n}{2}-(r\pm a)) G_{n+r\pm a}^{\pm}\,,\\[0pt]
    [J_n,J_m]&= \frac{c}{3}n\delta_{n+m,0}\,,\\[2pt]
    [J_n,G_{r\pm a}^\pm] &= \pm G_{n+r\pm a}^{\pm}\,,\\
    \{G_{r+a}^+,G_{s-a}^-\} &= 2 L_{r+s}+(r-s+2a)J_{r+s}+\frac{c}{3}((r+a)^2 -\frac{1}{4})\delta_{r+s,0}\,.
\end{align}
\end{subequations}
\endgroup
Here, $a=0$ gives the Ramond sector, while $a=\frac{1}{2}$ gives the Neveu-Schwarz sector. We can interpolate between the two by means of the spectral flow operator \hbox{$U_\eta = \exp({i\eta \sqrt{\frac{c}{3}}H})$}, where we bosonized the $U(1)$ current as $J = i\sqrt{\frac{c}{3}}\partial H$. We will assign even worldsheet fermion number to $L_m, J_m$ and odd worldsheet fermion number to $G_r^\pm$. The irreducible representations are labeled by highest weights with respect to the maximal commuting bosonic subalgebra, which is generated by $L_0,J_0$. The corresponding weights are therefore the conformal dimension $h$ and the $U(1)$ charge $q$. From now on, we will only consider unitary representations. Spectral flow acts on the modules by deforming their highest weights as
\begin{subequations}
\begin{align}
    h_\eta &= h +\eta q+ \eta^2 \frac{c}{6}\,, \label{eq:specFlh}\\
    q_\eta &= q+ \eta\frac{c}{3}\,.
    \label{eq:specFlq}
\end{align}
\end{subequations}
We can prove that in the NS sector $h\geqslant \frac{|q|}{2}$. Those states which saturate the inequality and have $q>0$ are called chiral, while those with $q<0$ are called anti-chiral. Chiral states $|\psi_c\rangle$ satisfy $G^+_{-{1}/{2}}|\psi_c\rangle =0$, while the anti-chiral ones $|\psi_a\rangle$ give $G^{-}_{-{1}/{2}}|\psi_a\rangle=0$. Under the spectral flow with $\eta=-1/2$, the chiral states map to the Ramond ground states, which are states with $h=\frac{c}{24}$. Indeed, for a general Ramond primary, we can prove that $h\geqslant \frac{c}{24}$.

\subsection*{Examples: free fields and minimal models}

Consider the theory of one complex free boson $X^\pm = (X^1 \pm i X^2 )/\sqrt{2}$ and one complex free fermion $\psi^\pm = (\psi^1\pm i \psi^2)/\sqrt{2}$. Then the currents
\begin{subequations}
\begin{align}
    T &= -\partial X^+\partial X^- -\psi^+ \partial \psi^- -\psi^-\partial \psi^+\,,\\
    J &= -\psi^-\psi^+\,,\\
    G^\pm &= i\psi^\pm\partial X^\mp
\end{align}
\end{subequations}
satisfy the $\mathcal{N}=2$ super-Virasoro algebra with $c=3$. It can be shown that this theory admits an infinite number of irreducible representations, i.e.\ it is non-rational. On the other hand, for
\begin{equation}
c_k = \frac{3k}{k+2}< 3\,,\quad k=1,2,3,\ldots
\end{equation}
we only obtain a finite number of $\mathcal{N}=2$ modules. These are the minimal models. It is convenient to work with the irreps with respect to the maximal bosonic subalgebra of the $\mathcal{N}=2$ super-Virasoro algebra, which is generated by all combinations of $\mathcal{N}=2$ generators with even fermion number. The irreps can be labeled by the triples $(l,m,s)$ where $0\leqslant l\leqslant k$, $m\in\mathbb{Z}_{2(k+2)}$, $s\in\mathbb{Z}_4$ with $l+m+s=0\,\text{mod 2}$ and modulo the field identification $(l,m,s)\sim(k-l,m+k+2,s+2)$. We will call this set $\mathcal{I}_\mathrm{b}$. The corresponding highest weights read
\begin{subequations}
\begin{align}
    h^{l}_{m,s} &= \frac{l(l+2)-m^2}{4(k+2)} + \frac{s^2}{8}\quad\text{mod 1}\,,\\
    q^{l}_{m,s} &= \frac{m}{k+2} -\frac{s}{2}\quad\text{mod 2}\,.
\end{align}
\end{subequations}
The states with $s=0,2$ belong to the NS sector, while those with $s=\pm 1$ bolong to the R sector. Using the formulae~\eqref{eq:specFlh} and~\eqref{eq:specFlq}, it can be readily derived that spectral flow by $\eta=+1/2$ maps $(l,m,s)$ to $(l,m-2,s-1)$. The bosonic irreps can be grouped into pairs $[l,m]=((l,m,s),(l,m,s+2))$ which form modules with respect to the full $\mathcal{N}=2$ super-Virasoro algebra. Let us define the Virasoro specialized characters $\chi^{l}_{m,s}(\tau) = \mathrm{Tr}_{\mathcal{H}^{l}_{m,s}} q^{L_0 - \frac{c}{24}}$ where $q=e^{2\pi i \tau}$ and $\mathcal{H}^{l}_{m,s}$ is the corresponding highest-weight module with respect to the bosonic subalgebra. Then
\begin{equation}
    \chi^{l}_{m,s}(-\tfrac{1}{\tau}) = \sum_{(l',m',s')\in\mathcal{I}_\mathrm{b}} S^{ll'}_{mm',ss'}\chi^{l'}_{m',s'}(\tau)\,,\label{eq:Stransf}
\end{equation}
where
\begin{equation}
S^{ll'}_{mm',ss'} = \frac{1}{k+2}\sin\left[\pi\frac{(l+1)(l'+1)}{k+2}\right]e^{i\pi\frac{mm'}{k+2}}e^{-i\pi\frac{ss'}{2}}\,.\label{eq:Smat}
\end{equation}
The characters $\chi^{l}_{m,s}(q)$ have the following $q$-expansion \cite{Eguchi:2001ip}
\begin{subequations}
\label{eq:qseries}
\begin{align}
\chi^{l}_{m,s}(q) &= \sum_{j=0}^{k-1}c^l_{m+4j-s}(q)\,\theta_{2m+(4j-s)(k+2),2k(k+2)}(q)\,,\\
\theta_{M,K} &=\sum_{n\in\mathbb{Z}}q^{K(n+\frac{M}{2k})^2}\,,
\end{align}
\end{subequations}
where $c^l_m$ are the $\widehat{\mathfrak{su}}(2)_k$ string functions, which satisfy $c^l_m = c^l_{-m}=c^{k-l}_{k\pm m}=c^l_{m+2k}$ with $c^l_{m}=0$ for $l+m\neq 2\mathbb{Z}$. 
For $k=1$ we have $c^l_m(q) = \eta(q)^{-1}$ while for $k=2$ we have
\begin{align}
c^{0}_{0}(q) &= \chi_0(q)\eta(q)^{-1}\,,\\
c^{2}_{0}(q) &= \chi_{1/2}(q)\eta(q)^{-1}\,,\\
2c^{1}_{1}(q) &= \chi_{1/16}(q)\eta(q)^{-1}\,,
\end{align}
where $\chi_0,\chi_{1/2},\chi_{1/16}$ are the critical Ising characters.

\subsection*{Gluing conditions}

There are two types of gluing conditions one may impose on the left- and right-moving modes of a $\mathcal{N}=2$ superconformal algebra: the A-type gluing conditions
\begin{subequations}
\begin{align}
(L_m-\overline{L}_{-m})\| b,\eta\rangle\!\rangle&=0\,,\\[+2pt]
(J_{m}-\overline{J}_{-m})\|b,\eta \rangle\!\rangle&=0\,,\\
(G_{r}^{\pm}+i\eta\overline{G}^{\mp}_{-r})\| b,\eta\rangle\!\rangle&=0\,,
\end{align}
\end{subequations}
and B-type gluing conditions
\begin{subequations}
\begin{align}
(L_m-\overline{L}_{-m})\| b,\eta\rangle\!\rangle&=0\,,\\[+2pt]
(J_{m}+\overline{J}_{-m})\|b,\eta \rangle\!\rangle&=0\,,\\
(G_{r}^{\pm}+i\eta\overline{G}^{\pm}_{-r})\| b,\eta\rangle\!\rangle&=0\,.
\end{align}
\end{subequations}
For theories whose chiral algebra can be written as a direct sum of several $\mathcal{N}=2$ SCAs, one can consider putting A- or B-type gluing conditions independently on each constituent chiral algebra. In addition, when the central charges of two or more of the constituent $\mathcal{N}=2$ SCAs agree, one can consider imposing the permutation gluing conditions \citep{Recknagel:2002qq}. Gluing conditions are then labeled by strings of As and Bs together with elements of the permutation group on the subset of SCAs whose central charges coincide. For instance, in a theory whose chiral algebra is the direct sum of six copies of a $\mathcal{N}=2$ SCA, a typical gluing condition will be encoded as $(1_\mathrm{A}2_\mathrm{A}4_\mathrm{B})(3_\mathrm{A}5_\mathrm{B})(6_\mathrm{B})$, which translates into the following gluing conditions on the $U(1)$ currents:
\begingroup
\allowdisplaybreaks
\begin{subequations}
\begin{align}
(J_{n}^{(1)}-\overline{J}_{n}^{(2)})\|b,\eta\rangle\!\rangle&=0\,,\\
(J_{n}^{(2)}-\overline{J}_{n}^{(4)})\|b,\eta\rangle\!\rangle&=0\,,\\
(J_{n}^{(4)}+\overline{J}_{n}^{(1)})\|b,\eta\rangle\!\rangle&=0\,,\\
(J_{n}^{(3)}-\overline{J}_{n}^{(5)})\|b,\eta\rangle\!\rangle&=0\,,\\
(J_{n}^{(5)}+\overline{J}_{n}^{(3)})\|b,\eta\rangle\!\rangle&=0\,,\\
(J_{n}^{(6)}+\overline{J}_{n}^{(6)})\|b,\eta\rangle\!\rangle&=0\,.
\end{align}
\end{subequations}
\endgroup

\section{RR-charged boundary states for general rational gluing conditions}
\label{app:class}

Here we summarize our results for general permutation gluing conditions which allow for boundary states carrying non-zero RR charges. All such boundary states are found to be stable and supersymmetric with $\mathcal{N}$ conserved supercharges. Their mass $\mathcal{M}$ is found to saturate the BPS bound for given RR charges $Z$, $Z^{mn}$, $Z^{mnrs}$. Also, the number $n$ of massless open string states (determined by computing mutual overlaps of the boundary states and S-transforming into the open string channel) in either the NS or R sector is found to agree with the ADHM formula $4kN+8$ with $kN$ given by \eqref{eq:kN}. Results for the $SU(3)^2$ 4-torus are shown in Table \ref{tab:classSU32}, while the results for the $SU(2)^4$ 4-torus are shown in Table \ref{tab:classSU24}.


{\small\begin{longtable}[htpb!]{cccccccccccc}
\caption{Mass $\mathcal{M}$, RR-charges $Z,Z^{mn},Z^{mnrs}$ and number $\mathcal{N}$ and $n$ of supercharges and massless boundary fields for stable Gepner-like boundary states on the $SU(3)^2$ 4-torus.}\label{tab:classSU32}\\
\toprule\toprule
gluing automorphism & $\mathcal{M}$ & $Z$ & $Z^{67}$ & $Z^{89}$ & $Z^{6789}$ & $Z^{68}$ & $Z^{69}$ & $Z^{78}$ & $Z^{79}$ & $\mathcal{N}$ & $n$\\
\midrule
\endfirsthead
\caption{\textbf{(cont'd)} Mass $\mathcal{M}$, RR-charges $Z,Z^{mn},Z^{mnrs}$ and number $\mathcal{N}$ and $n$ of supercharges and massless boundary fields for stable Gepner-like boundary states on the $SU(3)^2$ 4-torus.}\\
\toprule\toprule
gluing automorphism & $\mathcal{M}$ & $Z$ & $Z^{67}$ & $Z^{89}$ & $Z^{6789}$ & $Z^{68}$ & $Z^{69}$ & $Z^{78}$ & $Z^{79}$ & $\mathcal{N}$ & $n$\\
\midrule
\endhead
  \multirow{9}{*}{$\begin{array}{c}
  (1_\mathrm{B})(2_\mathrm{B})(3_\mathrm{B})(4_\mathrm{B})(5_\mathrm{B})(6_\mathrm{B})\\
  \text{and 8 others}
  \end{array}$}
 & 1 & $1$ & $0$ & $0$ & $0$ & $0$ & $0$ & $0$ & $0$ & 16 & 8\\
 & 1 & $0$ & $1$ & $0$ & $0$ & $0$ & $0$ & $0$ & $0$ & 16 & 8\\
 & 1 & $0$ & $0$ & $1$ & $0$ & $0$ & $0$ & $0$ & $0$ & 16 & 8\\
 & 1 & $0$ & $0$ & $0$ & $1$ & $0$ & $0$ & $0$ & $0$ & 16 & 8\\
 & 1 & $1$ & $-1$ & $0$ & $0$ & $0$ & $0$ & $0$ & $0$ & 16 & 8\\
 & 1 & $1$ & $0$ & $-1$ & $0$ & $0$ & $0$ & $0$ & $0$ & 16 & 8\\
 & 1 & $0$ & $-1$ & $0$ & $1$ & $0$ & $0$ & $0$ & $0$ & 16 & 8\\
 & 1 & $0$ & $0$ & $-1$ & $1$ & $0$ & $0$ & $0$ & $0$ & 16 & 8\\
 & 1 & $1$ & $-1$ & $-1$ & $1$ & $0$ & $0$ & 0& $0$ & 16 & 8\\
\midrule
 \multirow{9}{*}{$\begin{array}{c}
  (1_\mathrm{A}2_\mathrm{A})(3_\mathrm{A})(4_\mathrm{A}5_\mathrm{A})(6_\mathrm{A})\\
  \text{and 8 others}
  \end{array}$}
 & 1  & $0$ & $0$ & $0$ & $0$ & $1$ & $0$ & $0$ & $0$ & $16$ & $8$ \\
 & 1  & $0$ & $0$ & $0$ & $0$ & $0$ & $1$ & $0$ & $0$ & 16 & 8\\
 & 1  & $0$ & $0$ & $0$ & $0$ & $0$ & $0$ & $1$ & $0$ & 16 & 8\\
 & 1  & $0$ & $0$ & $0$ & $0$ & $0$ & $0$ & $0$ & $1$ & 16 & 8\\
 & 1  & $0$ & $0$ & $0$ & $0$ & $1$ & $-1$ & $0$ & $0$ & 16 & 8\\
 & 1  & $0$ & $0$ & $0$ & $0$ & $1$ & $0$ & $-1$ & $0$ & 16 & 8\\
 & 1  & $0$ & $0$ & $0$ & $0$ & $0$ & $-1$ & $0$ & $1$ & 16 & 8\\
 & 1  & $0$ & $0$ & $0$ & $0$ & $0$ & $0$ & $-1 $& $1$ & 16 & 8\\
 & 1  & $0$ & $0$ & $0$ & $0$ & $1$ & $-1$ & $-1$ & $1$ & 16 & 8\\
\midrule
\pagebreak
 \multirow{9}{*}{$\begin{array}{c}
  (1_\mathrm{B}2_\mathrm{B})(3_\mathrm{B})(4_\mathrm{B})(5_\mathrm{B})(6_\mathrm{B})\\
  \text{and 8 others}
  \end{array}$}
 & $\sqrt{3}$ & $1$ & $1$ & $0$ & $0$ & $0$ & $0$ & $0$ & $0$ & 16 & 8\\
 & $\sqrt{3}$ & $2$ & $-1$ & $0$ & $0$ & $0$ & $0$ & $0$ & $0$ & 16 & 8\\
 & $\sqrt{3}$ & $-1$ & $2$ & $0$ & $0$ & $0$ & $0$ & $0$ & $0$ & 16 & 8\\
 & $\sqrt{3}$ & $0$ & $0$ & $1$ & $1$ & $0$ & $0$ & $0$ & $0$ & 16 & 8\\
 & $\sqrt{3}$ & $0$ & $0$ & $2$ & $-1$ & $0$ & $0$ & $0$ & $0$ & 16 & 8\\
 & $\sqrt{3}$ & $0$ & $0$ & $-1$ & $2$ & $0$ & $0$ & $0$ & $0$ & 16 & 8\\
 & $\sqrt{3}$ & $1$ & $1$ & $-1$ & $-1$ & $0$ & $0$ & $0$ & $0$ & 16 & 8\\
 & $\sqrt{3}$ & $2$ & $-1$ & $-2$ & $1$ & $0$ & $0$ & $0$ & $0$ & 16 & 8\\
 & $\sqrt{3}$ & $1$ & $-2$ & $-1$ & $2$ & $0$ & $0$ & $0$ & $0$ & 16 & 8\\
  \midrule
 \multirow{9}{*}{$\begin{array}{c}
  (1_\mathrm{B})(2_\mathrm{B})(3_\mathrm{B})(4_\mathrm{B}5_\mathrm{B})(6_\mathrm{B})\\
  \text{and 8 others}
  \end{array}$}
 & $\sqrt{3}$ & $0$ & $1$ & $0$ & $1$ & $0$ & $0$ & $0$ & $0$ & 16 & 8\\
 & $\sqrt{3}$ & $0$ & $2$ & $0$ & $-1$ & $0$ & $0$ & $0$ & $0$ & 16 & 8\\
 & $\sqrt{3}$ & $0$ & $-1$ & $0$ & $2$ & $0$ & $0$ & $0$ & $0$ & 16 & 8\\
 & $\sqrt{3}$ & $1$ & $0$ & $1$ & $0$ & $0$ & $0$ & $0$ & $0$ & 16 & 8\\
 & $\sqrt{3}$ & $2$ & $0$ & $-1$ & $0$ & $0$ & $0$ & $0$ & $0$ & 16 & 8\\
 & $\sqrt{3}$ & $-1$ & $0$ & $2$ & $0$ & $0$ & $0$ & $0$ & $0$ & 16 & 8\\
 & $\sqrt{3}$ & $1$ & $-1$ & $1$ & $-1$ & $0$ & $0$ & $0$ & $0$ & 16 & 8\\
 & $\sqrt{3}$ & $2$ & $-2$ & $-1$ & $1$ & $0$ & $0$ & $0$ & $0$ & 16 & 8\\
 & $\sqrt{3}$ & $1$ & $-1$ & $-2$ & $2$ & $0$ & $0$ & $0$ & $0$ & 16 & 8\\
 \midrule
 \multirow{9}{*}{$\begin{array}{c}
  (1_\mathrm{A})(2_\mathrm{A})(3_\mathrm{A})(4_\mathrm{A} 5_\mathrm{A})(6_\mathrm{A})\\
  \text{and 8 others}
  \end{array}$}
 & $\sqrt{3}$  & $0$ & $0$ & $0$ & $0$& $1$ & $1$ & $0$ & $0$ & 16 & 8\\
 & $\sqrt{3}$  & $0$ & $0$ & $0$ & $0$& $2$ & $-1$ & $0$ & $0$ & 16 & 8\\
 & $\sqrt{3}$ & $0$ & $0$ & $0$ & $0$  & $-1$ & $2$ & $0$ & $0$& 16 & 8\\
 & $\sqrt{3}$  & $0$ & $0$ & $0$ & $0$ & $0$ & $0$ & $1$ & $1$& 16 & 8\\
 & $\sqrt{3}$  & $0$ & $0$ & $0$ & $0$ & $0$ & $0$ & $2$ & $-1$ & 16 & 8\\
 & $\sqrt{3}$  & $0$ & $0$ & $0$ & $0$ & $0$ & $0$ & $-1$ & $2$& 16 & 8\\
 & $\sqrt{3}$  & $0$ & $0$ & $0$ & $0$ & $1$ & $1$ & $-1$ & $-1$& 16 & 8\\
 & $\sqrt{3}$  & $0$ & $0$ & $0$ & $0$ & $2$ & $-1$ & $-2$ & $1$& 16 & 8\\
 & $\sqrt{3}$ & $0$ & $0$ & $0$ & $0$ & $1$ & $-2$ & $-1$ & $2$ & 16 & 8\\
  \midrule
 \multirow{9}{*}{$\begin{array}{c}
  (1_\mathrm{A}2_\mathrm{A})(3_\mathrm{A})(4_\mathrm{A})(5_\mathrm{A})(6_\mathrm{A})\\
  \text{and 8 others}
  \end{array}$}
 & $\sqrt{3}$  & $0$ & $0$ & $0$ & $0$ & $0$ & $1$ & $0$ & $1$ & 16 & 8\\
 & $\sqrt{3}$  & $0$ & $0$ & $0$ & $0$ & $0$ & $2$ & $0$ & $-1$ & 16 & 8\\
 & $\sqrt{3}$ & $0$ & $0$ & $0$ & $0$ & $0$ & $-1$ & $0$ & $2$ & 16 & 8\\
 & $\sqrt{3}$ & $0$ & $0$ & $0$ & $0$ & $1$ & $0$ & $1$ & $0$ & 16 & 8\\
 & $\sqrt{3}$ & $0$ & $0$ & $0$ & $0$ & $2$ & $0$ & $-1$ & $0$ & 16 & 8\\
 & $\sqrt{3}$ & $0$ & $0$ & $0$ & $0$ & $-1$ & $0$ & $2$ & $0$ & 16 & 8\\
 & $\sqrt{3}$  & $0$ & $0$ & $0$ & $0$ & $1$ & $-1$ & $1$ & $-1$& 16 & 8\\
 & $\sqrt{3}$  & $0$ & $0$ & $0$ & $0$ & $2$ & $-2$ & $-1$ & $1$& 16 & 8\\
 & $\sqrt{3}$  & $0$ & $0$ & $0$ & $0$ & $1$ & $-1$ & $-2$ & $2$& 16 & 8\\
\midrule
\pagebreak
 \multirow{9}{*}{$\begin{array}{c}
  (1_\mathrm{B}2_\mathrm{B})(3_\mathrm{B})(4_\mathrm{B}5_\mathrm{B})(6_\mathrm{B})\\
  \text{and 8 others}
  \end{array}$}
 & ${3}$ & $1$ & $1$ & $1$ & $1$ & $0$ & $0$ & $0$ & $0$ & 16 & 8\\
 & ${3}$ & $2$ & $2$ & $-1$ & $-1$ & $0$ & $0$ & $0$ & $0$ & 16 & 8\\
 & ${3}$ & $1$ & $1$ & $-2$ & $-2$ & $0$ & $0$ & $0$ & $0$ & 16 & 8\\
 & ${3}$ & $2$ & $-1$ & $2$ & $-1$ & $0$ & $0$ & $0$ & $0$ & 16 & 8\\
 & ${3}$ & $1$ & $-2$ & $1$ & $-2$ & $0$ & $0$ & $0$ & $0$ & 16 & 8\\
 & ${3}$ & $4$ & $-2$ & $-2$ & $1$ & $0$ & $0$ & $0$ & $0$ & 16 & 8\\
 & ${3}$ & $2$ & $-1$ & $-4$ & $2$ & $0$ & $0$ & $0$ & $0$ & 16 & 8\\
 & ${3}$ & $2$ & $-4$ & $-1$ & $2$ & $0$ & $0$ & $0$ & $0$ & 16 & 8\\
 & ${3}$ & $1$ & $-2$ & $-2$ & $4$ & $0$ & $0$ & $0$ & $0$ & 16 & 8\\
  \midrule
 \multirow{9}{*}{$\begin{array}{c}
  (1_\mathrm{A})(2_\mathrm{A})(3_\mathrm{A})(4_\mathrm{A})(5_\mathrm{A})(6_\mathrm{A})\\
  \text{and 8 others}
  \end{array}$}
 & ${3}$ & $0$ & $0$ & $0$ & $0$ & $1$ & $1$ & $1$ & $1$& 16 & 8\\
 & ${3}$ & $0$ & $0$ & $0$ & $0$ & $2$ & $2$ & $-1$ & $-1$ & 16 & 8\\
 & ${3}$ & $0$ & $0$ & $0$ & $0$ & $1$ & $1$ & $-2$ & $-2$& 16 & 8\\
 & ${3}$ & $0$ & $0$ & $0$ & $0$ & $2$ & $-1$ & $2$ & $-1$ & 16 & 8\\
 & ${3}$ & $0$ & $0$ & $0$ & $0$ & $1$ & $-2$ & $1$ & $-2$& 16 & 8\\
 & ${3}$ & $0$ & $0$ & $0$ & $0$ & $4$ & $-2$ & $-2$ & $1$ & 16 & 8\\
 & ${3}$ & $0$ & $0$ & $0$ & $0$ & $2$ & $-1$ & $-4$ & $2$& 16 & 8\\
 & ${3}$ & $0$ & $0$ & $0$ & $0$ & $2$ & $-4$ & $-1$ & $2$& 16 & 8\\
 & ${3}$ & $0$ & $0$ & $0$ & $0$ & $1$ & $-2$ & $-2$ & $4$& 16 & 8\\
 \midrule
 \multirow{9}{*}{$\begin{array}{c}
  (1_\mathrm{B}4_\mathrm{B})(2_\mathrm{B}5_\mathrm{B})(3_\mathrm{B}6_\mathrm{B})\\
  \text{and 17 others}
  \end{array}$}
 & $\sqrt{3}$ & $-1$ & $0$ & $0$ & $1$ & $-1$ & $1$ & $0$ & $-1$& 16 & 8\\
 & $\sqrt{3}$ & $-1$ & $1$ & $1$ & $0$ & $-1$ & $0$ & $1$ & $-1$ & 16 & 8\\
 & $\sqrt{3}$ & $0$ & $-1$ & $-1$ & $1$ & $0$ & $1$ & $-1$ & $0$& 16 & 8\\
 & $\sqrt{3}$ & $-1$ & $1$ & $1$ & $0$ & $0$ & $1$ & $-1$ & $0$ & 16 & 8\\
 & $\sqrt{3}$ & $0$ & $1$ & $1$ & $-1$ & $-1$ & $1$ & $0$ & $-1$& 16 & 8\\
 & $\sqrt{3}$ & $-1$ & $0$ & $0$ & $1$ & $1$ & $0$ & $-1$ & $1$ & 16 & 8\\
 & $\sqrt{3}$ & $0$ & $-1$ & $-1$ & $1$ & $-1$ & $0$ & $1$ & $-1$& 16 & 8\\
 & $\sqrt{3}$ & $-1$ & $0$ & $0$ & $1$ & $0$ & $-1$ & $1$ & $0$ & 16 & 8\\
 & $\sqrt{3}$ & $1$ & $-1$ & $-1$ & $0$ & $-1$ & $1$ & $0$ & $-1$& 16 & 8\\
\midrule
 \multirow{9}{*}{$\begin{array}{c}
  (1_\mathrm{A}4_\mathrm{A})(2_\mathrm{A}5_\mathrm{A})(3_\mathrm{A}6_\mathrm{A})\\
  \text{and 17 others}
  \end{array}$}
 & $\sqrt{3}$ & $-1$ & $1$ & $0$ & $-1$ & $-1$ & $1$ & $1$ & $0$& 16 & 8\\
 & $\sqrt{3}$ & $-1$ & $0$ & $1$ & $-1$ & $0$ & $1$ & $1$ & $-1$ & 16 & 8\\
 & $\sqrt{3}$ & $0$ & $1$ & $-1$ & $0$ & $-1$ & $0$ & $0$ & $1$& 16 & 8\\
 & $\sqrt{3}$ & $0$ & $1$ & $-1$ & $0$ & $0$ & $1$ & $1$ & $-1$ & 16 & 8\\
 & $\sqrt{3}$ & $-1$ & $1$ & $0$ & $-1$ & $1$ & $0$ & $0$ & $-1$& 16 & 8\\
 & $\sqrt{3}$ & $1$ & $0$ & $-1$ & $1$ & $-1$ & $1$ & $1$ & $0$ & 16 & 8\\
 & $\sqrt{3}$ & $-1$ & $0$ & $1$ & $-1$ & $-1$ & $0$ & $0$ & $1$& 16 & 8\\
 & $\sqrt{3}$ & $0$ & $-1$ & $1$ & $0$ & $-1$ & $1$ & $1$ & $0$ & 16 & 8\\
 & $\sqrt{3}$ & $-1$ & $1$ & $0$ & $-1$ & $0$ & $-1$ & $-1$ & $1$& 16 & 8\\
 \midrule
 \pagebreak
 \multirow{9}{*}{$\begin{array}{c}
  (1_\mathrm{B}4_\mathrm{B}2_\mathrm{B}5_\mathrm{B})(3_\mathrm{B}6_\mathrm{B})\\
  \text{and 17 others}
  \end{array}$}
 & ${3}$ & $-2$ & $1$ & $1$ & $1$ & $-1$ & $2$ & $-1$ & $-1$& 16 & 8\\
 & ${3}$ & $-1$ & $2$ & $2$ & $-1$ & $-2$ & $1$ & $1$ & $-2$ & 16 & 8\\
 & ${3}$ & $-1$ & $-1$ & $-1$ & $2$ & $1$ & $1$ & $-2$ & $1$ & 16 & 8\\
 & ${3}$ & $-1$ & $2$ & $2$ & $-1$ & $1$ & $1$ & $-2$ & $1$ & 16 & 8\\
 & ${3}$ & $1$ & $1$ & $1$ & $-2$ & $-1$ & $2$ & $-1$ & $-1$& 16 & 8\\
 & ${3}$ & $-2$ & $1$ & $1$ & $1$ & $2$ & $-1$ & $-1$ & $2$ & 16 & 8\\
 & ${3}$ & $-1$ & $-1$ & $-1$ & $2$ & $-2$ & $1$ & $1$ & $-2$& 16 & 8\\
 & ${3}$ & $-2$ & $1$ & $1$ & $1$ & $-1$ & $-1$ & $2$ & $-1$ & 16 & 8\\
 & ${3}$ & $1$ & $-2$ & $-2$ & $1$ & $-1$ & $2$ & $-1$ & $-1$& 16 & 8\\
 \midrule
 \multirow{9}{*}{$\begin{array}{c}
  (1_\mathrm{A}4_\mathrm{A}2_\mathrm{A}5_\mathrm{A})(3_\mathrm{A}6_\mathrm{A})\\
  \text{and 17 others}
  \end{array}$}
 & ${3}$ & $-1$ & $2$ & $-1$ & $-1$ & $-1$ & $2$ & $2$ & $-1$& 16 & 8\\
 & ${3}$ & $-2$ & $1$ & $1$ & $-2$ & $1$ & $1$ & $1$ & $-2$ & 16 & 8\\
 & ${3}$ & $1$ & $1$ & $-2$ & $1$ & $-2$ & $1$ & $1$ & $1$ & 16 & 8\\
 & ${3}$ & $1$ & $1$ & $-2$ & $1$ & $1$ & $1$ & $1$ & $-2$ & 16 & 8\\
 & ${3}$ & $-1$ & $2$ & $-1$ & $-1$ & $2$ & $-1$ & $-1$ & $-1$& 16 & 8\\
 & ${3}$ & $2$ & $-1$ & $-1$ & $2$ & $-1$ & $2$ & $2$ & $-1$ & 16 & 8\\
 & ${3}$ & $-2$ & $1$ & $1$ & $-2$ & $-2$ & $1$ & $1$ & $1$& 16 & 8\\
 & ${3}$ & $-1$ & $-1$ & $2$ & $-1$ & $-1$ & $2$ & $2$ & $-1$ & 16 & 8\\
 & ${3}$ & $-1$ & $2$ & $-1$ & $-1$ & $-1$ & $-1$ & $-1$ & $2$& 16 & 8\\
 \midrule
 \multirow{3}{*}{$\begin{array}{c}
  (1_\mathrm{B})(2_\mathrm{B})(3_\mathrm{B} 4_\mathrm{B})(5_\mathrm{B})(6_\mathrm{B})\\
  \text{and 80 others}
  \end{array}$}
 & $\sqrt{3}$ & $1$ & $0$ & $0$ & $-1$ & $0$ & $0$ & $0$ & $0$& 8 & 12\\
 & $\sqrt{3}$ & $1$ & $-1$ & $-1$ & $0$ & $0$ & $0$ & $0$ & $0$ & 8 & 12\\
 & $\sqrt{3}$ & $0$ & $-1$ & $-1$ & $1$ & $0$ & $0$ & $0$ & $0$ & 8 & 12\\
  \midrule
 \multirow{3}{*}{$\begin{array}{c}
  (1_\mathrm{B})(2_\mathrm{B})(3_\mathrm{A} 4_\mathrm{A})(5_\mathrm{B})(6_\mathrm{B})\\
  \text{and 80 others}
  \end{array}$}
 & $\sqrt{3}$ & $-1$ & $1$ & $0$ & $-1$ & $0$ & $0$ & $0$ & $0$& 8 & 12\\
 & $\sqrt{3}$ & $-1$ & $0$ & $1$ & $-1$ & $0$ & $0$ & $0$ & $0$ & 8 & 12\\
 & $\sqrt{3}$ & $0$ & $1$ & $-1$ & $0$ & $0$ & $0$ & $0$ & $0$ & 8 & 12\\
\midrule
 \multirow{3}{*}{$\begin{array}{c}
  (1_\mathrm{A}2_\mathrm{A})(3_\mathrm{A} 4_\mathrm{A})(5_\mathrm{A}6_\mathrm{A})\\
  \text{and 80 others}
  \end{array}$}
 & $\sqrt{3}$  & $0$ & $0$ & $0$ & $0$ & $1$ & $0$ & $0$ & $-1$& 8 & 12\\
 & $\sqrt{3}$  & $0$ & $0$ & $0$ & $0$& $1$ & $-1$ & $-1$ & $0$ & 8 & 12\\
 & $\sqrt{3}$ & $0$ & $0$ & $0$ & $0$  & $0$ & $-1$ & $-1$ & $1$  & 8 & 12\\
\midrule
 \multirow{3}{*}{$\begin{array}{c}
  (1_\mathrm{A}2_\mathrm{A})(3_\mathrm{B} 4_\mathrm{B})(5_\mathrm{A}6_\mathrm{A})\\
  \text{and 80 others}
  \end{array}$}
 & $\sqrt{3}$ & $0$ & $0$ & $0$ & $0$  & $-1$ & $1$ & $0$ & $-1$ & 8 & 12\\
 & $\sqrt{3}$  & $0$ & $0$ & $0$ & $0$& $-1$ & $0$ & $1$ & $-1$ & 8 & 12\\
 & $\sqrt{3}$ & $0$ & $0$ & $0$ & $0$  & $0$ & $1$ & $-1$ & $0$ & 8 & 12\\
\midrule
 \multirow{3}{*}{$\begin{array}{c}
  (1_\mathrm{B}2_\mathrm{B})(3_\mathrm{B} 4_\mathrm{B})(5_\mathrm{B})(6_\mathrm{B})\\
  \text{and 323 others}
  \end{array}$}
 & $3$ & $2$ & $-1$ & $-1$ & $-1$ & $0$ & $0$ & $0$ & $0$& 8 & 20\\
 & $3$ & $-1$ & $2$ & $2$ & $-1$ & $0$ & $0$ & $0$ & $0$ & 8 & 20\\
 & $3$ & $-1$ & $-1$ & $-1$ & $2$ & $0$ & $0$ & $0$ & $0$ & 8 & 20\\
  \midrule
 \multirow{3}{*}{$\begin{array}{c}
  (1_\mathrm{B}2_\mathrm{B})(3_\mathrm{A} 4_\mathrm{A})(5_\mathrm{B})(6_\mathrm{B})\\
  \text{and 323 others}
  \end{array}$}
 & $3$ & $-1$ & $2$ & $-1$ & $-1$ & $0$ & $0$ & $0$ & $0$& 8 & 20\\
 & $3$ & $2$ & $-1$ & $-1$ & $2$ & $0$ & $0$ & $0$ & $0$ & 8 & 20\\
 & $3$ & $-1$ & $-1$ & $2$ & $-1$ & $0$ & $0$ & $0$ & $0$ & 8 & 20\\
\midrule
\pagebreak
 \multirow{3}{*}{$\begin{array}{c}
  (1_\mathrm{A}2_\mathrm{A})(3_\mathrm{A} 4_\mathrm{A})(5_\mathrm{A})(6_\mathrm{A})\\
  \text{and 323 others}
  \end{array}$}
 & $3$  & $0$ & $0$ & $0$ & $0$ & $2$ & $-1$ & $-1$ & $-1$& 8 & 20\\
 & $3$  & $0$ & $0$ & $0$ & $0$& $-1$ & $2$ & $2$ & $-1$ & 8 & 20\\
 & $3$  & $0$ & $0$ & $0$ & $0$ & $-1$ & $-1$ & $-1$ & $2$& 8 & 20\\
\midrule
 \multirow{3}{*}{$\begin{array}{c}
  (1_\mathrm{A}2_\mathrm{A})(3_\mathrm{B} 4_\mathrm{B})(5_\mathrm{A})(6_\mathrm{A})\\
  \text{and 323 others}
  \end{array}$}
 & $3$ & $0$ & $0$ & $0$ & $0$  & $-1$ & $2$ & $-1$ & $-1$& 8 & 20\\
 & $3$  & $0$ & $0$ & $0$ & $0$ & $2$ & $-1$ & $-1$ & $2$ & 8 & 20\\
 & $3$ & $0$ & $0$ & $0$ & $0$  & $-1$ & $-1$ & $2$ & $-1$ & 8 & 20\\   
 \midrule
 \multirow{3}{*}{$\begin{array}{c}
  (1_\mathrm{B}2_\mathrm{B})(3_\mathrm{B} 4_\mathrm{B})(5_\mathrm{B}6_\mathrm{B})\\
  \text{and 242 others}
  \end{array}$}
 & $3\sqrt{3}$ & $3$ & $0$ & $0$ & $-3$ & $0$ & $0$ & $0$ & $0$& 8 & 44\\
 & $3\sqrt{3}$ & $3$ & $-3$ & $-3$ & $0$ & $0$ & $0$ & $0$ & $0$ & 8 & 44\\
 & $3\sqrt{3}$ & $0$ & $-3$ & $-3$ & $3$ & $0$ & $0$ & $0$ & $0$ & 8 & 44\\
  \midrule
 \multirow{3}{*}{$\begin{array}{c}
  (1_\mathrm{B}2_\mathrm{B})(3_\mathrm{A} 4_\mathrm{A})(5_\mathrm{B}6_\mathrm{B})\\
  \text{and 242 others}
  \end{array}$}
 & $3\sqrt{3}$ & $-3$ & $3$ & $0$ & $-3$ & $0$ & $0$ & $0$ & $0$& 8 & 44\\
 & $3\sqrt{3}$ & $-3$ & $0$ & $3$ & $-3$ & $0$ & $0$ & $0$ & $0$ & 8 & 44\\
 & $3\sqrt{3}$ & $0$ & $3$ & $-3$ & $0$ & $0$ & $0$ & $0$ & $0$ & 8 & 44\\
 \midrule
 \multirow{3}{*}{$\begin{array}{c}
  (1_\mathrm{A})(2_\mathrm{A})(3_\mathrm{A} 4_\mathrm{A})(5_\mathrm{A})(6_\mathrm{A})\\
  \text{and 242 others}
  \end{array}$}
 & $3\sqrt{3}$  & $0$ & $0$ & $0$ & $0$ & $3$ & $0$ & $0$ & $-3$& 8 & 44\\
 & $3\sqrt{3}$  & $0$ & $0$ & $0$ & $0$& $3$ & $-3$ & $-3$ & $0$ & 8 & 44\\
 & $3\sqrt{3}$ & $0$ & $0$ & $0$ & $0$  & $0$ & $-3$ & $-3$ & $3$  & 8 & 44\\
\midrule
 \multirow{3}{*}{$\begin{array}{c}
  (1_\mathrm{A})(2_\mathrm{A})(3_\mathrm{B} 4_\mathrm{B})(5_\mathrm{A})(6_\mathrm{A})\\
  \text{and 242 others}
  \end{array}$}
 & $3\sqrt{3}$ & $0$ & $0$ & $0$ & $0$  & $-3$ & $3$ & $0$ & $-3$ & 8 & 44\\
 & $3\sqrt{3}$  & $0$ & $0$ & $0$ & $0$& $-3$ & $0$ & $3$ & $-3$ & 8 & 44\\
 & $3\sqrt{3}$ & $0$ & $0$ & $0$ & $0$  & $0$ & $3$ & $-3$ & $0$ & 8 & 44\\
\bottomrule
  \end{longtable}}
\vspace{-2mm}

\clearpage

{\small\begin{longtable}{cccccccccccc}
\caption{Mass $\mathcal{M}$, RR-charges $Z,Z^{mn},Z^{mnrs}$ and number $\mathcal{N}$ and $n$ of supercharges and massless boundary fields for stable Gepner-like boundary states on the $SU(2)^4$ 4-torus.}\label{tab:classSU24}\\
\toprule\toprule
gluing automorphism & $\mathcal{M}$ & $Z$ & $Z^{67}$ & $Z^{89}$ & $Z^{6789}$ & $Z^{68}$ & $Z^{69}$ & $Z^{78}$ & $Z^{79}$ & $\mathcal{N}$ & $n$\\
\midrule
\endfirsthead
\caption{\textbf{(cont'd)} Mass $\mathcal{M}$, RR-charges $Z,Z^{mn},Z^{mnrs}$ and number $\mathcal{N}$ and $n$ of supercharges and massless boundary fields for stable Gepner-like boundary states on the $SU(2)^4$ 4-torus.}\\
\toprule\toprule
gluing automorphism & $\mathcal{M}$ & $Z$ & $Z^{67}$ & $Z^{89}$ & $Z^{6789}$ & $Z^{68}$ & $Z^{69}$ & $Z^{78}$ & $Z^{79}$ & $\mathcal{N}$ & $n$\\
\midrule
\endhead
  \multirow{16}{*}{$\begin{array}{c}
  (1_\mathrm{B})(2_\mathrm{B})(3_\mathrm{B})(4_\mathrm{B})\\
  \text{and 4 others}
  \end{array}$}
 & 1 & $1$ & $0$ & $0$ & $0$ & $0$ & $0$ & $0$ & $0$ & 16 & 8\\
 & 1 & $0$ & $1$ & $0$ & $0$ & $0$ & $0$ & $0$ & $0$ & 16 & 8\\
 & 1 & $0$ & $0$ & $1$ & $0$ & $0$ & $0$ & $0$ & $0$ & 16 & 8\\
 & 1 & $0$ & $0$ & $0$ & $1$ & $0$ & $0$ & $0$ & $0$ & 16 & 8\\
 & $\sqrt{2}$ & $1$ & $1$ & $0$ & $0$ & $0$ & $0$ & $0$ & $0$ & 16 & 8\\
 & $\sqrt{2}$ & $1$ & $-1$ & $0$ & $0$ & $0$ & $0$ & $0$ & $0$ & 16 & 8\\
 & $\sqrt{2}$ & $1$ & $0$ & $1$ & $0$ & $0$ & $0$ & $0$ & $0$ & 16 & 8\\
 & $\sqrt{2}$ & $1$ & $0$ & $-1$ & $0$ & $0$ & $0$ & $0$ & $0$ & 16 & 8\\
 & $\sqrt{2}$ & $0$ & $1$ & $0$ & $1$ & $0$ & $0$ & $0$ & $0$ & 16 & 8\\
 & $\sqrt{2}$ & $0$ & $-1$ & $0$ & $1$ & $0$ & $0$ & $0$ & $0$ & 16 & 8\\
 & $\sqrt{2}$ & $0$ & $0$ & $1$ & $1$ & $0$ & $0$ & $0$ & $0$ & 16 & 8\\
 & $\sqrt{2}$ & $0$ & $0$ & $-1$ & $1$ & $0$ & $0$ & $0$ & $0$ & 16 & 8\\
 & 2 & $1$ & $1$ & $1$ & $1$ & $0$ & $0$ & $0$ & $0$ & 16 & 8\\
 & 2 & $1$ & $-1$ & $-1$ & $1$ & $0$ & $0$ & $0$ & $0$ & 16 & 8\\
 & 2 & $1$ & $1$ & $-1$ & $-1$ & $0$ & $0$ & $0$ & $0$ & 16 & 8\\
 & 2 & $1$ & $-1$ & $1$ & $-1$ & $0$ & $0$ & $0$ & $0$ & 16 & 8\\
\midrule
  \multirow{16}{*}{$\begin{array}{c}
  (1_\mathrm{A})(2_\mathrm{A})(3_\mathrm{A})(4_\mathrm{A})\\
  \text{and 4 others}
  \end{array}$}
 & 1 & $0$ & $0$ & $0$ & $0$ & $1$ & $0$ & $0$ & $0$ & 16 & 8\\
 & 1 & $0$ & $0$ & $0$ & $0$ & $0$ & $1$ & $0$ & $0$ & 16 & 8\\
 & 1 & $0$ & $0$ & $0$ & $0$ & $0$ & $0$ & $1$ & $0$ & 16 & 8\\
 & 1 & $0$ & $0$ & $0$ & $0$ & $0$ & $0$ & $0$ & $1$ & 16 & 8\\
 & $\sqrt{2}$ & $0$ & $0$ & $0$ & $0$ & $1$ & $1$ & $0$ & $0$ & 16 & 8\\
 & $\sqrt{2}$ & $0$ & $0$ & $0$ & $0$ & $1$ & $-1$ & $0$ & $0$ & 16 & 8\\
 & $\sqrt{2}$ & $0$ & $0$ & $0$ & $0$ & $1$ & $0$ & $1$ & $0$ & 16 & 8\\
 & $\sqrt{2}$ & $0$ & $0$ & $0$ & $0$ & $1$ & $0$ & $-1$ & $0$ & 16 & 8\\
 & $\sqrt{2}$ & $0$ & $0$ & $0$ & $0$ & $0$ & $1$ & $0$ & $1$ & 16 & 8\\
 & $\sqrt{2}$ & $0$ & $0$ & $0$ & $0$ & $0$ & $-1$ & $0$ & $1$ & 16 & 8\\
 & $\sqrt{2}$ & $0$ & $0$ & $0$ & $0$ & $0$ & $0$ & $1$ & $1$ & 16 & 8\\
 & $\sqrt{2}$ & $0$ & $0$ & $0$ & $0$ & $0$ & $0$ & $-1$ & $1$ & 16 & 8\\
 & 2 & $0$ & $0$ & $0$ & $0$ & $1$ & $1$ & $1$ & $1$ & 16 & 8\\
 & 2 & $0$ & $0$ & $0$ & $0$ & $1$ & $-1$ & $-1$ & $1$ & 16 & 8\\
 & 2 & $0$ & $0$ & $0$ & $0$ & $1$ & $1$ & $-1$ & $-1$ & 16 & 8\\
 & 2 & $0$ & $0$ & $0$ & $0$ & $1$ & $-1$ & $1$ & $-1$ & 16 & 8\\
\midrule
\pagebreak
  \multirow{16}{*}{$\begin{array}{c}
  (1_\mathrm{B}4_\mathrm{B})(2_\mathrm{B}3_\mathrm{B})\\
  \text{and 4 others}
  \end{array}$}
 & 2 & $0$ & $1$ & $1$ & $0$ & $0$ & $1$ & $-1$ & $0$ & 16 & 8\\
 & 2 & $1$ & $0$ & $0$ & $-1$ & $-1$ & $0$ & $0$ & $-1$ & 16 & 8\\
 & 2 & $1$ & $0$ & $0$ & $-1$ & $1$ & $0$ & $0$ & $1$ & 16 & 8\\
 & 2 & $0$ & $-1$ & $-1$ & $0$ & $0$ & $1$ & $-1$ & $0$ & 16 & 8\\
  & 2 & $1$ & $0$ & $0$ & $-1$ & $0$ & $1$ & $-1$ & $0$ & 16 & 8\\
 & 2 & $0$ & $1$ & $1$ & $0$ & $1$ & $0$ & $0$ & $1$ & 16 & 8\\
  & 2 & $0$ & $1$ & $1$ & $0$ & $-1$ & $0$ & $0$ & $-1$ & 16 & 8\\
 & 2 & $-1$ & $0$ & $0$ & $1$ & $0$ & $1$ & $-1$ & $0$ & 16 & 8\\
& 2$\sqrt{2}$ & $1$ & $1$ & $1$ & $-1$ & $-1$ & $1$ & $-1$ & $-1$ & 16 & 8\\
 & 2$\sqrt{2}$ & $-1$ & $1$ & $1$ & $1$ & $1$ & $1$ & $-1$ & $1$ & 16 & 8\\ 
 & 2$\sqrt{2}$ & $1$ & $-1$ & $-1$ & $-1$ & $1$ & $1$ & $-1$ & $1$ & 16 & 8\\
 & 2$\sqrt{2}$ & $1$ & $1$ & $1$ & $-1$ & $1$ & $-1$ & $1$ & $1$ & 16 & 8\\
 & 2$\sqrt{2}$ & $1$ & $1$ & $1$ & $-1$ & $1$ & $1$ & $-1$ & $1$ & 16 & 8\\
 & 2$\sqrt{2}$ & $1$ & $-1$ & $-1$ & $-1$ & $-1$ & $1$ & $-1$ & $-1$ & 16 & 8\\
 & 2$\sqrt{2}$ & $-1$ & $1$ & $1$ & $1$ & $-1$ & $1$ & $-1$ & $-1$ & 16 & 8\\
 & 2$\sqrt{2}$ & $1$ & $1$ & $1$ & $-1$ & $-1$ & $-1$ & $1$ & $-1$ & 16 & 8\\
\midrule
  \multirow{16}{*}{$\begin{array}{c}
  (1_\mathrm{A}4_\mathrm{A})(2_\mathrm{A}3_\mathrm{A})\\
  \text{and 4 others}
  \end{array}$}
 & 2 & $0$ & $1$ & $-1$ & $0$ & $0$ & $1$ & $1$ & $0$ & 16 & 8\\
 & 2 & $-1$ & $0$ & $0$ & $-1$ & $1$ & $0$ & $0$ & $-1$ & 16 & 8\\
 & 2 & $1$ & $0$ & $0$ & $1$ & $1$ & $0$ & $0$ & $-1$ & 16 & 8\\
 & 2 & $0$ & $1$ & $-1$ & $0$ & $0$ & $-1$ & $-1$ & $0$ & 16 & 8\\
  & 2 & $0$ & $1$ & $-1$ & $0$ & $1$ & $0$ & $0$ & $-1$ & 16 & 8\\
 & 2 & $1$ & $0$ & $0$ & $1$ & $0$ & $1$ & $1$ & $0$ & 16 & 8\\
  & 2 & $-1$ & $0$ & $0$ & $-1$ & $0$ & $1$ & $1$ & $0$ & 16 & 8\\
 & 2 & $0$ & $1$ & $-1$ & $0$ & $-1$ & $0$ & $0$ & $1$ & 16 & 8\\
& 2$\sqrt{2}$ &  $-1$ & $1$ & $-1$ & $-1$ & $1$ & $1$ & $1$ & $-1$& 16 & 8\\
 & 2$\sqrt{2}$ & $1$ & $1$ & $-1$ & $1$ & $-1$ & $1$ & $1$ & $1$ & 16 & 8\\ 
 & 2$\sqrt{2}$ & $1$ & $1$ & $-1$ & $1$ & $1$ & $-1$ & $-1$ & $-1$ & 16 & 8\\
 & 2$\sqrt{2}$ & $1$ & $-1$ & $1$ & $1$ & $1$ & $1$ & $1$ & $-1$ & 16 & 8\\
 & 2$\sqrt{2}$ & $1$ & $1$ & $-1$ & $1$ & $1$ & $1$ & $1$ & $-1$ & 16 & 8\\
 & 2$\sqrt{2}$ & $-1$ & $1$ & $-1$ & $-1$ & $1$ & $-1$ & $-1$ & $-1$ & 16 & 8\\
 & 2$\sqrt{2}$ & $-1$ & $1$ & $-1$ & $-1$ & $-1$ & $1$ & $1$ & $1$ & 16 & 8\\
 & 2$\sqrt{2}$ & $-1$ & $-1$ & $1$ & $-1$ & $1$ & $1$ & $1$ & $-1$ & 16 & 8\\
\midrule
\pagebreak
  \multirow{4}{*}{$\begin{array}{c}
  (1_\mathrm{B})(2_\mathrm{B}3_\mathrm{B})(4_\mathrm{B})\\
  \text{and 16 others}
  \end{array}$}
 & $2\sqrt{2}$ & $1$ & $1$ & $1$ & $-1$ & $0$ & $0$ & $0$ & $0$ & 8 & 16\\
 & $2\sqrt{2}$ & $1$ & $-1$ & $-1$ & $-1$ & $0$ & $0$ & $0$ & $0$ & 8 & 16\\
 & 4 & $2$ & $0$ & $0$ & $-2$ & $0$ & $0$ & $0$ & $0$ & 8 & 24\\
 & 4 & $0$ & $2$ & $2$ & $0$ & $0$ & $0$ & $0$ & $0$ & 8 & 24\\
\midrule
  \multirow{4}{*}{$\begin{array}{c}
  (1_\mathrm{B})(2_\mathrm{A}3_\mathrm{A})(4_\mathrm{B})\\
  \text{and 16 others}
  \end{array}$}
 & $2\sqrt{2}$ & $1$ & $1$ & $-1$ & $1$ & $0$ & $0$ & $0$ & $0$ & 8 & 16\\
 & $2\sqrt{2}$ & $1$ & $-1$ & $1$ & $1$ & $0$ & $0$ & $0$ & $0$ & 8 & 16\\
 & 4 & $2$ & $0$ & $0$ & $2$ & $0$ & $0$ & $0$ & $0$ & 8 & 24\\
 & 4 & $0$ & $2$ & $-2$ & $0$ & $0$ & $0$ & $0$ & $0$ & 8 & 24\\
\midrule
  \multirow{4}{*}{$\begin{array}{c}
  (1_\mathrm{A})(2_\mathrm{A}3_\mathrm{A})(4_\mathrm{A})\\
  \text{and 16 others}
  \end{array}$}
 & $2\sqrt{2}$ & $0$ & $0$ & $0$ & $0$ & $1$ & $1$ & $1$ & $-1$ & 8 & 16\\
 & $2\sqrt{2}$ & $0$ & $0$ & $0$ & $0$ & $1$ & $-1$ & $-1$ & $-1$ & 8 & 16\\
 & 4 & $0$ & $0$ & $0$ & $0$ & $2$ & $0$ & $0$ & $-2$ & 8 & 24\\
 & 4 & $0$ & $0$ & $0$ & $0$ & $0$ & $2$ & $2$ & $0$ & 8 & 24\\
\midrule
  \multirow{4}{*}{$\begin{array}{c}
  (1_\mathrm{A})(2_\mathrm{B}3_\mathrm{B})(4_\mathrm{A})\\
  \text{and 16 others}
  \end{array}$}
 & $2\sqrt{2}$ & $0$ & $0$ & $0$ & $0$ & $1$ & $1$ & $-1$ & $1$ & 8 & 16\\
 & $2\sqrt{2}$ & $0$ & $0$ & $0$ & $0$ & $1$ & $-1$ & $1$ & $1$ & 8 & 16\\
 & 4 & $0$ & $0$ & $0$ & $0$ & $2$ & $0$ & $0$ & $2$ & 8 & 24\\
 & 4 & $0$ & $0$ & $0$ & $0$ & $0$ & $2$ & $-2$ & $0$ & 8 & 24\\
\bottomrule
\end{longtable}}

\bibliography{worldsheet}{}
\bibliographystyle{JHEP}

\end{document}